\documentclass{article_saj}
\usepackage{graphicx,saj,multicol,subeqnarray}
\usepackage{natbib}
\usepackage{float}
\usepackage{xcolor}
\usepackage{widetext}
\usepackage{url}
\usepackage{bm}
\usepackage{tikz} 
\usepackage{pifont} 
\usepackage{amsfonts}
\usepackage{amssymb}
\usepackage{amsmath,upgreek}
\usepackage{titlesec}
\titlelabel{\thetitle.\quad}
\definecolor{xlinkcolor}{cmyk}{1,0.6,0,0}
\usepackage[bookmarks=false,         
     pdfnewwindow=true,      
     colorlinks=true,    
     linkcolor=xlinkcolor,     
     citecolor=xlinkcolor,     
     filecolor=xlinkcolor,  
     urlcolor=xlinkcolor,      
final=true
]{hyperref}

\def\papertype{\ \hfill\ Editorial}

\def\udc{52}
\begin{document}
\parindent=.5cm
\baselineskip=3.8truemm
\columnsep=.5truecm
\newenvironment{lefteqnarray}{\arraycolsep=0pt\begin{eqnarray}}
{\end{eqnarray}\protect\aftergroup\ignorespaces}
\newenvironment{lefteqnarray*}{\arraycolsep=0pt\begin{eqnarray*}}
{\end{eqnarray*}\protect\aftergroup\ignorespaces}
\newenvironment{leftsubeqnarray}{\arraycolsep=0pt\begin{subeqnarray}}
{\end{subeqnarray}\protect\aftergroup\ignorespaces}
%


\markboth{\eightrm Solar Spectropolarimetric Inversions} 
{\eightrm I. MILI\'{c}}

\begin{strip}

{\ }

\vskip-1cm

\publ

\type

{\ }

\begin{center}
\title{Spectropolarimetric diagnostics of the lower solar atmosphere: \\
a review of inversion techniques}
\end{center}


\authors{I. Mili\'{c}$^{1,2,3}$}

\vskip3mm

\address{$^1$Institute for Solar Physics (KIS), Georges-K\"{o}hler-Allee 401A, 79110, Freiburg im Breisgau, Germany}

\address{$^2$Department of Astronomy, Faculty of Mathematics,
University of Belgrade\break Studentski trg 16, 11000 Belgrade,
Serbia}

\address{$^3$Astronomical Observatory, Volgina 7, 11060 Belgrade 38, Serbia}


\dates{December 3, 2025}{December 22, 2025}


\summary{The Sun, our host star, plays a fundamental role in astrophysics due to its vicinity. To understand the processes taking place in the outermost layers of the Sun, but also, indirectly, in its interior, we conduct high spectral and spatial resolution, high cadence observations of diagnostically important spectral lines. The intensity and the polarization of the electromagnetic radiation in the observed spectral lines are a result of the joint action of absorption, emission, and scattering processes, further altered by the Zeeman and Hanle effects. Spectropolarimetric inversions are the techniques devised for the inference of the physical parameters in the solar atmosphere through the interpretation of spectral line shapes and line polarization. This review aims to provide a pedagogical overview of the necessary radiative transfer formalism and cutting-edge techniques, and to help interested readers enter the field and use some of the concepts in their own research. The secondary aim is a critical assessment of the emerging directions for the development of spectropolarimetric inversions, namely, to discuss the friction between physical realism, numerical efficiency, and modeling of the instrumental effects.} 



\end{strip}

\tenrm


\section{INTRODUCTION}

\indent Compared to other stars, the Sun is extremely close to the Earth. This has direct implications for Earth's habitability and for solar influences on the magnetosphere and space weather \citep[e.g.][]{Temmer_2021_review}. At the same time, it means that the Sun plays a special role in stellar astrophysics, being the only star where we can study the interplay between partially ionized plasma, radiation, and magnetic field in a spatially resolved way. To illustrate that, consider the diffraction limit of the largest solar telescope, Daniel K. Inouye Solar Telescope \citep[DKIST, ][]{Rimmele_2020_DKIST}, which, according to the Rayleigh criterion: 

\begin{equation}
    \theta_{0} = 1.22\frac{\lambda}{D},
\end{equation}
can resolve details down to 20\,km on the solar surface thanks to its 4-m primary mirror\footnote{The solar radius is approximately $7\times10^5$\,km, and the Sun is about $1.5\times10^8$\,km away.}. Here, $\lambda$ is the wavelength of the observed light, and $D$ is the diameter of the primary aperture of the telescope. Just for the sake of comparison, a 10-m telescope, located at Proxima Centauri, would not be able to resolve any detail on the surface of the Sun. This simple argument should emphasize that the Sun is, indeed, a precious source of information about the structure of stars and their atmospheres. 

The focus of this review is the solar atmosphere, that is, the Sun's outermost layers that can be directly observed. While the atmosphere does take up a minuscule amount of solar mass (and, if we focus on the solar photosphere and chromosphere, of its radius), it is of primary interest because it hosts a plethora of interesting physical phenomena \citep[e.g., sunspots, flares, CMEs, various other energetic events, see e.g.][]{Fletcher_2011_SSRV, Carslsson_2019_review, Temmer_2021_review} but also because it is the only, albeit indirect, diagnostics of the events taking place in the solar interior, via the techniques of helioseismology \citep[for a classical review of the topic, see][]{CD_2002_review}. From the author's point of view, the solar atmosphere is also extremely interesting and valuable as a natural laboratory for testing radiative transfer models and developing comprehensive techniques for radiative diagnostics \citep[see e.g.][]{delaCruz_2017_rv}.

Apart from a few specific cases, the information on physical conditions in astrophysical bodies is obtained indirectly, employing some form of remote sensing to infer the temperature, plasma velocity, and magnetic fields from observations of the electromagnetic spectrum of the given body. Despite its vicinity, this is also true for the lower solar atmosphere. However, as the Sun is an extended source, we can interpret not only the received monochromatic flux, but also the angular distribution of the received monochromatic intensity, and consequently the spatial distribution of the emitted radiation. Additionally, solar observations typically exhibit excellent spectral (often of the order of $R = \lambda / \Delta \lambda = 10^5$ or more) and temporal resolution (seconds to minutes). 

The last few decades witnessed a series of successful solar missions and ground-based telescopes focused on observing visible, ultraviolet (UV), and infrared (IR) parts of the spectrum, which continuously advanced our understanding of the Sun and its atmosphere. In space, Hinode Solar Optical Telescope \citep[Hinode/SOT,][]{Tsuneta_2008_Hinode}, Solar Dynamics Observatory \citep[SDO,][]{Pesnell_2012_SDO}, Interface Region Imaging Spectrograph \citep[IRIS,][]{dePointieu_2014_IRIS}, and Solar Orbiter \citep[SO,][]{Mueller_2020_SO} provide an uninterrupted insight into the Sun's ever-changing atmosphere, while the large ground-based and balloon-borne telescopes provide exceptional spatial resolution and versatility in terms of instruments for analyzing the light. Telescopes like the Dunn Solar Telescope \citep[DST,][]{Dunn_1969_DST}, Swedish Solar Telescope \citep[SST,][]{Scharmer_2003_SST}, SUNRISE \citep{Solanki_2010_sunrise}, GREGOR \citep{Schmidt_2012_GREGOR}, and Goode Solar Telescope \citep{Cao_2010_GST} have continuously increased the resolving power, eventually culminating with the Daniel K. Inouye Solar Telescope \citep[DKIST,][]{Rimmele_2020_DKIST}, which is expected to further unveil specifics of energy transport and release in the atmosphere of the Sun. This excellent spatial resolution is complemented by the targeted experiments like the CLASP sounding rocket \citep{CLASP_2016}. The fundamental role in these observational developments is played by the wide range of imaging spectrographs and spectropolarimeters described in an excellent review by \citet{Iglesias_2019_pol}.

This review article aims to showcase the theoretical and numerical formalism needed to interpret these magnificent spectropolarimetric observations. The formalism that we use to model the radiation emerging from the solar atmosphere is known as the theory of polarized radiative transfer. Particular focus is on the spectral lines, regions of the spectrum that correspond to bound-bound transitions in atoms, ions, and molecules. The stark differences in the strengths of different lines, as well as the strong dependence of line opacity on wavelength, allow us to probe the solar atmosphere in depth (see the next section). At the same time, the spectral lines are sensitive to the magnetic field and plasma velocity, opening additional diagnostic approaches. Both of these advantages are naturally accounted for in the numerical methods known as spectropolarimetric inversions.

We want to provide a concise but complete introduction to polarized radiative transfer and spectropolarimetric inversions and to outline some of the future directions for the development of the field. The focus is on the interpretation of the observations with very high spatial and spectral resolution. Of course, the text represents the author's unavoidably biased opinion, and any omissions and mistakes are entirely their responsibility. An interested reader is also directed to other recent reviews on the interpretation of spectropolarimetric observations, like: \citet{delToroIniesta_2016_review}, \citet{delaCruz_2017_rv}, and \citet{TB_2022ARA&A_review}. 

\section{RADIATIVE TRANSFER FORMALISM}

The term radiative transfer denotes the formalism that describes the change of the radiation field as the radiation propagates through an absorbing, emitting, and scattering medium. Here, we use it to calculate the intensity and polarization in spectral lines from a given model of the atmosphere and to devise the techniques that fit a model atmosphere to the observed polarized spectrum. 

\subsection{Unpolarized Case}

\indent In the case of the Sun, which is an extended source, the observed quantity to be interpreted is the emergent specific monochromatic intensity \citep[e.g.][]{HubenyMihalas_2014_SAIII}: 
\begin{align}
    I(x,y,\lambda,t) &= \frac{d E}{dA\,dt\,d\Omega\,d\lambda\cos\theta}\nonumber 
    \label{eq:int}
\end{align}
where $dE$ is the radiative energy transported through the area $dA$ in time $dt$ in the cone $d\Omega=\sin \theta d\theta d\phi$ in the wavelength interval ($\lambda, \lambda+d\lambda$). The direction of the cone in the local reference frame is defined with the angles $\theta,\phi$. In the following discussion, we assume that the observed intensity can be measured and then compared with the numerically predicted one. The goal of \emph{forward modeling} is to predict the values of the observables from the given model atmosphere:
\begin{equation}
    T,p,\rho,\vec{v},\vec{B} (x,y,z,t) \rightarrow I(x,y,\lambda, t).
    \label{eq:mapping}
\end{equation}
This equation already suggests that the $z-$dependence ($z$ being the atmospheric normal) of the physical quantities translates into the wavelength dependence of the emergent intensity. For simplicity, we have assumed here that the observations are performed along the direction parallel to $z$, but the generalization is straightforward. Apart from depending on the physical structure of the atmosphere, the emergent intensity also depends on chemical abundances and atomic constants like the values of energy levels, oscillator strengths, and magnetic Land\'{e} factors. In the subsequent discussion, we will assume that these constants are known, but their knowledge is paramount. The elemental abundances, specifically, are the key to understanding the stellar and galactic evolution \citep[e.g.][]{Asplund_2006_NuPh, Grevesse_2007_composition}. Furthermore, high-resolution observations can be used to infer atomic constants like transition energies and line oscillator strengths \citep[see ][ for recent advances]{Trelles_2021_1.5mic, Vukadinovic_2024_globin}. 

The mapping described by Eq.~\ref{eq:mapping} includes several modeling stages: 
\begin{enumerate}
    \item Understanding the ionization and excitation state of the solar atmosphere, which is needed to calculate the so-called level populations. These are the number densities of the atoms (or molecules) responsible for the absorption and emission processes in the considered spectral range. 
    
    \item Modeling the relevant microphysics that describes absorption, emission, and photon scattering processes, as well as the atomic level splitting, and spectral line polarization introduced by the Zeeman \citep{Zeeman_1897_effect} and Hanle \citep{Hanle_1924_og} effects.
    
    \item Numerically solving the radiative transfer equation along the given lines of sight and at given wavelengths to calculate the emergent spectrum. While this point is conceptually the simplest, it can be coupled to the state of the gas through the so-called non-local thermodynamic equilibrium (non-LTE) effects. This substantially complicates the calculation of the spectra and especially their interpretation.
\end{enumerate}

Over the last century, great efforts have been dedicated to the modeling of the solar and stellar spectra, taking into account all the relevant atomic and molecular physics. We will not exhaustively describe neither the specific physics nor the numerical techniques used to solve the general problem of spectra formation. Instead, we direct the reader toward classical monographs of \citet{Athay_1972_book}, \citet{Rutten_2003_book}, or \citet{HubenyMihalas_2014_SAIII} for the discussion of atmospheric modeling and a general solution of non-LTE spectral line formation, and to \citet{Stenflo_1994_book}, \citet{delToroIniesta_2003_book}, and \citet{LL_2004_bible} for descriptions of the polarized radiative transfer and magnetic field diagnostics. From now on, we will refer to the problem of calculating the spatial and spectral distribution of the radiation intensity from the given atmospheric model as the \emph{forward problem}.


\subsection{Radiative transfer equation}

Mathematically, the emergent intensity is given as the solution of the radiative transfer equation (RTE), which, for the unpolarized light, reads: 
\begin{align}
    &\frac{\partial I(\vec{r}, \vec{n},\lambda,t)}{c \partial t} + \vec{n} \cdot \nabla I(\vec{r}, \vec{n},\lambda,t) = \nonumber \\ &j(\vec{r}, \vec{n},\lambda,t) - \chi(\vec{r}, \vec{n},\lambda,t) I(\vec{r}, \vec{n},\lambda,t).
    \label{eq:rte_general}
\end{align}
Here: 
\begin{equation}
\vec{r} = (x,y,z),
\end{equation}
and: 
\begin{equation}
\vec{n} = (\sin\theta \cos \varphi, \sin \theta \sin \varphi, \cos \theta).    
\end{equation}
The time derivative is usually omitted, assuming that the problem is stationary, i.e., that the light propagates fast compared to the typical timescale of changes in the solar atmosphere. In the upcoming discussion, some of the dependencies might be omitted for brevity, but should be clear from the context. Quantities $j$ and $\chi$ are the macroscopic coefficients of emission and absorption, respectively, and they can be calculated from the state of the gas by accounting for the relevant absorption, emission, and scattering processes (recall items 1 and 2 from the previous subsection).

The typical approach in many analyses is to treat each $(x,y)$ location (i.e., each pixel) in the image separately, and to solve the radiative transfer equation along the line of sight ($z$, but recall that other orientations are possible). Furthermore, Eq.\,\ref{eq:rte_general} is often divided by the opacity $\chi$ to yield:
\begin{equation}
    \mu \frac{dI_\lambda}{d\tau_\lambda} = I_\lambda - S_\lambda,
    \label{eq:rte_taulambda}
\end{equation}
where we have defined the source function as $S_\lambda = {j_\lambda}/{\chi_\lambda}$, monochromatic optical depth as $d\tau_\lambda = - \chi_\lambda dz$, and $\mu=\cos\theta$. The outgoing intensity is then obtained by numerically integrating the above equation. Most approaches rely on the so-called integral formal solution of the radiative transfer equation:
\begin{equation}
    I^+_\lambda = \int_0^{\infty} S_\lambda (\tau_\lambda) e^{-\tau_\lambda/\mu} ~d\tau_\lambda/\mu 
    \label{eq:rte_formal}
\end{equation}
where we have used superscript $+$ to denote the outgoing intensity at the top of the atmosphere. This formulation leads to the approach termed short characteristics, which can also be used for polarized radiation \citep[e.g.][]{OK_1987_SC1, delaCruz_2013_delo, Stepan_2013_PORTA}. Other approaches for integration of Eq\,\ref{eq:rte_formal} are also possible, as shown by \citet{Janett_2017_higherorder}, \citet{Janett_2018_formalsolution}, and other recent works by the same research group. This might seem like a lot of work to solve a simple one-dimensional integral, but we must recall that the radiation often strongly couples with the ionization and excitation state of the plasma, which requires a robust and fast formal solution of RTE \citep[as convincingly pointed out by][]{Auer_2003_fs}.

\begin{figure*}[h!]
    \centering
    \includegraphics[width=0.32\linewidth]{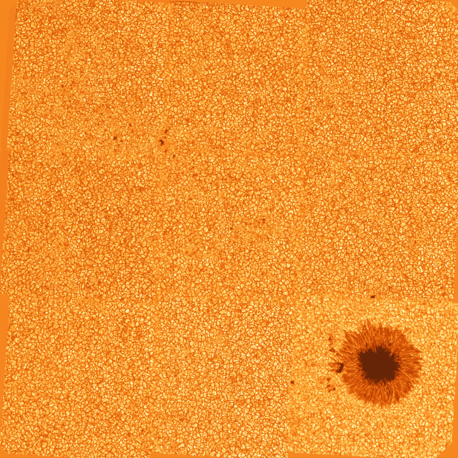}
    \includegraphics[width=0.32\linewidth]{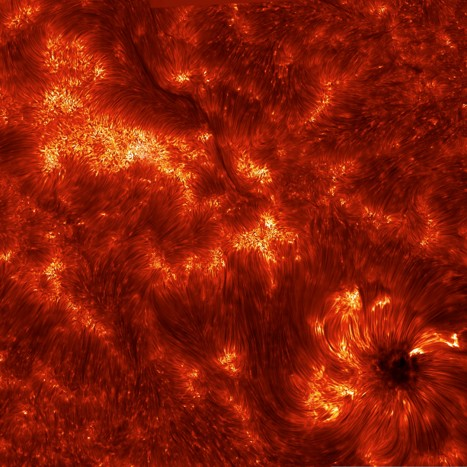}
    \includegraphics[width=0.3215\linewidth]{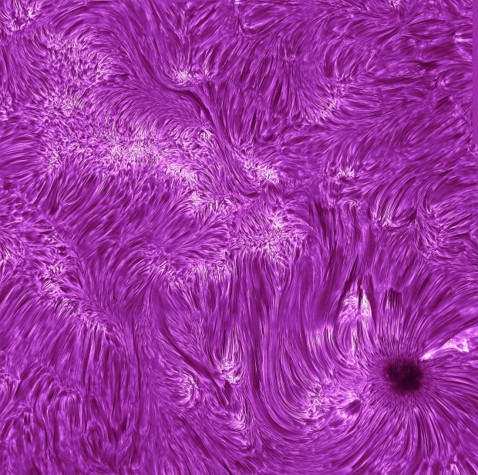}
    \caption{Observations of the solar photosphere and chromosphere at the Dunn solar telescope using the IBIS imaging polarimeter. Left: continuum image at around 430\,nm; Middle: image around the core of the Ca~II~854.2~nm (lower chromosphere); Right: image around the core of the H$\alpha$ spectral line (middle chromosphere). Each image is created from $3\times3$ mosaics and covers roughly $200\times200$ arcseconds.}
    \label{fig:ibis}
\end{figure*}

Before proceeding to discuss specific ingredients of RTE, let us show that our foreshadowing that $\lambda$ somehow maps onto $z$ (recall Eq.\,\ref{eq:mapping}) is justified by Eq.~\ref{eq:rte_formal}. Namely, if $S_\lambda$ slowly varies with $\tau_\lambda$ (say, linearly), the formal solution can be approximated with: 
\begin{equation}
    I^+_\lambda \approx S_\lambda (\tau_\lambda=1),
    \label{eq:eb}
\end{equation}
known as the Eddington-Barbier approximation \citep[e.g.,][]{HubenyMihalas_2014_SAIII}. It states that the outgoing intensity in an optically thick atmosphere can be approximated as the value of the source function at that wavelength, at the physical location where the monochromatic optical depth, again at that wavelength, is equal to one. Note that the solar atmosphere is essentially infinitely optically thick at all the wavelengths, i.e. $\tau_\lambda \gg 1$. Given that spectral line opacity varies dramatically with wavelength, intensity at different wavelengths effectively probes different geometrical locations in the solar atmosphere. While Eq.\,\ref{eq:eb} is not explicitly used in the interpretation, it provides us with a conceptual understanding of how spectral lines enable multi-height diagnostics. To illustrate these points further, consider the mosaics obtained by the Interferometric BIdimensional Spectrometer \citep[IBIS, ][]{Cavallini_2006_IBIS}, a Fabry-P\'{e}rot spectropolarimeter mounted at the DST \citep{Dunn_1969_DST} shown in Fig.~\ref{fig:ibis}. The first panel shows a continuum image that probes the quiet Sun photosphere and a sunspot. The next two panels show the intensity in the core of the chromospheric Ca\,II\,854.2\,nm and H$\alpha$ (656.3\,nm) spectral lines. Solar atmosphere is several orders of magnitude more opaque at these wavelengths than in the continuum, and thus the observations at the cores of these strong spectral lines probe layers some 1000-2000\,km above the photosphere (lower and middle chromosphere, respectively). The difference between these two panels and the first one, that is, between the chromospheric and photospheric structure, is evident and rather striking. See \citet{Cauzzi_2012_mosaic} and \citet{Reardon_2011_ibismagnetic} for more details on these specific observations.

\subsection{Spectral line absorption and emission}

In addition to the high opacity that allows the multi-height atmospheric sampling, the diagnostic power of spectral lines also comes from their sensitivity to temperature, pressure, velocity, and the magnetic field. To understand these sensitivities better, we will briefly recall the expressions for spectral line emissivity:
\begin{equation}
    j_\lambda = n_u A_{ul} \frac{hc}{4\pi \lambda} \phi_\lambda
    \label{eq:linem}
\end{equation}
and opacity:
\begin{equation}
    \chi_\lambda = \left (n_l B_{lu} - n_u B_{ul} \right ) \frac{hc}{4\pi \lambda} \phi_\lambda.
    \label{eq:lineop}
\end{equation}
Here $n_u$ and $n_l$ are the number densities of the atoms in the upper and lower state of the transition (level populations), respectively, $\phi_\lambda$ is the line absorption/emission profile, and $A_{ul}$, $B_{ul}$, and $B_{lu}$ are the Einstein coefficients for spontaneous emission, stimulated emission and absorption, respectively. Note that here, for simplicity, we have assumed that the absorption and emission profiles are identical (the so-called assumption of complete frequency redistribution, CRD). Some strong chromospheric spectral lines require a more general treatment: partial frequency redistribution \citep[PRD, ][]{Hummer_1962_PRD, PaletouA_1995_PRD, Uitenbroek_2001_RH}. The difference between the two can be specifically dramatic in modeling scattering line polarization (see below). In Eqs.~\ref{eq:linem} and \ref{eq:lineop}, Einstein coefficients must be theoretically calculated or measured, being constant for a specific transition.

Now, the sensitivity of the spectral line radiation to the physical parameters comes from the level populations and the line absorption profile $\phi_\lambda$. The latter depends on the temperature (via thermal broadening), pressure (via collisional damping), velocity (via systematic Doppler shifts), and magnetic field (via Zeeman splitting). For more details on the calculation of the line absorption coefficient, see standard textbooks like \citet{Rutten_2003_book}, \citet{delToroIniesta_2003_book}, \cite{Gray_2005_book}, or \citet{HubenyMihalas_2014_SAIII}. The level populations depend on the ionization and excitation state of the solar atmosphere, which are often jointly referred to as the equation of state. This is because one needs to solve for the ionization and excitation state of the gas to find the relationship between pressure, temperature, and mass density \citep[see ][ for more details]{HubenyMihalas_2014_SAIII}. We will now discuss briefly how to calculate the level populations.

\subsection{Ionization and excitation state of the solar atmosphere}
\label{ssec:eos}

In the state of thermodynamic equilibrium, the ionization and excitation state of the gas, as well as the velocity distribution of particles, are all determined by a single temperature and governed by the Saha, Boltzmann, and Maxwell equations, respectively. A common assumption in stellar atmosphere modeling is that such an equilibrium holds locally, yielding the term \emph{local thermodynamic equilibrium} (LTE). In such a state, the ionization and excitation are completely determined by the local values of temperature and pressure, while the radiation field (i.e., the specific monochromatic intensity) is still non-local\footnote{Some sources, somewhat misleadingly, state that in LTE, matter and radiation are in equilibrium. That is certainly not completely true, as the spectrum of such a star would then be a blackbody spectrum. In LTE, the local source function is assumed to be equal to the local Planck function ($B_\lambda$), which simplifies the calculations and interpretations, whereas the specific intensity is to be obtained as the solution of RTE (Eq.\,\ref{eq:rte_taulambda})}. That is: 
\begin{equation}
    n_i(\vec{r}) = f \left ( T(\vec{r}), p(\vec{r}) \right ).
\end{equation}

A far more realistic, hence more complex and, naturally, more interesting case is the state of the non-LTE. Non-LTE refers to any state of the gas that departs from LTE; that is, the non-LTE is defined by what it is not \citep{Rybicki_1991_ALI, Heinzel_1995_nlteprom, Rutten_2003_book}. The main cause of such a departure in a stellar atmosphere is the significant contribution of the radiation to the ionization and excitation processes. Namely, as the pressure drops approximately exponentially with height, at some point, the collisional rates that establish the LTE will drop below the radiative rates. This competition between the collisional and radiative transitions will then cause the decoupling between the local ionization and excitation, and the local temperature. Formally, in the interpretation of the observed (polarized) spectra, this coupling is treated through the assumption of statistical equilibrium (SE), which states that the population of a given level $i$ stays constant in time. This means that the total rate of transitions that populate the level $i$ is equal to the total rate of the transitions that depopulate it, that is: 
\begin{equation}
\sum_{j \ne i} \left( n_j R_{j i} + n_j C_{j i} \right)
-
n_i \sum_{j \ne i} \left( R_{i j} + C_{i j} \right)
= 0.
\label{eq:SE}
\end{equation}
where $n_i$ is the population of level $i$, and $R_{ij}, C_{ij}$ denote the radiative and collisional rates, respectively. For the case of spectral line (i.e., bound-bound) transitions, the radiative rates are:

\begin{align}
R_{ij} &= A_{ij} + B_{ij}\,\bar{J}_{ij} , \nonumber \\
R_{ji} &= B_{ji}\,\bar{J}_{ij} ,
\label{eq:rates}
\end{align}
where we have assumed that $i>j$. Finally: 
\begin{equation}
\bar{J}_{ij} = \bar{J}_{ji} = \oint \int \phi_{\lambda} J_{\lambda}\, d\lambda \sin \theta d\theta d\varphi 
\label{eq:scattering_integral}
\end{equation}
is referred to as the scattering integral. It is evident from Eqs.\,\ref{eq:SE} - \ref{eq:scattering_integral}, that, in non-LTE conditions, the intensity and the collisions jointly determine the level populations and, in turn, the absorption and emission coefficients. The absorption and emission coefficients, however, are needed to calculate the intensity and the scattering integral. We are now dealing with a two-way coupling between the matter and radiation, i.e., we need to solve the SE equations for all relevant species at all relevant atmospheric depths simultaneously with the radiative transfer equation. The main difficulty is that RTE now must be solved in directions and in the atomic transitions that could contribute to the level populations between which the observed line forms, which implies that we need to \emph{model even the transitions we are not observing}. This immediately leads us to the multilevel non-LTE radiative transfer problem. As an example, if one needs to interpret observations of the commonly analyzed Ca\,II\,854.2\,nm spectral line that probes the lower solar chromosphere \citep[see e.g.,][for some recent results]{Kianfar_2020_fibrils, Morosin_2022_heating}, to model the line behavior, one must also model the formation of the two blue resonant lines of ionized calcium (Ca\,II H\&K), even though these two lines were, perhaps, not even observed, or, in general, might not even be observable with the given instrument. 

Here, we analyzed a simplified case of the spectral lines only. In general, the non-LTE conditions also extend to continuum processes \citep[see e.g.,][]{Rybicki_1991_ALI}. That is, the ionization state of the gas also departs from the LTE, which implies that we must also model the continuum processes in detail to interpret the formation of some lines. For example, neutral iron in the photosphere exhibits over-ionization \citep[][]{Shchukina_2005_fenlte, Smitha_2020_NLTEFEI}, which has implications for the strength of neutral iron lines. This again necessitates simultaneous modeling of the observed transition with a comprehensive set of other emissivity and opacity sources, to get the correct ionization state \citep[this complex modeling is sometimes approximated by using the so-called opacity fudge, e.g.,][]{Busa_2001_blanket,  2022A&A...664A.182V}. 

This coupling between the matter and radiation has another important consequence. Namely, in the diagnostics process, when the observed spectrum is assumed to be formed under LTE, the solution of RTE (Eq.\,\ref{eq:rte_general}) is done along the line-of-sight (LOS), and, due to the assumption of LTE, only the physical conditions along the line of sight contribute to the observed radiation. In non-LTE, the calculation of the scattering integral requires a calculation of the specific monochromatic intensity in directions other than the line-of-sight, which requires that we account for the contribution of other points in the atmosphere apart from those lying on the LOS. This then leads to a three-dimensional coupling of the matter and radiation, which is notoriously hard to solve in a general case \citep[see ][]{Leenaarts_2009_multi3d, Stepan_2013_PORTA, Osborne_2025_dexRT}. The importance of these 3D effects for the quantitative interpretation of the observations of the non-LTE lines is evident, but somewhat underexplored and not often quantified \citep[for a pioneering effort, see][]{delaCruz_2012_3D1D}.

The solution of the non-LTE problem has been explored in detail over the last several decades and presents a fascinating synthesis of techniques and concepts from atomic, molecular, and quantum physics, linear algebra, numerical calculus, and optimization methods. Groundbreaking work was laid down in the 1970s and 1980s, \citep[see, e.g.][]{Rybicki_1972, Scharmer_1985_NLTE}, followed up by works of \citet{Rybicki_1991_ALI}, \citet{TBFB_1995_GS}, \citet{Atanackovic_1997_fbili}, and \citet{Uitenbroek_2001_RH}, to name a few. It is probably fair to say that we currently understand the non-LTE spectral line formation to a great detail, but that computational limitations still significantly hamper our possibilities to routinely solve the non-LTE line formation in 3D.

Determining the ionization and excitation state of the atmosphere to calculate the opacity and emissivity and to solve the radiative transfer equation allows us to calculate the spectrum in a given direction at given wavelengths, and to compare the result to the observations. The following two subsections will introduce us to the physical phenomena responsible for polarization of the spectrum, while the next section will deal with techniques that routinely extract physical information by comparing the observed and calculated spectra.

\subsection{Polarized radiative transfer}

The polarization state of the observed light is described by the Stokes formalism, introduced to radiative transfer calculations by \citet{Chandrasekhar_1950_book}. The specific monochromatic intensity is now replaced with the monochromatic Stokes vector: 
\begin{equation}
    \vec{I}_\lambda = \begin{pmatrix}
I \\ Q \\ U \\ V
\end{pmatrix}_\lambda
\end{equation}
where $I$ component describes the total intensity of the light, $Q$ and $U$ components describe the linear polarization, and $V$ the circular polarization \citep[for more details, reader is directed to standard optics textbooks, e.g., ][]{Born_1999_book}. The radiative transfer equation now turns into a system of four coupled differential equations \citep{delToroIniesta_2003_book, LL_2004_bible}:
\begin{equation}
\frac{d\vec{I}_\lambda}{d\tau_\lambda}
=
-\,\hat{K}_\lambda\,\vec{I}_\lambda
+ \vec{S}_\lambda ,
\label{eq:rte_polarized}
\end{equation}
where $\hat{K}_\lambda$ is the $4\times4$ polarized propagation matrix, and $\vec{S}_\lambda$ is the polarized source function. The specific terms of $\hat{K}_\lambda$ and $\vec{S}_\lambda$ depend on the specific physics processes responsible for absorption and generation of polarized light. In the lower solar atmosphere, at visible, UV, and IR wavelengths, these physical mechanisms are the Zeeman effect (or, in a more general case, Paschen-Back effect), scattering line polarization, and the Hanle effect.

\subsection{Zeeman effect}

The Zeeman effect \citep{Zeeman_1897_effect} describes the splitting of atomic energy levels in the presence of an external magnetic field. A level with total angular momentum quantum number $J$ splits into $2J+1$ components, where each component exhibits an energy shift of: 
\begin{equation}
    \Delta E = \mu_{\mathrm{B}}\, g_L\, m\, B.
\end{equation}
In this expression, $\mu_{\mathrm{B}} = \frac{e\hbar}{2m_e}$ is the Bohr magneton, 
$g_L$ is the Land\'{e} factor of the level, $m$ is the magnetic quantum number 
associated with the projection of the total angular momentum onto the 
quantization axis, and $B$ denotes the magnetic field strength. 

\begin{figure}
    \centering
    \includegraphics[width=\linewidth]{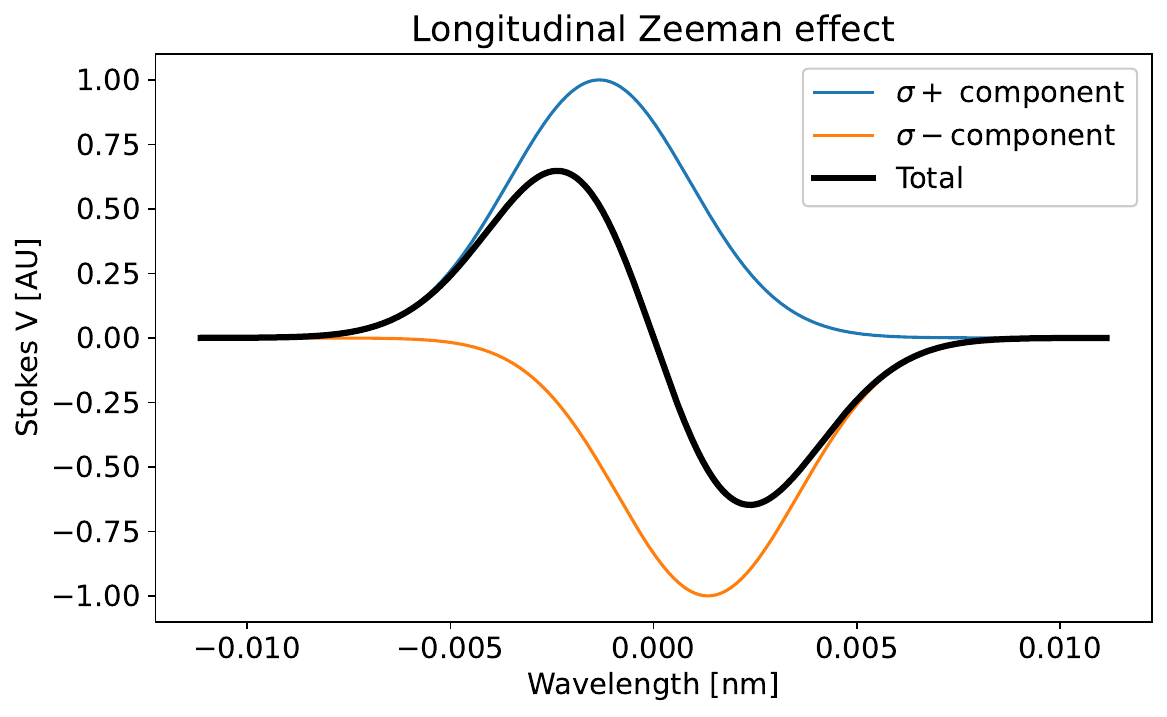}\\
    \includegraphics[width=\linewidth]{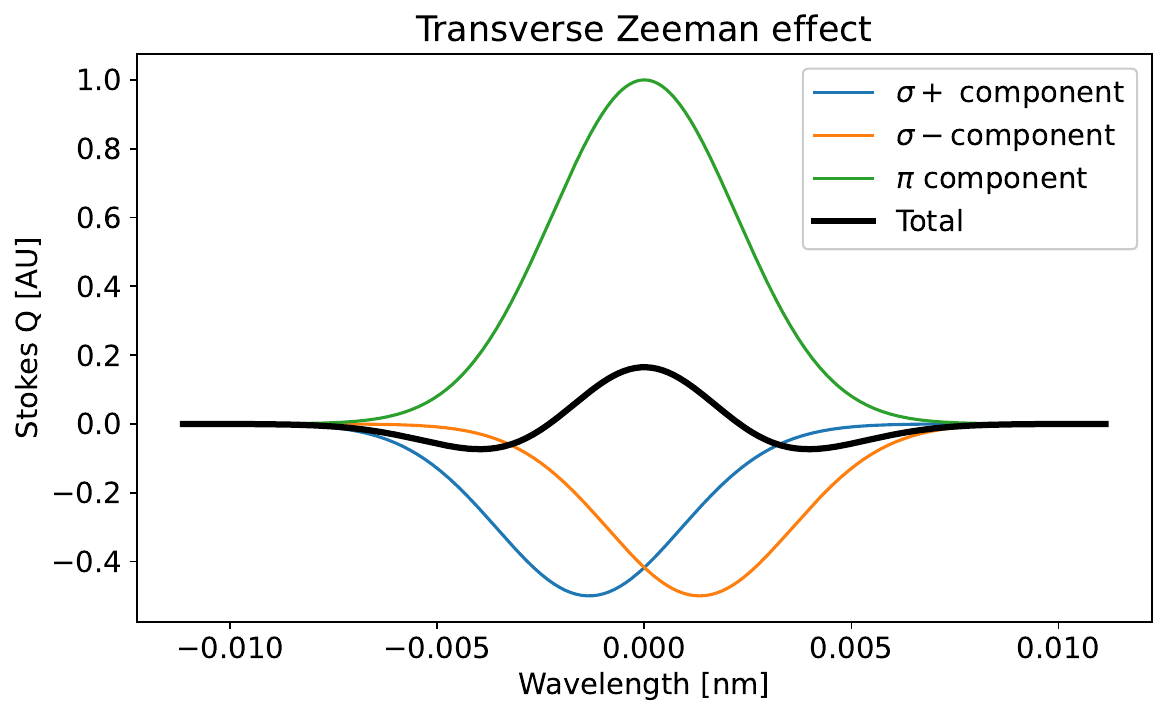}
    \caption{Top: Formation of the Stokes $V$ profile as the sum of two shifted, oppositely polarized Stokes $V$ profiles resulting from the $\sigma+$ and $\sigma -$ components. Bottom: Formation of the Stokes $Q$ profile as the sum of three Stokes $Q$ profiles resulting from the $\pi, \sigma+,\sigma-$ components. The $\sigma$ components are shifted, negatively polarized, and have a two times smaller amplitude than the $\pi$ component.}
    \label{fig:zeeman_formation}
\end{figure}

In the general case, both the upper and the lower levels of the spectral line transition are split into multiple components, enumerated with $m$ and often referred to as the Zeeman sub-levels. Only the transitions between those sub-levels with $\Delta m = 0,\pm 1$ are allowed. These three cases are typically denoted as three components: $\pi$, $\sigma+$, and $\sigma-$, and correspond to different polarizations of the absorbed/emitted photon. In the case where the levels are populated according to LTE, and there is no Zeeman splitting, the spectral line transitions do not change the polarization state of the light. Introduction of the magnetic field then shifts the transitions corresponding to different polarization states, thus resulting in spectral line polarization. Fig.\,\ref{fig:zeeman_formation} depicts the formation of a circular polarization profile (longitudinal Zeeman effect) and a linear polarization profile (transversal Zeeman effect), as a sum of multiple components with different signs of polarization. This figure already suggests that the circular polarization (Stokes $V$) is sensitive to the LOS component of the magnetic field, while the linear polarization (Stokes $Q$ and $U$) is sensitive to the component of the field perpendicular to the LOS. 

The action of the Zeeman effect is introduced through the elements of the absorption matrix $\vec{K}_\lambda$, which has the following form \citep{delToroIniesta_2003_book, LL_2004_bible}:

\begin{equation}
\vec{K}_\lambda =
\begin{pmatrix}
\eta_{I,\lambda} & \eta_{Q,\lambda} & \eta_{U,\lambda} & \eta_{V,\lambda} \\
\eta_{Q,\lambda} & \eta_{I,\lambda} & \rho_{V,\lambda} & -\rho_{U,\lambda} \\
\eta_{U,\lambda} & -\rho_{V,\lambda} & \eta_{I,\lambda} & \rho_{Q,\lambda} \\
\eta_{V,\lambda} & \rho_{U,\lambda} & -\rho_{Q,\lambda} & \eta_{I,\lambda}
\end{pmatrix},
\label{eq:K_form}
\end{equation}
where the $\eta$ terms describe the selective absorption terms: 
\begin{align}
\eta_{I,\lambda} &= \eta_0 \Big[
  \phi_{p,\lambda}\,\sin^2\theta +
  \tfrac12 \big(\phi_{b,\lambda} + \phi_{r,\lambda}\big) \nonumber \\ 
  &\times (1+\cos^2\theta)
  \label{eq:etaI}
\Big], \\[4pt]
\eta_{Q,\lambda} &= \eta_0 \Big[
  \phi_{p,\lambda} -
  \tfrac12 \big(\phi_{b,\lambda} + \phi_{r,\lambda}\big)
\Big] \sin^2\theta\,\cos 2\chi, \\[4pt]
\eta_{U,\lambda} &= \eta_0 \Big[
  \phi_{p,\lambda} -
  \tfrac12 \big(\phi_{b,\lambda} + \phi_{r,\lambda}\big)
\Big] \sin^2\theta\,\sin 2\chi, \\[4pt]
\eta_{V,\lambda} &= \eta_0 \Big[
  \phi_{r,\lambda} - \phi_{b,\lambda}
\Big]\cos\theta,
\end{align}
and the $\rho$ terms describe the so-called anomalous dispersion effects: 
\begin{align}
\rho_{Q,\lambda} &= \eta_0 \Big[
  \psi_{p,\lambda} -
  \tfrac12 \big(\psi_{b,\lambda} + \psi_{r,\lambda}\big)
\Big]\sin^2\theta\,\cos 2\chi, \\[4pt]
\rho_{U,\lambda} &= \eta_0 \Big[
  \psi_{p,\lambda} -
  \tfrac12 \big(\psi_{b,\lambda} + \psi_{r,\lambda}\big)
\Big]\sin^2\theta\,\sin 2\chi, \\[4pt]
\rho_{V,\lambda} &= \eta_0 \Big[
  \psi_{r,\lambda} - \psi_{b,\lambda}
\Big]\cos\theta.
\end{align}

Here $\phi$ and $\psi$ denote the absorption and dispersion profiles (given by the Voigt function and its first derivative), and $\theta$ and $\chi$ are, respectively, the inclination and azimuth of the magnetic field with respect to the LOS. Subscripts $p,b,r$ denote transitions with $\Delta m = 0, +1, -1$, respectively. 

\begin{figure*}
    \centering
    \includegraphics[width=0.95\textwidth]{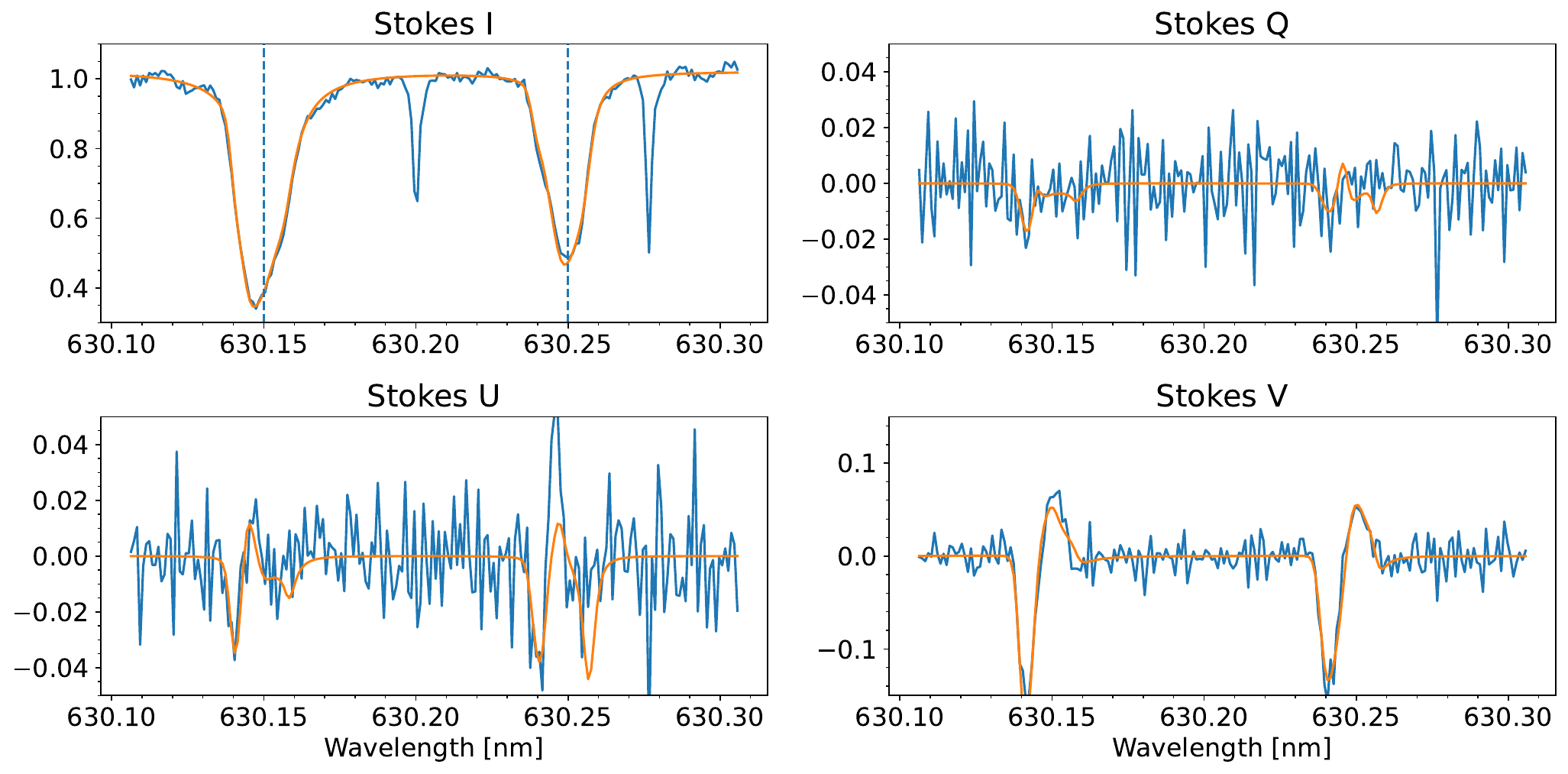}
    \caption{Stokes profiles of neutral iron lines at 630.15 and 630.25~nm, in a solar plage photosphere.  The observations are obtained using the MiHI integral field unit at SST and exhibit very high spatial and spectral resolution and fast cadence. Blue line: observations; Orange line: fits obtained with the FIRTEZ code. Dashed vertical lines denote nominal centers of the lines.}
    \label{fig:zeeman_examples}
\end{figure*}

Fig.\,\ref{fig:zeeman_formation} displays a significantly simplified case where only selective absorption is present. Full modeling (and interpretation) treatment must take into account the full form of $\hat{K}$ (that is, Eq.\,\ref{eq:K_form}) and numerically solve the polarized RTE (Eq.\,\ref{eq:rte_polarized}) to obtain the emergent polarized spectrum from a given model atmosphere. To illustrate the complexity of an observed Stokes profile, in Fig.\,\ref{fig:zeeman_examples} we show observations conducted with the Swedish Solar Telescope \citep{Scharmer_2003_SST}, using Microlensed Hyperspectral Imager \citep[MiHI, ][]{vanNoort2022A&AMihiInstrument, vanNoort2022MiHI3}. These spectropolarimetric observations contain two magnetically sensitive spectral lines of neutral iron around 630\,nm and have been recently presented and analyzed by \citet{Liu_2025_paperI}. The Stokes profiles exhibit notable asymmetries in all four components. Also, note the significantly larger amplitude of the Stokes $V$ component compared to the Stokes $Q$ and $U$. 

We shall digress here a bit and remark on the amount of photon noise present in these observations. Despite the Sun's vicinity, the noise is unavoidable in the observations with very high spatial and spectral resolution, and high cadence, especially when image restoration techniques are used \citep[see, e.g. ][]{vanNoort_2005_momfbd, vanNoort2022MiHI3}. The combination of different sensitivities of different Stokes components and the presence of photon noise leads to notable biases when inferring orientation of the magnetic field pervading the solar atmosphere \citep[see, e.g.,][]{Borrero_2011_photon_noise, Borrero_2012_QS}. These noise effects can be significantly mitigated by employing spatial and temporal regularization (see the discussion in Section\,\ref{sec:developments}).

\subsection{Scattering line polarization and Hanle effect}
\label{ssec:scatpolhanle}

The Zeeman effect leaves an imprint on the polarization through the energy shift of the Zeeman sub-levels. It affects both the spectral lines formed in the LTE and non-LTE conditions and often implicitly assumes that the Zeeman sub-levels are populated according to their equilibrium distribution\footnote{In the non-LTE case, it is often assumed that the fractional distribution of populations of the sub-levels is the same as in LTE.}. Scattering line polarization, on the other hand, is a consequence of an uneven population of the Zeeman sub-levels. This can only happen in non-LTE, when the absorbing/emitting atoms are illuminated by an anisotropic radiation field \citep[often referred to as the anisotropic radiation pumping, e.g.][]{TrujilloBueno_2003_review}. The fully-fledged approach requires the so-called density matrix formalism, which can be found in quantum physics textbooks \citep[e.g.][]{Blum2012texbook, LandiDeglInnocenti2014book} and is presented comprehensively in the ``Bible'' of the solar spectral line polarization, the monograph of \citet{LL_2004_bible}. Here, we opt to capture the gist of the problem following the simplified, pedagogic approaches from \citet{Rees_1987nrt} and \citet{TrujilloBueno_2006_gentle}. 

\begin{figure}
    \centering
    \includegraphics[width=0.4\textwidth]{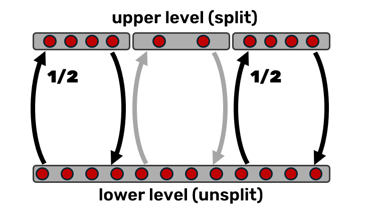}
    \includegraphics[width=\linewidth]{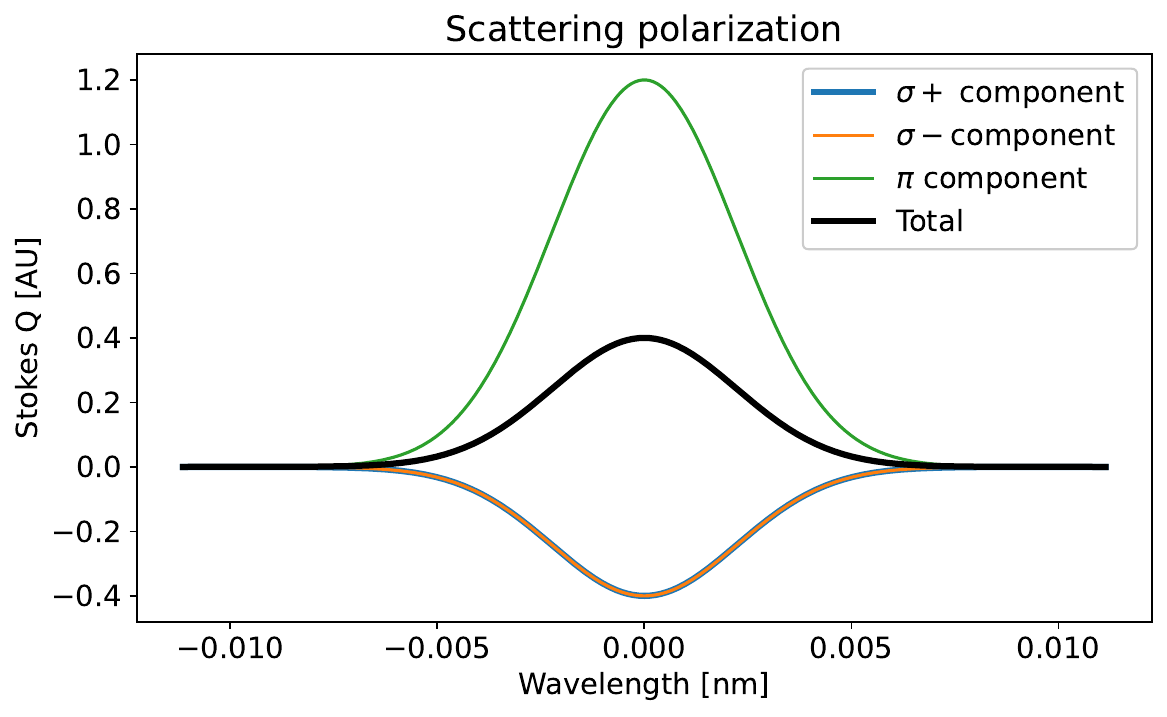}
    \caption{Formation of the linearly polarized profile due to spectral line scattering. Top: Uneven population of the upper level due to the selective absorption of an anisotropic radiation field. Bottom: A plot analogous to the Fig.\,\ref{fig:zeeman_formation}, but instead of the shift of Zeeman sub-levels, we consider their uneven population.}
    \label{fig:scatpol_formation}
\end{figure}

We consider a simple model for the scattering line polarization in the absence of a magnetic field. Upper panel of Fig.\,\ref{fig:scatpol_formation} shows the process of the so-called selective absorption due to the law of conservation of angular momentum. Namely, the photons traveling along the quantization axis for the angular momentum carry the angular momentum of either $-1$ or $1$. This, in turn, means that the absorption of the photon can only excite the $\Delta m = \pm 1$ transitions\footnote{This can already be seen from Eq.\,\ref{eq:etaI}, where substituting $\theta=0$ leads to the presence of only two components.}. This selection rule leads to an overpopulation of specific excited states. In the case of LTE, the collisions are frequent enough to redistribute electrons in the excited state and remove this uneven population. In non-LTE, however, the emission rate competes with the collisional rates, and the population remains uneven, leading to preferential emission of photons with specific polarization states. This, in turn, causes the linear polarization of the photons emitted at an angle with respect to the quantization axis. An example of the resulting profile is given in the lower panel of Fig.\,\ref{fig:scatpol_formation}. Note that, in the absence of the magnetic field, the natural choice for the quantization axis is the $z$ axis, because, in one-dimensional atmosphere models, the radiation field is axially symmetric with respect to $z$. When we introduce the magnetic field to the problem, the magnetic field vector becomes the natural choice for the quantization axis.

Notice that the Zeeman effect only influences the polarized propagation matrix $\vec{K}$ and the thermal emissivity. In the pure Zeeman case, the emission is such that the source function is unpolarized. Interestingly, the emission is still polarized, but fulfills Kirchhoff's law. At first, this can be somewhat confusing (refer to Eq.\,\ref{eq:j_Q} below). The scattering polarization in spectral lines, on the other hand, stems from the anisotropy of the scattered radiation, similar to classical Rayleigh scattering \citep{Rayleigh1871}. This will result in a non-zero source function for polarized terms $Q$ and $U$ in Eq.\,\ref{eq:rte_polarized}. Building up on the example from Fig.\,\ref{fig:scatpol_formation}, we can write the polarized emission term for Stokes $Q$ component as \citep{TrujilloBueno_2006_gentle}: 
\begin{align}
j_Q
&= \frac{hc}{4\pi\lambda} A_{ul} \frac{3}{2}
\left[
    \rho_1(0,0)\,\phi_{p,\lambda}
\right.
\nonumber \\[4pt]
&\qquad\left.
    -
    \frac{
        \rho_1(-1,-1)\,\phi_{r,\lambda}
        + 
        \rho_1(1,1)\,\phi_{b,\lambda}
    }{2}
\right]
\sin^2\theta
\label{eq:j_Q}
\end{align}
where $\rho(M,M)$ is the population of the upper Zeeman sub-level with magnetic quantum number $M$ (note that we swapped from $m$ to $M$, to stay consistent with the literature). In the case when levels are equally populated and profiles are unshifted, the polarized emission is zero. The equal population of the shifted profiles leads to the polarized emission in concordance with Kirchhoff's law \citep{delToroIniesta_2003_book, TrujilloBueno_2006_gentle}. Unequal population with unshifted profiles leads to the rise of scattering line polarization. 

In the more general case, $\rho(M,M')$ are the elements of the atomic density matrix, which describes the so-called orientation and alignment of the atomic levels. It can be decomposed using the multipole components $\rho_Q^K$ \citep{LandiDeglInnocenti2014book}:
\begin{align}
\rho(M,M')
&=
\sum_{K,Q}
(-1)^{J-M}\,
\sqrt{2K+1}\,
\begin{pmatrix}
J & J & K \\
M & -M' & -Q
\end{pmatrix} \nonumber
\\[6pt]
&\qquad\qquad\times\;
\rho^{K}_{Q}(J) \, .
\end{align}
where the tensors $\rho^K_Q$ depend on the irreducible spherical tensors of the radiation field, $J_Q^K$, which describe the anisotropy of the radiation field in the given point and given transition. Finally, the spherical tensors of the radiation field are given by:

\begin{equation}
J^{K}_{Q}
=
\frac{1}{4\pi}
\int \phi_\lambda d\lambda \oint d\Omega
\sum_{i=0}^{3}
T^{K}_{Q}(i,\Omega)\, I_{i,\lambda}(\Omega).
\label{eq:j_qk}
\end{equation}
Here $T_Q^K$ are the spherical tensors for polarimetry \citep[see][]{LL_2004_bible}, angle $\Omega$ describes the direction of propagation of radiation, and $I_i$ counts the components of the Stokes vector of the local radiation field. 

Here, we witness a kind of generalization of the non-LTE problem defined in Section\,\ref{ssec:eos}. Indeed, Eq.\,\ref{eq:j_qk} resembles Eq.\,\ref{eq:scattering_integral} with the same difficulties in modeling. This problem is known as the non-LTE problem of the second kind \citep{LandiDeglInnocenti1983_linepolI, MansoSainz_2003_multilevel2ndkind}. Calculation of $J_Q^K$ values requires knowledge of the full angular dependence of the whole Stokes vector, which in turn often requires a detailed 3D radiative transfer modeling \citep{Stepan_2013_PORTA}. These calculations can often become prohibitively expensive and require efficient ways to solve the polarized radiative transfer equation and to iterate the huge system of equations toward a stable solution \citep[e.g.][]{Benedusi_2021_krylov}. An interesting consequence of the 3D effects is that scattering polarization in inhomogeneous media can be non-zero even for forward scattering geometries, i.e., at the disk center. Accurate modeling was performed by \citet{delpino_2018_sr4607}, and several observational works attempted to confirm these predictions \citep{Zeuner_2018_fsp, Zeuner_2020ApJ_fspII, Zeuner_2025_DKIST}.

The presence of an external magnetic field further alters the scattering polarization. This effect is known as the Hanle effect \citep{Hanle_1924_og}, and it opens interesting new diagnostic techniques. The magnetic field alters the $\rho_Q^K$ quantities described above \citep{TrujilloBueno2009_2nd}:

\begin{equation}
\rho^{K}_{Q}
=
\frac{1}{1 + i\,Q\,\Gamma_u}
\bigl[ \rho^{K}_{Q} \bigr]_{B=0}.
\label{eq:hanle_rhos}
\end{equation}
Here: 
\begin{equation}
\Gamma_u = 8.79 \times 10^6 \frac{B g_u} {A_{ul}}    
\label{eq:hanleparameter}
\end{equation}
is known as the Hanle parameter, and it parametrizes the line's sensitivity to the magnetic field $B$. $A_{ul}$ is the Einstein coefficient of spontaneous emission and $g_u$ is the Land\'{e} factor of the upper level. Looking at Eq.\,\ref{eq:hanle_rhos}, we can surmise that the Hanle effect will decrease the values of $\rho_Q^K$ and thus decrease the amount of polarization. A less obvious result is that it will also \emph{rotate} the plane of polarization. Finally, the Hanle effect can \emph{create} polarization in some scattering geometries that would not exhibit the polarization otherwise, such as the forward scattering \citep[polarization at the disk center, observed and discussed for example by ][]{Bianda_2011_fwd, Belluzzi_2024_ca4227}. Note that the scattering polarization is also affected by depolarizing collisions \citep[e.g.][]{Derouich_2003_coll}, which affects the sensitivity of the line to the Hanle effect. Furthermore, the lines can be, at the same time, affected by both the Zeeman and the Hanle effect. Particularly interesting case is the modulation of the scattering polarization of the wing via the so-called magneto-optical (MO) effects, which opens interesting diagnostic capabilities, as pointed out by \citet{AlsinaB_2016_MO}. Another interesting aspect is the sensitivity of the scattering polarization to the plasma velocities. These effects can dramatically change the anisotropy of radiation illuminating the scattering atoms and thus significantly change the amount of scattering polarization \citep[e.g.,][]{Carlin_2012_vel}.

The general treatment of scattering polarization and the Hanle effect requires an approach that takes into account polarization of both the lower and the upper level, a multi-level or a multi-term treatment, that also takes into account PRD where needed \citep[e.g.,][]{Bommier_1997_masterII, Belluzzi_2012_HPRD}. Note that the general case of polarized line formation in a multi-level atom with PRD is still not satisfactorily described. For a recent development of a unified polarization formalism for electric and magnetic multipole interactions, see \citet{Casini_2025_unified}. Finally, 3D radiative transfer effects are known to be critical, both in the photosphere \citep{delpino_2018_sr4607} and in the chromosphere \citep{Stepan_2016_CaIIscatpol}.

\subsection{Concluding remarks on radiative \\
transfer modeling}

Before we move to diagnostics and spectropolarimetric inversions, it is important to note that, while the problem of the spectral line formation is relatively well understood, its numerical solution in a general case is still not routinely performed. The reason for that is a strong non-linearity in the radiative transfer problem, which often leads to instabilities in numerical solutions. These instabilities are particularly severe when synthesizing spectra from large, high-resolution numerical simulations, which often experience strong vertical gradients. Furthermore, atoms with many levels are often needed (for example, to describe ionization and optical pumping properly), leading to slow and poorly-converging solutions. The other difficulty is the lack of reliable atomic data, such as the Einstein coefficients, and line broadening or depolarizing parameters. Finally, detailed time-dependent radiative transfer calculations that go beyond the assumption of statistical equilibrium \citep[e.g.][]{Leenaarts_2020_rmhd} are also not routinely done. We should keep these difficulties in mind when we discuss the spectropolarimetric inversions and other diagnostics.

\section{INVERSION APPROACHES}
\label{sec:invgen}

\indent The previous section formulated the \emph{forward problem}. That is, the calculation of generally polarized spectra from the three-dimensional structure of the solar atmosphere:
\begin{equation}
\vec{I}_\lambda^+(x,y) = \mathcal{F} \left [ T,p,\rho, \vec{v}, \vec{B}.. = f (x,y,z) \right 
].
\label{eq:forward}
\end{equation}
Here, the functional $\mathcal{F}$ encompasses the whole process of radiative transfer modeling we have just described. We will sometimes refer to it as the \emph{generative model}. Note that some of the atmospheric parameters can, in principle, be calculated from the others (e.g., the mass density can be calculated from the temperature and pressure), or that some parameters might not be needed at all (for example, we might only care about the line-of-sight component of the velocity). The complexity of the atmosphere model also might vary: from a single ``slab'' to 3D atmosphere models. Furthermore, the specific physics involved in $\mathcal{F}$ might vary from case to case. For some spectral lines, the LTE approximation and Zeeman effect will suffice to describe the line polarization satisfactorily, while others might require complex three-dimensional non-LTE calculations with scattering polarization treatment and the Hanle effect. 

The instrumental effects deserve a special mention. Given that Eq.\,\ref{eq:forward} describes the generation of our observations, $\mathcal{F}$ must also involve modeling of changes of the observed intensity introduced by our instruments. The spatial smearing of the intensity, caused by the point spread function (PSF) of the telescope and the Earth's atmosphere, and the spectral smearing of the intensity caused by the limited spectral resolution of imaging spectropolarimeters are probably the most important. A few other, field-specific terms are usually encountered in the discussions on the spectropolarimetric diagnostics. Far wings of the PSF, that are hard to eliminate with adaptive optics or image restoration approaches, are usually termed \emph{stray light} (sometimes also called \emph{false light}). The observed intensity is then polluted by the stray light, a wavelength-dependent contribution to the observed intensity, that slowly varies in the $(x,y)$ plane. If observations contain large, unipolar regions, the stray light can even be polarized. \emph{Scattered light}, on the other side, is usually a wavelength-independent offset in spectra caused by the imperfections inside the wavelength discriminator. Contrary to the stray light, this effect is often relatively straightforward to remove. Note that the usage of these two terms varies across the literature, so here we adopt the particular choice used by the author. Another important observational aspect is the photon noise (and of course systematic measurement errors, which in the most general sense stem from an imperfect knowledge of the generative model $\mathcal{F}$).  For more detailed discussion of observational aspects of high-resolution spectropolarimetry, we refer the reader to the recent review by \citet{Iglesias_2019_pol}.

\subsection{Parameter inference}

We will take a somewhat unorthodox approach to define spectropolarimetric inversions by starting from probabilistic inference. Inference of physical parameters from the observed polarized spectra involves some probabilistic statement on their values, given the observations and their uncertainties. Given a set of observations $\vec{I}^{\rm obs}_\lambda(x,y)$ and a set of model parameters $\vec{\theta}(x,y,z)$, where $\vec{\theta}$ encompasses all the relevant atmospheric quantities, we are looking for the probability density function: 
\begin{equation}
    p \left (\vec{\theta}(x,y,z)| \vec{I}_\lambda(x,y) \right ).
    \label{eq:apost}
\end{equation}
The vertical line in the above equation denotes conditional probability, in the Bayesian sense \citep[see, e.g.][]{Jaynes2003Probability}, while the probability density function $p$ is called the posterior probability of the model parameters, given the data, or simply the posterior. Given the high resolution of today's solar observations, this probability density function is obviously highly dimensional. This renders the traditional sampling methods unusable and significantly limits the application of probabilistic inference in high-resolution solar spectropolarimetry. 

A common approach is  to apply Eq.\,\ref{eq:apost} to one pixel at a time, and then apply the Bayes' theorem \citep{bayes1763essay}, to get:

\begin{equation}
    p\left (\vec{\theta}(z) | \vec{I}^{\rm obs}_\lambda \right ) = \frac{p\left ( \vec{I}^{\rm obs}_\lambda |  \vec{\theta}(z) \right ) p\left (\vec{\theta}(z)  \right )}{p\left ( \vec{I}^{\rm obs}_\lambda \right )}.
    \label{eq:1pixbayest}
\end{equation}
Here, the first term in the numerator is called the likelihood (probability of the data given the model), the second part is the prior probability, and the denominator is referred to as the evidence. If we are interested only in the shape of the posterior, we can neglect the evidence. Calculating the evidence still might be needed for the model selection \citep[e.g.,][]{mackay2003information}. We often assume that $p(\vec{\theta}(z)  )={\rm const}$, and look only for the maximum of the posterior. In that case, the maximum of the posterior and the maximum of the likelihood coincide, and it is enough to maximize the likelihood for the given realization of the data.

In the presence of the uncorrelated Gaussian noise (a good approximation for the Poisson noise when the counts are high), this likelihood has a Gaussian shape. Writing likelihood for one specific Stokes parameter $s$ at one specific wavelength $l$, we get:
\begin{align}
p ( I^{\rm obs}_{s,l} | \vec{\theta}(z) ) =& \frac{1}{\sqrt2{\pi} \sigma_{s,l}} \nonumber \\
\times &\exp{\left [-\frac{\left ( I^{\rm obs}_{s,l} - I^{\rm syn}_{s,l} (\vec{\theta}(z)) \right )^2}{2\sigma_{s,l}^2}\right ]},
\end{align}
where we have emphasized that the synthetic (predicted) intensity $I^{\rm syn}_{s,l}$ depends on values of the model parameters. The likelihood for the whole set of observations ($N_S$ Stokes parameters and $N_L$ wavelengths) is then the product of individual likelihoods. Given that we are usually interested in the model parameters $\vec{\theta}$ that maximize the likelihood, we typically take the logarithm of it, leading to the definition of the well-known $\chi^2$ metric:
\begin{equation}
    \chi^2(\vec{\theta}) = \sum_{s,l} \frac{\left (I^{\rm obs}_{s,l} - I^{\rm syn}_{s,l} (\vec{\theta}) \right )^2}{\sigma_{s,l}^2}.
    \label{eq:chisq}
\end{equation}\
Therefore, under the assumption of uniform priors and Gaussian noise, maximizing the likelihood is equivalent to minimizing $\chi^2$.  Discussing the justification for these two critical assumptions goes somewhat beyond the scope of this paper, but the assumptions should be reconsidered and scrutinized whenever possible. Except for a few notable cases, most of the solar spectropolarimetric diagnostics techniques rely on a minimization of the $\chi^2$ metric for a chosen generative model. Eq.\,\ref{eq:chisq} shows that the $\chi^2$ metric is a multi-dimensional function of the parameter values. The goal is then to find the global minimum of this metric in the parameter space. For linear problems, this minimization can be done analytically, while for non-linear ones it is often achieved through derivative-based optimization schemes \citep[e.g., Levenberg-Marquardt minimization, see][]{Levenberg1944, Marquardt63}, which are never guaranteed to find the global minimum. Examples of some other minimization techniques are genetic algorithms \citep[used e.g. by][]{Lagg_2004_HELIXplus}, the Nelder-Mead simplex algorithm \citep[used by][]{2012A&A...547A..38M} and, very recently, the trust-region reflective algorithm used by \citet{2025A&A...701A..80J}.

While $\chi^2$ minimization can be very efficient, it does not provide meaningful parameter uncertainties or any insight into the degeneracy between the parameters \citep[for an illuminating discussion on the $\chi^2$ approaches, see][]{Andrae_2010_chisq}. A fully Bayesian approach based on sampling would remedy most of these problems \citep[for a primer on model-fitting see][]{Hogg_2010_fitting}, but these techniques are generally too numerically demanding to be applied to spectropolarimetric inversion. One of the inspiring exceptions is an application of hierarchical Bayesian modeling, which makes use of parametrized priors, presented in \citet{AAR_2014_qs} and \citet{AAR_2017_bayesianfibrils}. In Section\,\ref{sec:developments} we will discuss the Bayesian approaches further in the light of contemporary machine learning methods. 

A key ingredient of Eq.\,\ref{eq:chisq} is the measurement uncertainty $\sigma_{s,l}$, simply referred to as the noise. Noise, in principle, should follow from the square root of the number of photon counts for a given Stokes parameter and wavelength. This means that the relative signal-to-noise level in Stokes $Q,U,V$ should be lower than in Stokes $I$ because the polarimetric efficiency is always below unity \citep{Iglesias_2019_pol}. This is seldom the case in practical applications, mostly because of various systematic errors in the measurement of the Stokes $I$ component. These systematic errors often cancel out for other Stokes components, as these components are measured differentially. Furthermore, various diagnostic approaches additionally weight specific Stokes components in Eq.\,\ref{eq:chisq}, to give more emphasis to specific polarization states. This approach is, strictly speaking, not justified, but given the imperfect knowledge of measurement uncertainties and sometimes preference for inferring specific physical parameters (e.g., magnetic field vector), implementation of weights can significantly help the spectropolarimetric diagnostics.

To summarize: the majority of inference techniques in solar spectropolarimetry are based on a choice of a suitable model that generates the observables and depends on a set of physical quantities. Such a model is then fit to the data using a $\chi^2$ minimization technique. We will now discuss a selected set of approaches, ranging from simple models with only a handful of free parameters to more complex, depth-stratified, or even three-dimensional approaches.
 
\subsection{Simple diagnostics}

We are often faced with the situation where we are interested only in a single quantity,  or where the data exhibits large uncertainties or pronounced instrumental effects (e.g., poor spectral resolution or spectral sampling). It can also happen that the formation of the observed spectral lines is so physically complex that the best we can do is to fix, approximate, or parametrize as many aspects as we can. In such cases, we cannot reliably use comprehensive diagnostics that require fully-fledged radiative transfer modeling, described in the previous section, and we revert to simple solutions. We will briefly review some of these simple approaches and mention them again in Section\,\ref{sec:developments}, in the context of machine learning-based approaches. Simplified diagnostics usually focus on the inference of the LOS velocity or the magnetic field vector. 

\subsubsection{Velocity inference}

A common way to infer the velocity in a simplified way is to determine the location of the line core, and calculate the LOS velocity from the Doppler effect formula, as: 
\begin{equation}
    v_{\rm LOS} = \frac{\Delta \lambda}{\lambda_0} c,
    \label{eq:doppler_los_basic}
\end{equation}
where $\Delta \lambda$ is the shift of the line core position with respect to the unshifted line core $\lambda_0$, and $c$ is the speed of light. This inference implicitly assumes that the intensity at the line core is formed at a specific height (the so-called line formation height) in the solar atmosphere, and the location of the line core determines the velocity at that height. This approach is obviously extremely fast and simple, but it neglects many intricacies of the spectral line formation. For example, the line formation height is often ill-defined. This approach is often used to infer the velocities of secondary absorption or emission peaks coming from the observed spectral line, the so-called ``components''. 

Other common approaches to infer the LOS velocity are the center of gravity (CoG) method and bisector analysis \citep[also used in stellar physics, see, e.g.][]{Gray_2005_book}. The latter, specifically, provides some insight into the depth dependence of the LOS velocity in the atmosphere. Again, such depth dependence stems from an analysis where each specific line intensity level is formed in some representative part of the assumed model of the atmosphere. Note that none of these diagnostics explicitly solves the radiative transfer equation.

\subsubsection{Magnetic field inference}

The simplest approaches to estimate the magnetic field from spectral lines are based on a direct measurement of the splitting in the Stokes $I$ component. Needless to say, such an approach works only for very strong magnetic fields \citep[e.g.][]{Okamoto_2018_strong}. Furthermore, the Zeeman effect is degenerate with other sources of spectral line broadening. Still, a rather unique approach was proposed by \citet{AAR_2014_meta} who was able to infer the magnetic fields down to 100\,G magnitude, by analyzing broadening of many spectral lines simultaneously. All other approaches we will discuss rely on the observations and analysis of spectral line polarization.

The weak field approximation (WFA) assumes that the Zeeman broadening $\Delta\lambda_B$ of the observed spectral line is much smaller than the Doppler width $\Delta\lambda_D$. This often happens for broad chromospheric lines where the Zeeman broadening is also small, both due to relatively small Land\'{e} factors and because of weak magnetic fields in the chromosphere. If the field is weak and depth-independent, it can be shown that: 
\begin{equation}
V_\lambda = -4.69 \times 10^{-13} \lambda_0^2 g_l B_{LOS} \frac{\partial I_\lambda}{\partial \lambda}    
\end{equation}
where $\lambda$ is given in \AA, and line-of-sight component of the magnetic field, $B_{LOS}$, in Gauss. Here, $g_l$ is the effective Land\'{e} factor for the considered spectral line. The similarity between the derivative of Stokes $I$, and Stokes $V$ is evident in Fig.\,\ref{fig:zeeman_examples}. WFA can also be extended to linear polarization (Stokes $Q$ and $U$), which is related to the component of the magnetic field perpendicular to the line of sight, $B_\bot$, and its azimuth. These expressions are defined separately for the line core and the line wing \citep[see, e.g.][]{Centeno_2018_caIIwfa}. WFA diagnostics suffer from the lack of depth context: usually, it is assumed that the magnetic field retrieved by the WFA is either the field at some specific depth, or some kind of average field over a line-forming height range. An attempt to formalize the determination of the formation height of the WFA signal was recently undertaken by \citet{2022A&A...664A.182V}.

WFA ceases to be valid already around 1kG for magnetically sensitive photospheric spectral lines. The more suitable approximations, then, are the strong-field approximation and the center of gravity (CoG) method for the magnetic field \citep[e.g.,][]{delToroIniesta_2003_book}. Finally, \citet{Auer_1977_vector} provided expressions for approximate calculations of inclination and azimuth of the magnetic field, directly from the observed Stokes vector. This technique is very robust and is comparable with the results provided by the Milne-Eddington approximation (see the Subsection\,\ref{ssec:ME} below).


\subsubsection{Slab models}

A slab is typically a one-dimensional object with finite optical depth and constant temperature (or the source function), velocity, and the magnetic field vector. The slab model accounts for rudimentary radiative transfer within the slab, through the solution of the radiative transfer equation for a constant source function, which for unpolarized radiation reads:
\begin{equation}
    I^+_\lambda = I^0_\lambda e^{-\tau_\lambda} + S_\lambda (1-e^{-\tau_\lambda}).
    \label{eq:slab}
\end{equation}
A typical slab model is characterized by a total optical depth, source function, LOS velocity, line Doppler width, and sometimes the line damping coefficient. Slab geometry is the basis for the so-called cloud model that was used to interpret $H\alpha$ observations \citep[e.g.][]{Becker_1964_phd, Mein_1996_cloudmodel_new}. A critical aspect of the cloud model, when interpreting on-disk observations, is the choice of the boundary condition $I_\lambda^0$, as pointed out by \citet{Mein_1988_diffcloud}. Note that slab models with much more comprehensive radiative transfer approaches exist, and are used to diagnose plasma conditions in the prominences \citep[e.g.][]{Peat_2021_inversion} or other chromospheric structures \citep[][]{Tziotziou_2007_rev}.

Another very successful implementation of the slab model is the Hanle and Zeeman Light (HAZEL), spectropolarimetric inversion code developed by \citet{AAR_2008_Hazel}. In this approach, the slab model is complemented by the magnetic field vector, and the source function also contains the polarized terms, as described in Section~\ref{ssec:scatpolhanle}. This allowed the authors to model the spectral line formation of 587.76~nm (D3) and 1083.0\,nm lines of neutral helium, taking into account polarization due to the Zeeman and Hanle effects. The solution of the polarized RTE is again, analytical:
\begin{equation}
\vec{I}_\lambda
= e^{-\hat{K}_\lambda \tau}\,\vec{I}_{\lambda}^0
  + \hat{K}_\lambda^{-1}\left( 1 - e^{-\hat{K}_\lambda \tau} \right)\vec{S}_\lambda,
\end{equation}
where $\tau$ is now the optical depth in the line core, and the polarized source function $\vec{S}_\lambda$ is calculated from the angular variation of the radiation illuminating the slab from the photosphere (considered to be fixed for the given height of the slab). HAZEL combines simple geometry with detailed atomic physics and has proven to be an outstandingly versatile and robust code to interpret both Hanle and Zeeman signals in filaments, prominences, spicules, and active regions.

\subsubsection{Milne-Eddington atmosphere}
\label{ssec:ME}

This approach gets its name from the Milne-Eddington approximation, in which the wavelength-independent source function varies linearly with the optical depth in the continuum:
\begin{equation}
    S(\tau_c) = S_0 + S_1 \tau_c,
    \label{eq:me_scalar}
\end{equation}
where $\tau_c$ denotes the source function in the continuum. This solution is motivated by the first analytical solutions to the Milne problem \citep[problem of finding the structure of a gray atmosphere in radiative equilibrium, e.g.][]{Mihalas_1978_SAII}. Spectral lines can be modeled in this approximation by introducing the line strength $\eta$, and expressing monochromatic optical depth as: 
\begin{equation}
    \tau_\lambda = \tau_c(1+ \eta \phi_\lambda),
    \label{eq:me_op}
\end{equation}
where the line absorption profile $\phi_\lambda$ depends on the line Doppler width, dimensionless damping coefficient, and constant LOS velocity. One of the first applications to the inference of the magnetic fields in the photosphere was presented by \citet{Skumanich_1987_inv}. They assumed that the magnetic field is also constant with height. The solution for the emergent Stokes vector is then analytical. This allowed them to fit the nine-parameter model ($B,\theta_B,\chi_B,v_{\rm LOS}, \Delta v_D, a, \eta, S_0, S_1$) to the observations made by the High Altitude Observatory (HAO) scanning Stokes polarimeter \citep{Baur_1980_HAOsp} using a least-squares based approach.

Milne-Eddington (ME) spectropolarimetric inversion is robust and very efficient. It has no limitations in terms of magnetic field strength or orientation, as the Zeeman splitting and the elements of the propagation matrix are calculated exactly. The main limitation is the assumption of constant LOS velocity and the magnetic field vector. Therefore, it is often used in the photosphere, where the spectral lines are formed over a narrower height range. Velocities inferred by ME inversions are usually noticeably lower than the velocities inferred by the depth-stratified inversions (see Section\,\ref{ssec:full_inv}). Similar to other approaches where constant properties with depth are assumed, it is often said that the parameters inferred by ME inversions represent either a mean value over some depth or a value at some specific depth. To better understand the results retrieved by ME inversions, it is recommended to perform a benchmark study for a given spectral line(s) and instrument \citep[see, e.g.][]{Borrero_2014_MEMHD}.

Thanks to its stability and performance, ME is often chosen for pipelines that automatically invert the data for a specific instrument. Some of the examples are MERLIN, which evolved from the original \citet{Skumanich_1987_inv} approach, MILOS \citep{Orozco_2007_milos}, and VFISV \citep{Borrero_2011_VFISV}. The analytical nature of the solution is important not only because the forward model can be computed quickly, but also because the calculation of the derivatives of the observed spectrum with respect to the model parameters is very fast \citep{Orozco_2007_milos}. These derivatives are called \emph{response functions} and are used to drive the $\chi^2$ minimization techniques.

\subsection{Response functions}

A response function $\mathcal{R}^{q}_{s,\lambda}(z)$ of the given Stokes parameter $s$ to the given atmospheric quantity $q$ (temperature, pressure, velocity, magnetic field), is defined as a function that maps a small, generally depth-dependent perturbation of the given parameter to the perturbation of the emergent intensity in the given Stokes parameter. That is:
\begin{equation}
    \delta I^+_{s,\lambda} = \int_0^{\infty} \mathcal{R}^{q}_{s,\lambda}(z) \delta q(z) dz,
\end{equation}
where $\delta_q(z)$ is the perturbation of the given atmospheric parameter, and we have assumed that the integration is done in $z$. One of the first formulations of the response functions (RFs) are due to \citet{Mein_1971_rfs} and \citet{Beckers_1975_RFs}, with subsequent extension to the polarized line formation by \citet{landideglinnocenti_1977_rfs}. 

\begin{figure}
    \centering
    \includegraphics[width=0.99\linewidth]{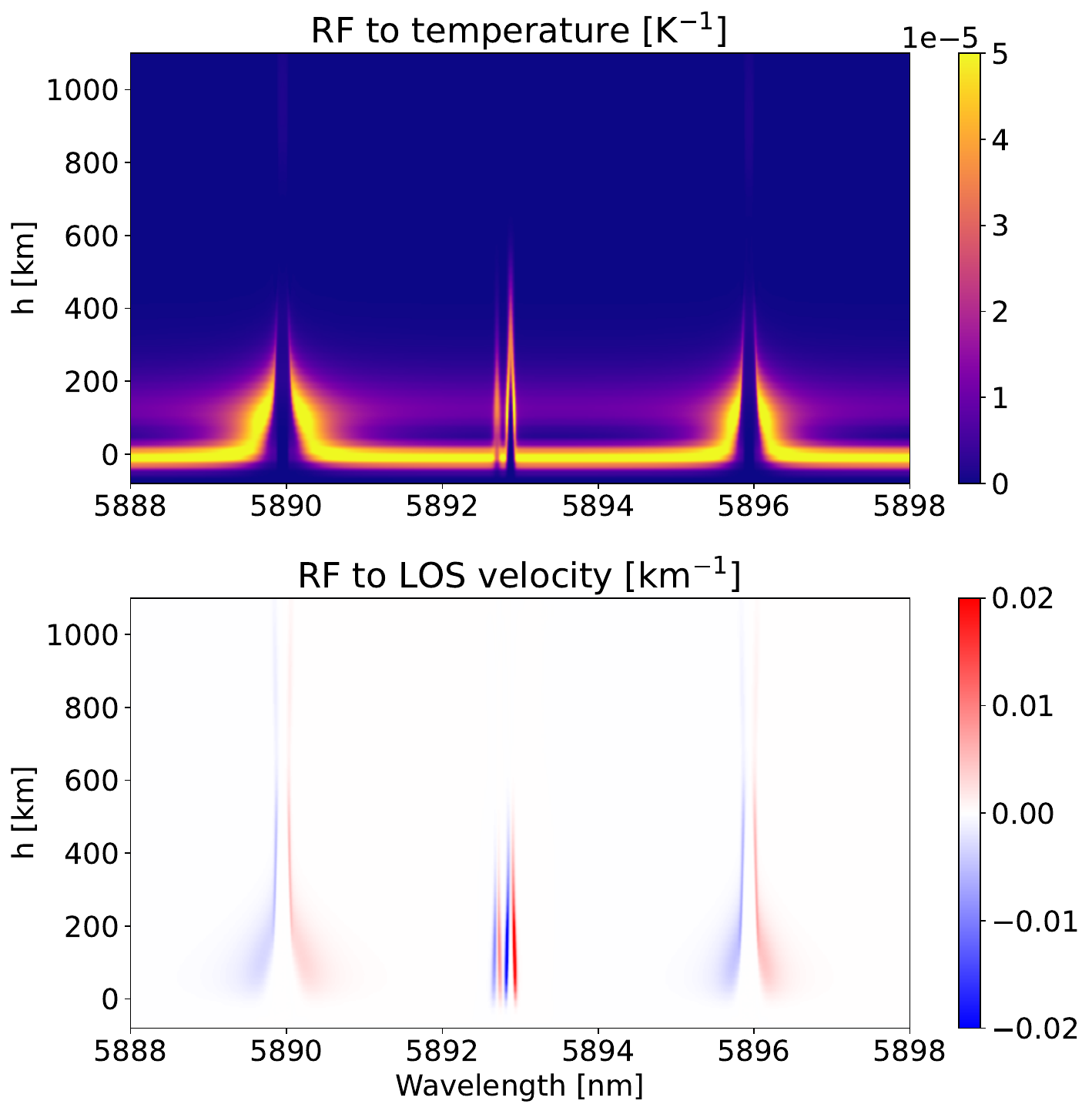}
    \caption{Response functions of the Stokes $I$ component in the spectral region between 588.8\,nm and 589.8\,nm to perturbations in temperature (up) and LOS velocity (down), for the FALC model atmosphere. Note the lack of sensitivity of the Na~I~D lines at 589.0~nm and 589.6~nm to temperature in the upper layers, due to non-LTE effects.}
    \label{fig:rfexamples}
\end{figure}

Conceptually, the response functions quantify the response of the emergent spectrum to the physical parameters at various depths in the atmosphere. They provide an intuitive understanding of how important various parameters at various depths are for the formation of the spectral line. Response functions are usually visualized as 2D plots in the $\lambda,z$ plane (or, more often, in the $\lambda,\tau_c$ plane, as the continuum optical depth $\tau_c$ is often a proxy for height). An example of a response function of the Stokes $I$ component in the 589\,nm region to the perturbations of temperature and the LOS velocity is given in Fig.\,\ref{fig:rfexamples}. The calculations are done with the SNAPI code \citep{2018A&A...617A..24M}, for the FAL-C \citep{Avrett_1993_FALC} model atmosphere. The spectral region contains two strong spectral lines of neutral sodium (Na\,I\,D1 and D2) and two weaker lines of neutral iron and cobalt. All heights are given with respect to the base of the photosphere ($\tau=1$ in the continuum). Note that, even though the Na\,I lines are much stronger than the photospheric ones, its response function drops to zero very fast above $h=200$\,km. This is because the Na\,I\,D lines exhibit dramatic non-LTE effects and their source function decouples from the local temperature very low in the atmosphere. One way to interpret this is the following: Even though the Na\,I\,D1 line is formed around the temperature minimum (say, $h=600$\,km), that radiation is essentially originating in the lower layers and reaching the observer through spectral line scattering processes. Note that the sensitivity of the line to the LOS velocity reaches much higher. This is because the response to the velocity is local, even though the spectral line is formed in non-LTE. This property can be understood better by analyzing Eqs.\,\ref{eq:linem} and \ref{eq:lineop}: the non-LTE effects significantly alter the level populations, but the response of the line absorption/emission coefficient is almost entirely local. Note that the shape of the response function can dramatically depend on the model atmosphere used. 

To discuss the calculation of the response functions, we will cast them in the discrete form, recalling that in practical calculations, the depth variation of physical quantities in the atmosphere is given on a discrete depth grid. We will follow the formulation by \citet{2017A&A...601A.100M}, who cast the response function of the Stokes parameter $s$ at the wavelength index $l$ to the physical parameter $q$ at the depth $d$ in the discrete form as: 
\begin{equation}
    R_{s,l,q_d} = \frac{\partial I^+_{s,l}}{\partial q_d}.
    \label{eq:rfmilic}
\end{equation}
One way to calculate the response function numerically would be to calculate the intensity for a given atmospheric model, perturb the quantity $q$ at the depth $d$ by a small amount $\delta q_d$, and then approximate Eq.\,\ref{eq:rfmilic} as: 
\begin{equation}
    R_{s,l,q_d} = \frac{I^+_{s,l}(\vec{\theta} + \delta q_d) - I^+_{s,l}(\vec{\theta})}{\delta q_d},
    \label{eq:rffindid}
\end{equation}
where $\vec{\theta}$, again, denotes the full set of atmospheric quantities at all the depths, describing the model atmosphere. Alternatively, Eq.\,\ref{eq:rffindid} can be cast in a symmetric form \citep[see, e.g.][]{delaCruz_2017_rv}. These approaches, based on a finite-difference scheme, are very robust but also notoriously numerically demanding. This is especially true for the lines formed in non-LTE, where the forward problem must be solved many times to calculate the needed response functions. In LTE, the response functions can be calculated analytically, with the extension to non-LTE lines formed in the CRD approximation with LTE electron density given by \citet{2017A&A...601A.100M}. We follow their approach and outline the main steps in the calculation of the response functions below. For a more detailed discussion of the LTE case, see also \citet{RuizCobo_1992_sir, ruizcobo1994_sensitivity, delToroIniesta_2016_review}.

For simplicity, we consider the unpolarized case. Eq.\,\ref{eq:rte_formal} can be cast in a discrete form as: 
\begin{equation}
    I^+_{l} = \sum_{d'} w_{l,d'} S_{l,d'}
    \label{eq:numformal}
\end{equation}
where index $d'$ again denotes specific depths. Coefficients $w_{l,d'}$ follow from a specific form of the numerical formal solution \citep[see ][]{OK_1987_SC1, delaCruz_2013_delo, Janett_2018_formalsolution}. The formal solution of RTE depends on the values of opacity and emissivity at the given wavelength at each depth point $d'$. That means that we can calculate the derivative of the emergent intensity $I^+_l$ to the perturbation of quantity $q$ at the depth $d$ as:
\begin{equation}
    \frac{\partial I^+_{l}}{\partial q_d} = \sum_{d'} \left ( \frac{\partial I^+_l}{\partial \chi_{l,d}} \frac{\partial \chi_{l,d}}{\partial q_d} + \frac{\partial I^+_l}{\partial j_{l,d}} \frac{\partial j_{l,d}}{\partial q_d} \right ).
    \label{eq:numformal_deriv}
\end{equation}
Now, the derivatives of the emergent intensity with respect to the values of opacity and emissivity can be found from the specific form of the numerical solution, while the derivatives of the opacity and emissivity to $q_d$ can be found from the exact expressions for the relevant contributors to the opacity and emissivity at the given wavelength. For example, consider the expression for line emissivity given in Eq.~\ref{eq:linem}. We then get:
\begin{equation}
    \frac{\partial j_{l,d}}{\partial q_d} = A_{ul} \frac{hc}{4\pi \lambda} \left ( n_u \frac{\partial \phi_{l,d'}}{\partial q_d} + \frac{\partial n_{u,d'}}{\partial q_d} \phi_{l,d'} \right ).
    \label{eq:deremqd}
\end{equation}
The term $\frac{\partial \phi_{l,d'}}{\partial q_d}$ is evaluated in a relatively straightforward way, as it is reasonable to assume that the response of the profile shape only depends on the local values of physical quantities. The term $\frac{\partial n_{u,d'}}{\partial q_d}$ is referred to as the response function of the level populations. In LTE, it is also strictly local and can be exactly analytically calculated by taking analytical derivatives of the Saha and Boltzmann equations. In non-LTE, this response is non-local (i.e., the population at the depth $d'$ depends on physical quantities at each other depth $d$). Approximate approaches are based on ignoring this non-locality and scaling the LTE responses according to the so-called departure coefficients \citep[fixed departure coefficients, ][]{Socas_1998_nlte, ruizcobo_2022_desire}. \citet{2017A&A...601A.100M} developed a semi-analytical approach based on using the derivative of the statistical equilibrium equation to couple intensity and level population responses and find the level population responses directly from the converged non-LTE populations. The solution is cumbersome, but straightforward, and an interested reader is directed to the original publication for the derivation and benchmarks. A sketch of the extension to response functions for scattering line polarization is given in \citet{2019ASPC..526..179M}.

To summarize this part, the calculation of the response function of the emergent polarized radiation reduces, for the Zeeman case, to the calculation of the responses of level populations to perturbations of the depth-dependent physical quantities. This problem is challenging in the non-LTE case and has not really been explored for more general cases of scattering polarization and the Hanle effect. We will return to the response functions later, and now move to the use of the response functions for depth-stratified inversions.

\subsection{Depth-stratified inversions}
\label{ssec:full_inv}

The depth-stratified inversions aim to infer a full depth variation of temperature, pressure, velocity, and magnetic field from the observed polarized spectrum. Assume we have a guess atmosphere, which we have used to calculate a test spectrum $I^+_{s,l}$. To improve the solution iteratively, we can then use the Gauss-Newton algorithm \citep[e.g.,][]{bjorck1996numerical}:
\begin{equation}
    \sum_j J_{ij} \delta x_j = \delta y_i = y^{\rm obs}_i - y_i
\end{equation}
where the right-hand side corresponds to the residuals between the observed value and current prediction of the model, $x_j$ are the model parameters, and $J_{ij}$ is the Jacobian, i.e., the partial derivative of $y_i$ with respect to $x_i$. The solution for $\delta x_j$ is obtained by solving the generally non-square linear system above. The solution minimizes the sum of squares of the residuals and thus can be used for $\chi^2$ minimization. The Levenberg-Marquardt algorithm mentioned earlier is an improved variant of the Gauss-Newton method. Now, considering our specific problem of polarized line formation, the residuals on the right-hand side are differences between the observed and predicted spectra, and $\delta x_i$ are the desired corrections to the model atmosphere, $\delta q_d$. The Jacobian $J_{ij}$ is then the response function. Therefore, we get:
\begin{equation}
    \sum_{q,d} R_{s,l,q,d} \delta q_d  = \delta I_{s,l}.
    \label{eq:rf_to_solve_system}
\end{equation}
For an alternative approach to deriving Eq.~\ref{eq:rf_to_solve_system}, see \citet{RuizCobo_1992_sir} or \citet{delToroIniesta_2003_book}. 

The system above has $N_S\times N_L$ equations (number of Stokes components times number of wavelengths), and $N_P \times N_D$ unknowns (number of considered physical parameters times the number of depth points). Typically, considered parameters involve temperature, the LOS velocity, and the magnetic field vector. Note that the gas pressure is necessary to solve the Saha equation and obtain level populations, but the pressure is usually calculated from the temperature stratification using the hydrostatic equilibrium \citep[see for example, ][ for an example of this procedure]{Bohm_1989_book}. Also, it is possible to consider additional parameters like microturbulent velocity (additional broadening of the line absorption and emission profile) or macroturbulent velocity (additional broadening of the emergent line intensity profile, usually assumed to be depth-independent). Some approaches can even consider line oscillator strengths as free model parameters \cite[see][]{Vukadinovic_2024_globin}.

Given typical model atmospheres with $50-100$ depth points, the system described by Eq.\,\ref{eq:rf_to_solve_system} is almost guaranteed to be ill-posed. Usually, a regularization approach is employed, which is, in practice, most commonly a reduction of the number of free parameters by using \emph{nodes}. Nodes are fixed points in the atmosphere where the parameters can vary freely. The number of nodes is typically an order of magnitude smaller than the number of depth points, and from given node positions and their values, the full atmosphere can be reconstructed. One must then replace the full depth-dependent response functions in Eq.~\ref{eq:rf_to_solve_system} with the equivalent response functions to the node values \citep{RuizCobo_1992_sir}. Practically all codes for spectropolarimetric inversion in the photosphere and chromosphere employ nodes to simplify the atmosphere model and make the inversion feasible. We briefly summarize the developments over the last three decades below and direct the reader to the reviews by \citet{delToroIniesta_2016_review} and \citet{delaCruz_2017_rv} for more details.

The way toward depth-stratified inversions was paved by the Stokes Inversion based on Response Functions (SIR) code by \citet{RuizCobo_1992_sir}. The first attempts to extend these inversions were done by \citet{Socas_1998_nlte} and \citet{Socas_2000_nltezeeman}. The SPINOR code by \citet{frutiger_2000_spinor} was designed for LTE lines, but it solved the response functions to node values using the finite-differences method. The Stockholm Inversion Code (STiC) \citet{delacruz_2019_stic}, took the same approach, but considered a much more general case of non-LTE spectral lines formed in partial frequency redistribution. STiC used a modified version of the RH code \citep{Uitenbroek_2001_RH} as its engine. Spectropolarimetric non-LTE Analytically Powered Inversion (SNAPI) code by \citet{2018A&A...617A..24M} used the semi-analytical approximation to accelerate the calculation of response functions in non-LTE without using finite differences, but cannot account for the PRD effects. The Departure coefficient aided Stokes Inversion based on Response functions (DeSIRE) code of \citet{ruizcobo_2022_desire} revisits the idea of fixed departure coefficients of \citet{Socas_1998_nlte} to effectively integrate SIR and RH codes and provide another approach to non-LTE inversions. The idea of analytical response functions of LTE and the concept of fixed departure coefficients is also used in the FIRTEZ code \citep{PastorYabar_2019_firtez, Borrero_2019_firtez_I}, the only code so far formulated in the $z$ scale (see the next section for the implications of this choice). All the codes mentioned above are restricted to the spectral line polarization due to the Zeeman effect. The Tenerife Inversion Code (TIC) by \citet{Li_2022_tic} considered a much more general case of partial frequency redistribution and arbitrary magnetic field, taking into account scattering polarization and Hanle effect. A variant of this approach, limited to two-level and two-term atoms, which also accounts for the angle-dependent PRD case, was recently presented by \citet{2025A&A...701A..80J}.

\begin{figure*}
    \centering
    \includegraphics[width=0.85\textwidth]{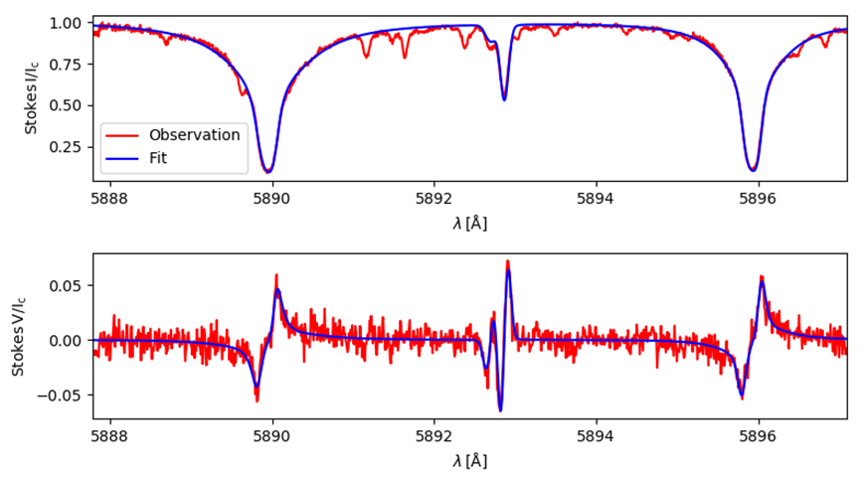}
    \caption{Observed (red) and fitted (blue) spectrum of the four spectral lines found in the region around the Sodium D lines (Left to right: Na~I~D2, Fe~I, Ni~I, Na~I~D1). The upper panel shows the Stokes $I$ and the lower panel the Stokes $V$ spectra. These spectra correspond to an example single pixel from Fig.\,\ref{fig:invmaps}, and an example response function is provided in Fig.\,\ref{fig:rfexamples}.}
    \label{fig:na_spectrum}
\end{figure*}

\begin{figure*}
    \centering
    \includegraphics[width=\textwidth]{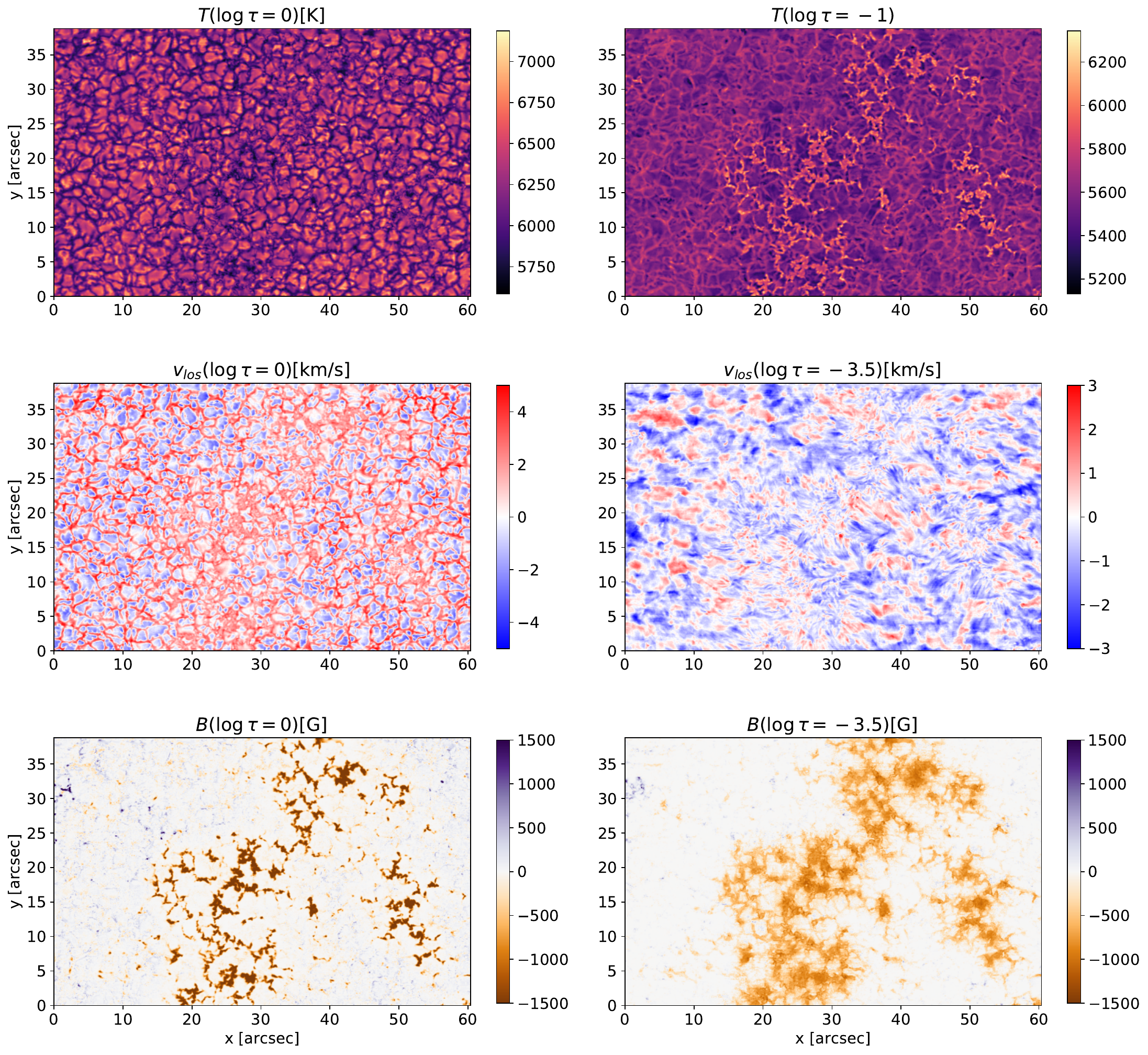}
    \caption{Example of depth-stratified inversion applied to the spectral region showcased in Fig.\,\ref{fig:na_spectrum}. From top to bottom: Temperature, LOS velocity, LOS magnetic field. Left panels showcase the base of the photosphere $\log\tau_c = 0$, while the right panels show $\log\tau_c=-1$ for temperature and $\log\tau_c=-3.5$ for the velocity and the magnetic field.}
    \label{fig:invmaps}
\end{figure*}

An example of a spectropolarimetric inversion applied to the spectral region between 588.8 and 589.8\,nm is shown in Fig.\,\ref{fig:na_spectrum}. The observations have been obtained at the SST using the slit-reconstruction technique developed by \citet{vannoort_2017_specrest}. The blue line is the synthetic polarized profile from the best-fit, depth-dependent model where the temperature, LOS velocity, and LOS magnetic field are allowed to vary with depth. The inversion is performed with the SNAPI inversion code, taking into account non-LTE effects in the Na~I lines. The application to a larger field of view, containing a solar plage is shown in Fig.~\ref{fig:invmaps}. Note the dramatic difference in the spatial structure between the lower (left-hand side panels) and the higher atmospheric layers (right-hand side panels).

Note that the approach presented in this section pertained to one spectrum, containing one or more wavelength regions, possibly with multiple spectral lines, and one atmosphere model. We have implicitly assumed that we are relating one spectrum in a given $x,y$ position to the underlying one-dimensional atmosphere, i.e., we are performing inversions one pixel at a time. There are numerous limitations and pitfalls in this approach, and we move beyond this approximation in the next section. We will now discuss contemporary approaches to spectropolarimetric inversion, many of which are currently under active development.


\section{ONGOING AND FUTURE \\
DEVELOPMENTS}
\label{sec:developments}

The inversion approaches that have been under development over the last decade typically aim to frame the inversion problem as inherently 3D, i.e., to solve the inversion of Eq.~\ref{eq:rte_general} as a coupled, multi-dimensional problem, sometimes even taking time dependence into account. There are multiple reasons why it is advantageous to take into account all observed $x,y$ locations simultaneously. First, the spatial positions are coupled due to the limited resolution of our telescopes. Then, they are bound to be coupled due to the specific physical laws, and, finally, they can be coupled due to the effects of 3D radiative transfer. We discuss these three aspects separately and, at the end, review the machine learning approaches to spectropolarimetric inversion.

\subsection{Spatio-temporal regularization 
\\and coupling}

We already mentioned the spatial coupling between the observed spectra when we discussed the stray light in Section~\ref{sec:invgen}. Namely, the observed spectrum in the given point $\vec{I}_\lambda(x,y)$ can be written as the convolution between the original spectrum $\vec{I}^0_\lambda(x,y)$ and the telescope point spread function (PSF):
\begin{equation}
    \vec{I}_\lambda(x,y) = \int \int \vec{I}^0_\lambda(x',y') PSF(x-x',y-y') dx'dy'.
    \label{eq:psf}
\end{equation}
For space-based telescopes, the PSF is roughly constant and depends only on the shape of the telescope's primary aperture. For ground-based observations, the situation is more complicated: a part of the PSF is removed by adaptive optics \citep[for a review of the topic, see][]{Rimmele_2011_AO}, while another part can be removed via image restoration techniques \citep[e.g.,][]{vanNoort_2005_momfbd}. We first discuss the approaches where only the PSF of the telescope matters, i.e., space-based data. 

The PSF contains a significant non-local contribution, which means that the physical parameters $\vec{\theta}$ at physical locations different than $(x,y)$ also contribute to observations in the given location. This coupling was approximately modeled using various multi-component models and filling factor approaches, but the first one to self-consistently take it into account was \citet{vanNoort_2012_spatiallyc}. The idea is to write Eq.~\ref{eq:rf_to_solve_system} for all observed Stokes spectra simultaneously, taking into account coupling due to the telescope PSF. That is, for the pixel $(i,j)$, we have:
\begin{equation}
    \sum_{i',j'} \sum_{q,d} PSF_{ii'jj'} R_{s,l,i',j',q,d} \delta q_{i',j',d} = \delta I_{s,l,i,j}.
\end{equation}
This results in a very large linear system of dimension $N_P \times N_x \times N_y$ where $N_P$ is the total number of model parameters (in this specific case, the total number of nodes used for depth-stratified polarimetric inversion), and $N_x$ and $N_y$ denote the number of pixels in the image in the $x$ and $y$ dimension, respectively. \citet{vanNoort_2012_spatiallyc} originally solved this problem using iterative solutions of the linear systems, akin to the solution of the non-LTE problem \citep[][]{Olson_1986_ALO}, and implemented the inversion using the framework of SPINOR inversion code \citep{frutiger_2000_spinor}. 

The result is the inversion code that effectively spatially deconvolves the data in the inversion process. The applications were restricted to Hinode/SOT/SP data, given that it is the only high-resolution space-based telescope with spectropolarimetric capabilities. \citet{Danilovic_2016_qs} analyzed the quiet Sun inversions and found that this spatially coupled approach dramatically changes the distribution of inferred weak fields, specifically the biases pointed out by \citet{Borrero_2011_photon_noise} are largely mitigated. Another notable highlight is the detection of super-strong (6~kG and stronger) fields in sunspot penumbrae \citep{vanNoort_2013_penumbra, SiuTapia_2017_coupled} and lightbridges \citep{Castellanos_2025_bridges}. While it is expected that the inversions accounting for the instrument PSF do retrieve stronger fields, these results have not yet been confirmed by high-resolution ground-based observations. While the PSF is only exactly known for space-based instruments, an attempt to apply spatially coupled inversions to ground-based data was recently undertaken by \citet{Borrero_2025_eq}, who achieved significant improvements when inverting the data from the GREGOR telescope.

The spatially coupled approach taking into account the instrument PSF has recently been implemented in a Milne-Eddington inversion framework by \citet{delaCruz_2019_multires}. While the physical model is a bit simplified, the code implements an explicit spatial regularization on top of the spatial coupling. The motivation for the spatial coupling comes from the fact that the presence of photon noise and other inversion degeneracies can create many possible solutions for the spatial distribution of model parameters $\vec{\theta}(x,y)$. In that case, spatially smooth solutions are preferred to those with small-scale fluctuations \citep[this can be interpreted through the sparsity of the model, see ][]{AAR_2015_sparse}. Regularization is usually implemented by minimizing the spatial derivative of physical parameters (i.e., Tikhonov regularization). That is, we define a regularization metric, which has the form:
\begin{equation}
    \chi^2_{\rm reg} = \sum_{i,j,p} \alpha_p\left (  (\frac{\partial \theta_p}{\partial x})_{i,j}^2 + (\frac{\partial \theta_p}{\partial y})_{i,j}^2 \right ).
\end{equation}
Here, the coefficients $\alpha_p$ individually scale the regularization term for each parameter $p$. This regularization term is then added on top of the $\chi^2$ defined by Eq.~\ref{eq:chisq}. The minimization scheme then attempts to find the solution that produces both good fits and relatively spatially smooth solutions. Given that spatial derivatives are non-local quantities, such inversion is also spatially coupled, but the ensuing coupling matrix is much sparser than in the case of the spatial coupling due to the telescope PSF. Other variants of spatial regularization are also possible, for example, penalizing deviations from a reference value. Note that the STiC inversion code \citep{delacruz_2019_stic} employs such regularization schemes also in depth. 

Going back to standard depth-stratified inversions, notice that the nodes are essentially a way to spatially regularize the model in depth. It is curious that the spatial regularization in the $x,y$ plane appeared only some 25 years after the concept of nodes was introduced by \citet{RuizCobo_1992_sir}. The obstacle to its routine implementation is still the necessity of storing and inverting very large matrices (easily $10^6 \times 10^6$ or more). \citet{delacruz_2024_spatemp} recently presented an extension of the spatially coupled approach to the time domain. This is especially relevant with the advent of integral field solutions like the MiHI prototype \citep{vanNoort2022A&AMihiInstrument}. 

The key realization in this aspect of contemporary inversion approaches has been made by \citet{delaCruz_2019_multires}, who pointed out that the model and the observations do not, in fact, have to have identical spatial resolution in the $x,y$ plane. While this realization is perhaps not useful when considering a dataset in a single spectral range from a single instrument, it proves critical when attempting to perform a simultaneous inversion of datasets with different spatial resolutions and spatial samplings. It is then useful to consider a physical model $\vec{\theta}(x,y,z)$ that generates the observable as: 
\begin{equation}
    \vec{I}_\lambda(x,y) = \mathcal{F} [\vec{\theta}(x,y,z)].
\end{equation}
This intensity is then spatially and spectrally sampled by different instruments to provide different datasets that can be, at the same time, used to constrain the properties of the model. In this paradigm, the spatial and spectral sampling of the observations is simply determining the number and location of data points where we test the prediction of our model, while the model resolution can be, strictly speaking, completely independent. These techniques prove critical when interpreting the data from significantly different wavelength domains. For example, when using ALMA millimeter-range observations to better constrain the chromospheric temperatures \citep[for examples, see][]{daSilvaSantos2020_alma, Hofmann_2022_alma}. In the Section~\ref{ssec:3Dinv} we will see how the multi-resolution concept can be developed further.

\subsection{Physics-based regularization}

In the previous subsection, we argued for spatial regularization as it encourages the smoothness of the inferred atmospheric parameters. Another way to choose favorable values of the model parameters is to require that the inferred model atmosphere fulfill specific physical laws. We already encountered one kind of physics-based regularization: depth-stratified inversions eliminate pressure through the (one-dimensional) equation of hydrostatic equilibrium. In some approaches, this equation can also include magnetic or turbulent pressure.  \citet{Riethmuller_2017_MHD} attempted to make the inversion results consistent with 3D MHD (magneto-hydrodynamic) simulations by matching the observed spectra to the simulated ones, starting a simulation from the inverted atmosphere, and running it until relaxation. 

The tension between the physical laws (say, MHD equations) and the inversion problem lies in the fact that the former are cast on a geometric grid, while the radiative transfer equation is cast on an optical depth grid. Also, no inversion approach can establish a common reference point for height over the whole field of view. This can be readily understood if one remembers that an offset of the whole atmosphere, by any height $\Delta z$ does not change the shape of the spectrum at all. One must then develop an inversion code that works in the geometrical height scale and formulate a physical relationship between the neighboring points in the $(x,y)$ plane to establish a common $z-$grid.

The first attempt to develop such inversion code started with the development of the FIRTEZ code by \citet{PastorYabar_2019_firtez}, who formulated the inversion in $z-$scale. The following paper by \citet{Borrero_2019_firtez_I} presented an iterative method that alternates pixel-by-pixel inversion in $z$ scale with an adjustment of the atmospheric structure so that the atmosphere satisfies the magneto-hydrostatic (MHS) equilibrium:
\begin{equation}
    \nabla P_{\mathrm{g}} = -\rho\, g 
    + \frac{1}{4\pi} (\nabla \times \vec{B}) \times \vec{B}.
\end{equation}
Given that the term $\nabla \times \vec{B}$ contains the derivatives of the magnetic field on a Cartesian grid, such a formulation couples the nearby pixels, regularizes the solution, and, similarly to the Tikhonov regularization, makes the problem global. Application of the code to the Hinode/SOT/SP data, presented in \citet{Borrero_2021_mhsII}, demonstrated a robust inference of Wilson depression in sunspots, which is a canonical example for going from $\tau-$scale to $z-$scale. Most recently, \citet{Borrero_2024_mhsiv} extended the inversion approach so that the condition $\nabla \cdot \vec{B} =0 $ is fulfilled. Note that the FIRTEZ approach performs an iterative atmospheric reconstruction by alternating between the inversion steps and physical regularization (i.e., MHS steps). In principle, the physics-based regularization can be formulated similarly to the Tikhonov regularization. For example, the equation of MHS equilibrium can be formulated as a loss by writing:
\begin{equation}
    \chi^2_{\rm phys} = \left (\nabla P_{\mathrm{g}} + \rho\, g 
    - \frac{1}{4\pi} (\nabla \times \vec{B}) \times \vec{B} \right )^2.
\end{equation}
Such a loss can then be added on top of the standard $\chi^2$ and other regularization terms. 

The implementation of the physics constraints to the inversion problem is of paramount importance for obtaining robust and physically meaningful inversion results. We can argue that the path forward is the formulation of the problem less in the form of standard inversion and more in the framework of a data-driven simulation. This is especially true once the time dependence of the observations is taken into account.

\subsection{Inversions with 3D radiative transfer}
\label{ssec:3Dinv}

Finally, the most complex and most interesting reasons for coupling pixels in the observed field of view are due to the radiative transfer. Strictly speaking, any non-LTE treatment requires a 3D approach as the solar atmosphere is inhomogeneous in all three dimensions. As soon as the line is scattering-dominated and the photon mean free path is comparable to the size of the pixel, it can be assumed that the observed radiation, in fact, comes from the nearby pixels and is being scattered toward the observer by the atoms residing in the observed pixel. This effect is particularly important for modeling very strong chromospheric lines \citep[e.g.][]{Bjorgen_2019_3D} and isolated atmospheric structures such as solar prominences \citep{Heinzel_2001_2DRT, Jenkins_2024_25D}. Contrary to the previous two cases, where the coupling was relatively straightforward to formulate, in the case of 3D radiative transfer, the coupling manifests through the non-LTE problem, which means that it is highly non-local and non-linear. The coupling is particularly complex for the case of scattering line polarization. We can now write: 
\begin{equation}
    \vec{I}_\lambda(x,y) = \mathcal{F}_{3D} [\vec{\theta}(x,y,z)],
\end{equation}
where the operator $\mathcal{F}_{3D}$ denotes the whole solution of the 3D non-LTE problem. Assume, similarly to the standard, one-dimensional, depth-stratified case, that we have a current estimate of $\vec{\theta}(x,y,z)$ and we want to use a gradient-based approach to drive the solution toward a suitable local minimum. This requires finding the derivative of the emergent intensity in each $x,y$ location to each physical parameter $\vec{\theta}$ at each physical location in $x',y',z'$. In a one-dimensional approach, these derivatives are zero whenever $(x,y) \neq (x',y')$. In 3D, non-local coupling implies this is no longer the case. Furthermore, the analytical derivative of the operator $\mathcal{F}_{\rm 3D}$ does not exist, and we would have to rely on some finite-difference approach. This is completely unfeasible as it would require a huge number of 3D calculations for each step of the inversion. \citet{Stepan_2022_3Dinv} presented a completely novel approach to tackle this problem, where they break with tradition in several insightful ways. We present the main ideas below and direct the interested reader to the original publication and to the recent new insights published in \citet{Stepan_2025_3Dnlteerrors}. 

First, the approach immediately casts the atmospheric model in a compressed way. Namely, instead of having the atmosphere represented as a set of completely independent discrete values on an $x,y,z$ mesh, the variation of the physical quantities in space (and, in the more general case, time) is represented through a set of basis functions, for example, Legendre polynomials. This presents a generalization of the node concept, which was described earlier. It is important to realize that there is no necessity that the model parameters have to correspond to actual physical values. As long as we define a set of model parameters, $\vec{\alpha}$, that can generate the physical parameters of the atmosphere through some well-defined mapping: 
\begin{equation}
    \vec{\alpha} \rightarrow \vec{\theta}(x,y,z),
    \label{eq:modelparmap}
\end{equation}
we can formulate the forward problem as: 
\begin{equation}
    \vec{I}_\lambda(x,y) = \mathcal{F}'_{3D} [\vec{\alpha}].
\end{equation}

This approach follows the idea of \citet{AAR_2015_sparse} (who used the wavelets or cosine transform for their basis), and significantly reduces the number of free parameters of the model. The smoothness of the atmosphere is ensured, and the change of one model parameter $\alpha_i$ now, technically, changes the physical parameters everywhere in the atmosphere, thus increasing the sensitivity of the emergent spectrum to the model parameters. 

Another deeply insightful aspect of this approach is that the non-LTE problem is not solved explicitly, but rather as another regularization constraint. The level populations are not treated as a quantity that follows from the solution of the non-LTE problem, but as a free parameter that needs to satisfy a physics-based loss cast through the equations of statistical equilibrium (in the general case, for polarized radiation). This means that, contrary to standard non-LTE inversions, the statistical equilibrium equations are not converged in each iteration, but simultaneously with the inversion. The framework then also implements additional physics-based regularizations on top, for example, conservation equations, or $\nabla \cdot \vec{B} = 0$.

Finally, and perhaps most importantly, the method is mesh-free. It is important to realize that, for given model parameters $\vec{\alpha}$, we can retrieve physical parameters of the atmosphere at any point $(x,y,z)$. That means that, for physics-based constraints, the model can be tested at any point in the considered volume \citep[][call them pivot points]{Stepan_2022_3Dinv}. Similarly, given a set of observations at the locations ($x_o, y_o$), we can test which one of these to use to test our model and calculate the $\chi^2$ between the observed and predicted spectra in each iteration. The authors report that the random choice of several test points in each iteration leads to the best convergence properties. The difference in the resolution of the model and the resolution of the observations is reminiscent of the approach of \citet{delaCruz_2019_multires}, which we discussed earlier. 

Such an approach represents a major leap forward in our design of 3D inversion techniques. However, there are still computational limitations, especially if the method is to be extended to the photosphere, where the small-scale structuring of the atmosphere would necessitate too complicated basis to make the model tractable. 

\subsection{Machine learning approaches}

Machine learning techniques experienced a surge in solar physics over the last decade \citep[for a recent comprehensive review, see][]{AAR_2023LRSP}. Given the substantial computational requirements of the spectropolarimetric inversion techniques and the huge amount of data high-resolution solar telescopes can deliver, it is reasonable to expect that machine learning methods, developed to tackle large datasets, will facilitate the interpretation of spectropolarimetric datasets and our understanding of the solar atmosphere. 

\subsubsection{Supervised approaches}

An obvious approach to machine-learning-based spectropolarimetric inversions is to use a supervised method and infer physical parameters from Stokes spectra more quickly and, hopefully, more robustly. A detailed description of the supervised techniques goes beyond the scope of this paper, so we will discuss just the main concept and some specific approaches. The goal of supervised learning is to have the method ``learn'' from a large number of representative examples. The canonical case for such learning is recognition of handwritten digits from a large number of samples that have already been identified by a human. Recent approaches for inversions based on supervised learning were almost exclusively based on neural networks (NNs). An NN consists of a series of neurons that apply a linear transformation and then an activation function. The simplest example is the multi-layer perceptron (MLP). Given an input vector $\vec{x}_i$, we produce the output vector $\vec{x}_{i+1}$ as:
\begin{equation}
    \vec{x}_{i+1} = \sigma( \hat{w} \vec{x_i} + \vec{b}).
\end{equation}
Here, matrix $\hat{w}$ and vector $\vec{b}$ are known as weights and biases, and $\sigma$ denotes a simple, but non-linear activation function which allows the network to capture nonlinearities in the relationship between the input and output. The training process, then, is essentially finding the weights and biases that best reproduce the given training set provided to the network. The recent popularity of neural network applications started with the advent of ``deep'' learning, which stacks a large number (tens or even hundreds) of these neurons (layers) to produce a large and flexible network. Other possibilities for neuron architecture are also used, with the most popular ones being the convolutional layers. 

Application of the NNs to the spectropolarimetric inversion is based on providing a large number of examples of the relationship: 
\begin{equation}
    \vec{I}_\lambda(x,y) = \mathcal{F} [\vec{\theta}(x,y,z)]
    \label{eq:forward_mapping}
\end{equation}
to the network and then assuming that the application to the new data will yield meaningful results. The training examples are chosen either by calculating spectra from numerical simulations of the Sun's atmosphere or by using examples obtained by standard spectropolarimetric inversions. The obvious advantage of this approach is an immense increase in the speed, up to 5 or 6 orders of magnitude. Usually cited disadvantages are the lack of interpretability of the network and the lack of a reliable training set. A way toward overcoming these disadvantages is probably the use of neural networks to accelerate only the non-LTE part of the spectrum synthesis, as shown by \citet{Andreu_2022_GNNs} and \citet{Andreu_2026_GNNs3D}.

Another issue we would like to point out is that, currently, the number of supervised approaches being devised is larger than the number of actual applications where neural networks have produced meaningful results. We refer the reader to the review by \citet{AAR_2023LRSP} for a more thorough review of the topic and point out the works that the author considers to have employed supervised machine learning in a manner that goes beyond standard spectropolarimetric inversion. The most critical aspect of these studies is their focus on the ill-posedness of the inversion process. This means that multiple models can yield almost identical spectra, which poses a fundamental problem in the inversion process. Furthermore, they put substantial effort into providing the uncertainties of the result.

First, \citet{Osborne_2019_radynv} tackled the problem of inverting flare spectra using a deep invertible neural network. Flares are highly energetic and non-stationary phenomena taking place in the solar atmosphere. They are most likely very far away from being in hydrostatic equilibrium, and their structure changes so fast that not even the standard non-LTE approach based on statistical equilibrium is sufficient. To attempt to go beyond this approach, \citet{Osborne_2019_radynv} trained an invertible neural network on example flare spectra coming from RADYN \citet{Carlsson_1997_RADYN} simulations. RADYN is a one-dimensional radiative-hydrodynamics code that takes into account fully time-dependent non-LTE radiative transfer and has been extensively used to model wave propagation and flares in the solar atmosphere. Apart from employing these highly complex models, which cannot be handled by standard inversions, \citet{Osborne_2019_radynv} also recognized the degeneracy of spectropolarimetric inversion and tackled it using an invertible neural network approach in which the space of observables is complemented by an additional, unobservable, latent space that is randomly sampled in the inversion process. This approach allows us to estimate uncertainties of the inversion, which is an aspect that is severely lacking in applications of conventional inversion methods.

\begin{figure*}
    \centering
    \includegraphics[width=0.9\textwidth]{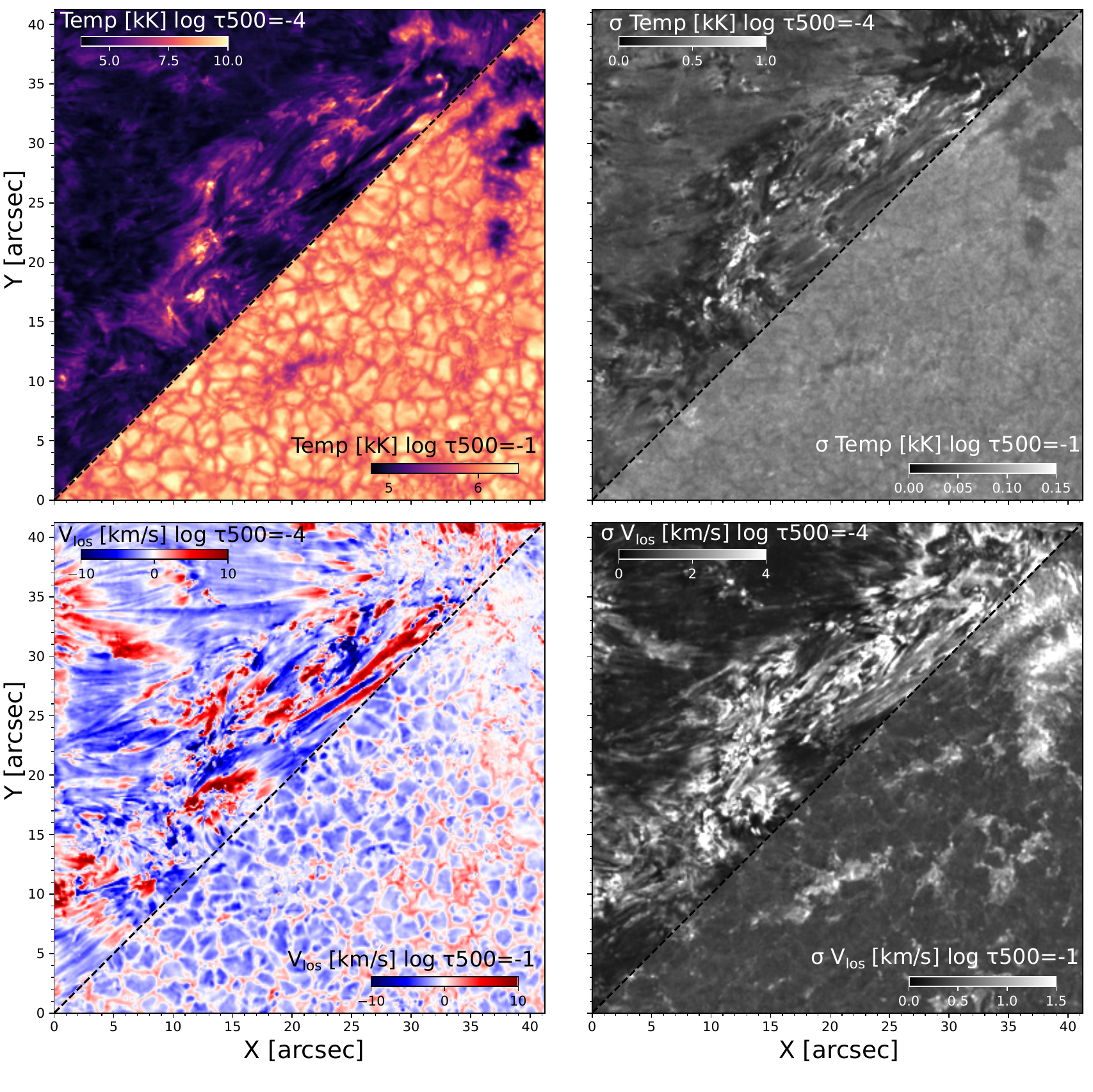}
    \caption{Application of normalizing flows to multi-line observations done with CRisp Imaging Spectro-Polarimeter \citep[CRISP,][]{Scharmer_2008_CRISP}. Top left: Temperature in the chromosphere ($\log\tau_{500}=-4$) and the photosphere ($\log\tau_{500}=-4$); Top right: Map of temperature uncertainties in the chromosphere and the photosphere. Bottom: Same but for the LOS velocity. Adapted with permission from \citet{Diaz_Baso_2022_normflows}.}
    \label{fig:nflows}
\end{figure*}

Another way to use the neural networks to invert solar spectra and obtain reliable uncertainties was proposed by \citet{Diaz_Baso_2022_normflows}, who used the normalizing flow architecture to perform inversion in a Bayesian way. Normalizing flows are NN architectures that map distribution to distribution (instead of value to value). This way, we can devise a supervised NN approach that maps the observations and their uncertainties to the output posterior distribution of model parameters. While the problems with the generalization and training set remain, this approach solves one of the fundamental disadvantages of conventional inversions: we can estimate meaningful uncertainties for large fields of view and study the degeneracies between the parameters using the marginalized posterior distributions. An example of inferred temperature and velocity at two atmospheric optical depths, and their uncertainties, is given in Fig.~\ref{fig:nflows}. Also, there is no fundamental difference in the training and application between spectral lines formed in different physical conditions, as long as a suitable training set is provided. Given that normalizing flows have been rather popular in cosmological parameter inference \citep[e.g.][]{Dai_2022_flows}, it would be interesting to see them used more widely in spectropolarimetric inversions. Note another two very recent ML approaches, designed to tackle the inversion problem while providing meaningful uncertainties, using transformers \citep{AAR_2025_neuraltranslation} and flow matching \citep{AAR_2026_flowmatching}. 

\subsubsection{Physically-Informed Neural Networks (PINNs)}

The allure of supervised training is obvious. We can generate example training data much more quickly and reliably than performing inversions. Furthermore, the degree of physical realism we can include in forward modeling is larger than the spectropolarimetric inversion can account for. However, the fact that the result of such supervised inference always carries the imprint of the training set often causes distrust in the community. The approach of supervised learning can be summed up as: given a large enough sample of $\vec{\theta}$, the network can be trained to perform an inverse mapping of Eq.~\ref{eq:forward_mapping}, where the implicit assumption is that such a mapping is represented for the network for \emph{any given case}. We now present an approach based on physically-informed neural networks \citep[PINNs][]{RAISSI2019686}, which, instead of learning the general mapping $\mathcal{F}^{-1}$, learns only one specific model atmosphere $\vec{\theta}$, constrained by observables $\vec{I}_\lambda(x,y)$ and by a set of appropriate physical laws. Contrary to the supervised techniques, the forward mapping $\mathcal{F}$ is directly implemented in the network.

In the PINN paradigm, the neural network $\mathcal{N}$, parametrized by a set of values $\vec{\alpha}$ (weights, biases, etc.) encodes the physical parameters of the atmosphere as: 
\begin{equation}
    \mathcal{N}(x,y,z,t| \vec{\alpha}) = \vec{\theta}(x,y,z,t).
\end{equation}
That is, given a set of coordinates, the network returns the values of physical parameters in the atmosphere at these coordinates. The network is then trained by providing a set of points where we test the validity of physical laws and a set of observables where we test the prediction of our model $\vec{\theta}$. We then assess how well we have constrained our model by calculating the metric (so-called loss) by adding these two tests in a weighted way and minimizing the loss with respect to the NN parameters $\vec{\alpha}$. Such a network, also known as a neural field, can be extremely flexible and powerful, enabling seamless combination of standard inversion and physics-based regularization. Furthermore, reducing the complexity of the network automatically imposes a spatial regularization. The method was pioneered in solar physics by \citet{Jarolim_2023_pinnex}, who used it for magnetic field extrapolation. It was also used by \citet{AAR_2023_nfs}, who used it to reconstruct the coronal structure from intensity and polarization observations, and \citet{Diaz_Baso_2025_neuralfields} and \citet{Jarolim_2025_pinnme}, who implemented WFA and ME inversions into a neural field framework. 

\begin{figure*}
    \centering
    \includegraphics[width=0.95\textwidth]{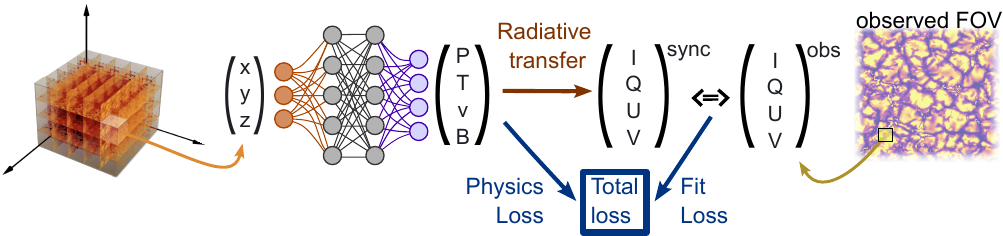}
    \caption{Flowchart of a possible neural field approach to spectropolarimetric inversion: A neural network maps the $(x,y,z)$ coordinates of the observations onto the physical parameters. The radiative transfer module predicts the Stokes spectrum, which is then compared to the observed spectrum. The difference between them is then used, together with physical constraints, to tune the network and reconstruct the model parameters.}
    \label{fig:pinn}
\end{figure*}

An example scheme of a depth-dependent PINN inversion approach is given in Fig.~\ref{fig:pinn}. The neural field maps $(x,y,z)$ coordinates to the values of the physical parameters. That is, the neural field embeds the 3D model of the observed atmosphere. The Stokes profiles are then calculated from the model one $(x,y)$ location at the time, and compared with the observation. Training loss is then constructed from the standard $\chi^2$ metric, and additional physics loss, which reflects the physical consistency of the model. Training the network on this loss then reconstructs the atmospheric structure. Recent attempts to employ PINNs found a substantial increase in the fidelity of the inversions compared to the pixel-by-pixel case. Note that the neural fields achieve this increase through spatio-temporal regularization thanks to sparsity introduced by the network. Conceptually, this is expected to yield similar benefits to standard spatial regularization, but with significantly smaller memory requirements (thanks to the lower number of parameters).

PINNs are also mesh-free, and in that sense can be considered to be an extension of the method by \citet{Stepan_2022_3Dinv}. Their specific advantage is that they make use of the well-tested, optimized, and widely-used frameworks for machine learning, like PyTorch \citep{paszke2019pytorch} and Jax \citep{jax2018github}, which also seamlessly run on graphical processing units (GPUs). An advantage of such frameworks is that the calculation of the derivative of the loss function with respect to model parameters, and in turn, with respect to network parameters $\vec{\alpha}$, is calculated automatically, using the so-called automatic differentiation approach. This eliminates the need for developing a response function and allows a straightforward implementation of various instrumental effects. Currently, the only missing ingredient for a full-fledged, depth-stratified, 3D spectropolarimetric inversion is a robust implementation of polarized spectral line synthesis in a framework like PyTorch or Jax. Also, an extension of these automatic differentiation schemes to non-LTE problems is not obvious, given that non-LTE spectral line synthesis is an iterative procedure in itself. 

\section{CONCLUSIONS}

We presented a condensed but hopefully useful theoretical introduction to the spectropolarimetric diagnostics of the lower solar atmosphere (photosphere and chromosphere). The techniques presented here apply to the observations from space-based, balloon and rocket-borne, and ground-based solar telescopes and instruments that provide spectroscopic and spectropolarimetric data in visible, near-IR, and near-UV domains. The theory presented aims to connect the three-dimensional state of the observed patch of the solar atmosphere with the spatially resolved, polarized spectra observed with an imaging spectropolarimeter. The main challenge lies in finding an inverse of this mapping, i.e., in solving the inverse problem. We presented diagnostic techniques with increasing levels of complexity, culminating in depth-stratified spectropolarimetric inversions. We then turned toward current developments, where we outlined several main directions for improvements: spatial regularization, physics-based regularization, and 3D radiative transfer. We provided a critical review of machine learning techniques, where we paid special attention to the approaches that can provide conceptual improvements over the existing inversion methods. We note that even such a ``solved'' field, like the spectral line formation and diagnostics, experienced a dramatic surge in methods over the last decade.

We end this review with some words of caution. First, despite the Sun's proximity, the diffraction-limited observations with larger and larger telescopes are bound to have lower and lower nominal signal-to-noise ratios. This is because a larger telescope implies larger collecting power, but also smaller resolution elements and shorter exposures, to avoid image smearing due to flows in the solar atmosphere. Furthermore, we may need to apply more aggressive image restoration techniques to achieve the diffraction limit. Therefore, modern inversion techniques must be spatially coupled by design to minimize the effects of this photon noise. Increasing the number of spectral lines considered will also help offset these noise constraints \citep{Riethmuller_2019_multiline, Hoelken_2026_multi}. Second, given improvements in instrumentation and ever-increasing data rates, the upcoming inversion approaches must be built by default for high-performance computing (HPC) facilities. GPUs are bound to play an important role here, as they are a more efficient, more affordable, and more equitable path toward HPC. For more community-related comments on the development of spectropolarimetric inversions, we point the reader toward the recent white paper by \citet{Reardon_2023_white}. Finally, we should always keep in mind the physical assumptions we are making and the physical processes we are unable to take into account. To understand flares, for example, it is required that we move away from hydrostatic equilibrium and toward time-dependent radiative diagnostics, even if it is at the cost of abandoning sophisticated radiative transfer calculations. We must also turn our attention toward the estimation of physical quantities inferred from the inversion results, like the radiative losses \citep{Yadav_2022_radiative}, acoustic fluxes \citep{2024ApJ...976...21D}, magnetic energy \citep{Kaithakkal_2023_ellerman}, or Lorentz force \citep{Borrero_2025_eq}. 

It is unlikely that spectropolarimetric inversions alone will be a path toward understanding specific solar phenomena, but they are, without doubt, the best path forward toward solid, quantitative information on these phenomena. Discussing inversion techniques is always an excellent excuse to paraphrase the famous quote by \citet{Box_1976_SS}: ``All models are wrong (but some are useful)''.



\acknowledgements{I thank my students Teodor Kosti\'{c}, {\fontencoding{T1}\selectfont {\DJ or\dj e} Mijovi\'c} and David \v{Z}ivkovi\'{c}, and Prof. Olga Atanackovi\'{c} for a critical reading of the manuscript. I am indebted to Gioele Janett and Du\v{s}an Vukadinovi\'{c} for their help with specific parts of the text. I gained immeasurable insight from the discussions I had with all my colleagues at the Institute for Solar Physics, National Solar Observatory, and Max Planck Institute for Solar System Research, and especially with all the participants of informal non-LTE workshops organized in 2019, 2023, and 2025. I acknowledge the financial support from the Serbian Ministry of Science and Technology through the grants 451-03-136/2025-03/200104 and 451-03-136/2025-03/200002.
}



\newcommand\eprint{in press }

\bibsep=0pt

\bibliographystyle{aa_url_saj}

{\small

\bibliography{bib_im}

@ARTICLE{Andrae_2010_chisq,
       author = {{Andrae}, Rene and {Schulze-Hartung}, Tim and {Melchior}, Peter},
        title = "{Dos and don'ts of reduced chi-squared}",
      journal = {arXiv e-prints},
         year = 2010,
        month = dec,
          eid = {arXiv:1012.3754},
        pages = {arXiv:1012.3754},
          doi = {10.48550/arXiv.1012.3754},
archivePrefix = {arXiv},
       eprint = {1012.3754},
 primaryClass = {astro-ph.IM},
       adsurl = {https://ui.adsabs.harvard.edu/abs/2010arXiv1012.3754A},
          url = {}
}

@ARTICLE{Asplund_2006_NuPh,
       author = {{Asplund}, Martin and {Grevesse}, Nicolas and {Jacques Sauval}, A.},
        title = "{The solar chemical composition}",
      journal = {\nphysa},
         year = 2006,
        month = oct,
       volume = {777},
        pages = {1-4},
          doi = {10.1016/j.nuclphysa.2005.06.010},
archivePrefix = {arXiv},
       eprint = {astro-ph/0410214},
 primaryClass = {astro-ph},
       adsurl = {https://ui.adsabs.harvard.edu/abs/2006NuPhA.777....1A}
}

@ARTICLE{AAR_2014_meta,
       author = {{Asensio Ramos}, A.},
        title = "{A meta-analysis of the magnetic line broadening in the solar atmosphere}",
      journal = {\aap},
         year = 2014,
        month = mar,
       volume = {563},
          eid = {A114},
        pages = {A114},
          doi = {10.1051/0004-6361/201323166},
archivePrefix = {arXiv},
       eprint = {1401.8111},
 primaryClass = {astro-ph.SR},
       adsurl = {https://ui.adsabs.harvard.edu/abs/2014A&A...563A.114A}
}

@ARTICLE{AAR_2014_qs,
       author = {{Asensio Ramos}, A. and {Mart{\'\i}nez Gonz{\'a}lez}, M.~J.},
        title = "{Hierarchical analysis of the quiet-Sun magnetism}",
      journal = {\aap},
         year = 2014,
        month = dec,
       volume = {572},
          eid = {A98},
        pages = {A98},
          doi = {10.1051/0004-6361/201423860},
archivePrefix = {arXiv},
       eprint = {1410.5953},
 primaryClass = {astro-ph.SR},
       adsurl = {https://ui.adsabs.harvard.edu/abs/2014A&A...572A..98A}
}

@ARTICLE{AAR_2015_sparse,
       author = {{Asensio Ramos}, A. and {de la Cruz Rodr{\'\i}guez}, J.},
        title = "{Sparse inversion of Stokes profiles. I. Two-dimensional Milne-Eddington inversions}",
      journal = {\aap},
         year = 2015,
        month = may,
       volume = {577},
          eid = {A140},
        pages = {A140},
          doi = {10.1051/0004-6361/201425508},
archivePrefix = {arXiv},
       eprint = {1503.07666},
 primaryClass = {astro-ph.SR},
       adsurl = {https://ui.adsabs.harvard.edu/abs/2015A&A...577A.140A}
}

@ARTICLE{AAR_2017_bayesianfibrils,
       author = {{Asensio Ramos}, A. and {de la Cruz Rodr{\'\i}guez}, J. and {Mart{\'\i}nez Gonz{\'a}lez}, M.~J. and {Socas-Navarro}, H.},
        title = "{Inference of the chromospheric magnetic field orientation in the Ca II 8542 {\r{A}} line fibrils}",
      journal = {\aap},
         year = 2017,
        month = mar,
       volume = {599},
          eid = {A133},
        pages = {A133},
          doi = {10.1051/0004-6361/201629755},
archivePrefix = {arXiv},
       eprint = {1612.06088},
 primaryClass = {astro-ph.SR},
       adsurl = {https://ui.adsabs.harvard.edu/abs/2017A&A...599A.133A}
}

@ARTICLE{AlsinaB_2016_MO,
       author = {{Alsina Ballester}, E. and {Belluzzi}, L. and {Trujillo Bueno}, J.},
        title = "{The Magnetic Sensitivity of the Mg II k Line to the Joint Action of Hanle, Zeeman, and Magneto-optical Effects}",
      journal = {\apjl},
         year = 2016,
        month = nov,
       volume = {831},
       number = {2},
          eid = {L15},
        pages = {L15},
          doi = {10.3847/2041-8205/831/2/L15},
archivePrefix = {arXiv},
       eprint = {1610.00649},
 primaryClass = {astro-ph.SR},
       adsurl = {https://ui.adsabs.harvard.edu/abs/2016ApJ...831L..15A}
}

@ARTICLE{AAR_2023LRSP,
       author = {{Asensio Ramos}, Andr{\'e}s and {Cheung}, Mark C.~M. and {Chifu}, Iulia and {Gafeira}, Ricardo},
        title = "{Machine learning in solar physics}",
      journal = {Living Reviews in Solar Physics},
         year = 2023,
        month = dec,
       volume = {20},
       number = {1},
          eid = {4},
        pages = {4},
          doi = {10.1007/s41116-023-00038-x},
archivePrefix = {arXiv},
       eprint = {2306.15308},
 primaryClass = {astro-ph.SR},
       adsurl = {https://ui.adsabs.harvard.edu/abs/2023LRSP...20....4A}
}

@ARTICLE{AAR_2023_nfs,
       author = {{Asensio Ramos}, Andr{\'e}s},
        title = "{Tomographic Reconstruction of the Solar K-Corona Using Neural Fields}",
      journal = {\solphys},
         year = 2023,
        month = nov,
       volume = {298},
       number = {11},
          eid = {135},
        pages = {135},
          doi = {10.1007/s11207-023-02226-2},
       adsurl = {https://ui.adsabs.harvard.edu/abs/2023SoPh..298..135A}
}

@ARTICLE{AAR_2008_Hazel,
       author = {{Asensio Ramos}, A. and {Trujillo Bueno}, J. and {Landi Degl'Innocenti}, E.},
        title = "{Advanced Forward Modeling and Inversion of Stokes Profiles Resulting from the Joint Action of the Hanle and Zeeman Effects}",
      journal = {\apj},
         year = 2008,
        month = aug,
       volume = {683},
       number = {1},
        pages = {542-565},
          doi = {10.1086/589433},
archivePrefix = {arXiv},
       eprint = {0804.2695},
 primaryClass = {astro-ph},
       adsurl = {https://ui.adsabs.harvard.edu/abs/2008ApJ...683..542A}
}

@ARTICLE{AAR_2025_neuraltranslation,
       author = {{Asensio Ramos}, A. and {de la Cruz Rodr{\'\i}guez}, J.},
        title = "{Neural translation for Stokes inversion and synthesis}",
      journal = {\aap},
         year = 2025,
        month = nov,
       volume = {703},
          eid = {A55},
        pages = {A55},
          doi = {10.1051/0004-6361/202556195},
archivePrefix = {arXiv},
       eprint = {2507.00594},
 primaryClass = {astro-ph.IM},
       adsurl = {https://ui.adsabs.harvard.edu/abs/2025A&A...703A..55A}
}

@ARTICLE{AAR_2026_flowmatching,
       author = {{Asensio Ramos}, A. and {Yang}, K.~E. and {Martinez Gonzalez}, M.~J. and {Dodds}, S. Curt and {Sun}, X.},
        title = "{3DSTokesFlow: simulation-based inference for 3D Stokes profiles using flow matching}",
      journal = {arXiv e-prints},
         year = 2026,
        month = jun,
          eid = {arXiv:2606.13004},
        pages = {arXiv:2606.13004},
          doi = {10.48550/arXiv.2606.13004},
archivePrefix = {arXiv},
       eprint = {2606.13004},
 primaryClass = {astro-ph.SR},
       adsurl = {https://ui.adsabs.harvard.edu/abs/2026arXiv260613004A}
}

@ARTICLE{Auer_1977_vector,
       author = {{Auer}, L.~H. and {Heasley}, J.~N. and {House}, L.~L.},
        title = "{The determination of vector magnetic fields from Stokes profiles.}",
      journal = {\solphys},
         year = 1977,
        month = nov,
       volume = {55},
       number = {1},
        pages = {47-61},
          doi = {10.1007/BF00150873},
       adsurl = {https://ui.adsabs.harvard.edu/abs/1977SoPh...55...47A}
}

@ARTICLE{Andreu_2022_GNNs,
       author = {{Vicente Ar{\'e}valo}, A. and {Asensio Ramos}, A. and {Esteban Pozuelo}, S.},
        title = "{Accelerating Non-LTE Synthesis and Inversions with Graph Networks}",
      journal = {\apj},
         year = 2022,
        month = apr,
       volume = {928},
       number = {2},
          eid = {101},
        pages = {101},
          doi = {10.3847/1538-4357/ac53b3},
archivePrefix = {arXiv},
       eprint = {2111.10552},
 primaryClass = {astro-ph.SR},
       adsurl = {https://ui.adsabs.harvard.edu/abs/2022ApJ...928..101V}
}

@ARTICLE{Andreu_2026_GNNs3D,
       author = {{Ar{\'e}valo}, A. Vicente and {Asensio Ramos}, A. and {D{\'\i}az Baso}, C.~J.},
        title = "{Accelerating 3D Non-LTE Synthesis with Graph Neural Networks}",
      journal = {arXiv e-prints},
         year = 2026,
        month = may,
          eid = {arXiv:2605.09543},
        pages = {arXiv:2605.09543},
          doi = {10.48550/arXiv.2605.09543},
archivePrefix = {arXiv},
       eprint = {2605.09543},
 primaryClass = {astro-ph.SR},
       adsurl = {https://ui.adsabs.harvard.edu/abs/2026arXiv260509543A}
}

@BOOK{Athay_1972_book,
      address = {Dordrecht},
    publisher = {D. Reidel},
       volume = {1},
       series = {Geophysics and Astrophysics Monographs},
       author = {{Athay}, R. Grant},
        title = "{Radiation Transport in Spectral Lines}",
         year = 1972,
       adsurl = {https://ui.adsabs.harvard.edu/abs/1972rtsl.book.....A}
}

@ARTICLE{Atanackovic_1997_fbili,
       author = {{Atanackovi{\'c}-Vukmanovi{\'c}}, O. and {Crivellari}, L. and {Simonneau}, E.},
        title = "{A Forth-and-Back Implicit {\ensuremath{\Lambda}}-Iteration}",
      journal = {\apj},
         year = 1997,
        month = oct,
       volume = {487},
       number = {2},
        pages = {735-746},
          doi = {10.1086/304626},
       adsurl = {https://ui.adsabs.harvard.edu/abs/1997ApJ...487..735A}
}

@INPROCEEDINGS{Auer_2003_fs,
       author = {{Auer}, L.},
        title = "{Formal Solution: EXPLICIT Answers}",
    booktitle = {Stellar Atmosphere Modeling},
         year = 2003,
       editor = {{Hubeny}, Ivan and {Mihalas}, Dimitri and {Werner}, Klaus},
       series = {Astronomical Society of the Pacific Conference Series},
       volume = {288},
        month = jan,
        pages = {3},
       adsurl = {https://ui.adsabs.harvard.edu/abs/2003ASPC..288....3A}
}

@ARTICLE{Avrett_1993_FALC,
       author = {{Fontenla}, J.~M. and {Avrett}, E.~H. and {Loeser}, R.},
        title = "{Energy Balance in the Solar Transition Region. III. Helium Emission in Hydrostatic, Constant-Abundance Models with Diffusion}",
      journal = {\apj},
         year = 1993,
        month = mar,
       volume = {406},
        pages = {319},
          doi = {10.1086/172443},
       adsurl = {https://ui.adsabs.harvard.edu/abs/1993ApJ...406..319F}
}

@ARTICLE{PaletouA_1995_PRD,
       author = {{Paletou}, F. and {Auer}, L.~H.},
        title = "{A new approximate operator method for partial frequency redistribution problems.}",
      journal = {\aap},
         year = 1995,
        month = may,
       volume = {297},
        pages = {771},
       adsurl = {https://ui.adsabs.harvard.edu/abs/1995A&A...297..771P}
}

@article{bayes1763essay,
  author  = {Bayes, Thomas},
  title   = {An essay towards solving a problem in the doctrine of chances},
  journal = {Philosophical Transactions of the Royal Society of London},
  volume  = {53},
  pages   = {370--418},
  year    = {1763},
  doi     = {10.1098/rstl.1763.0053}
}

@ARTICLE{Belluzzi_2024_ca4227,
       author = {{Belluzzi}, Luca and {Riva}, Simone and {Janett}, Gioele and {Guerreiro}, Nuno and {Riva}, Fabio and {Benedusi}, Pietro and {del Pino Alem{\'a}n}, Tanaus{\'u} and {Ballester}, Ernest Alsina and {Bueno}, Javier Trujillo and {{\v{S}}t{\v{e}}p{\'a}n}, Ji{\v{r}}{\'\i}},
        title = "{Accurate modeling of the forward-scattering Hanle effect in the chromospheric Ca I 4227 {\r{A}} line}",
      journal = {\aap},
         year = 2024,
        month = nov,
       volume = {691},
          eid = {A278},
        pages = {A278},
          doi = {10.1051/0004-6361/202450178},
archivePrefix = {arXiv},
       eprint = {2404.00104},
 primaryClass = {astro-ph.SR},
       adsurl = {https://ui.adsabs.harvard.edu/abs/2024A&A...691A.278B}
}

@ARTICLE{Carlin_2012_vel,
       author = {{Carlin}, E.~S. and {Manso Sainz}, R. and {Asensio Ramos}, A. and {Trujillo Bueno}, J.},
        title = "{Scattering Polarization in the Ca II Infrared Triplet with Velocity Gradients}",
      journal = {\apj},
         year = 2012,
        month = may,
       volume = {751},
       number = {1},
          eid = {5},
        pages = {5},
          doi = {10.1088/0004-637X/751/1/5},
archivePrefix = {arXiv},
       eprint = {1203.4438},
 primaryClass = {astro-ph.SR},
       adsurl = {https://ui.adsabs.harvard.edu/abs/2012ApJ...751....5C}
}

@ARTICLE{Baur_1980_HAOsp,
       author = {{Baur}, T.~G. and {House}, L.~L. and {Hull}, H.~K.},
        title = "{A spectrum scanning Stokes polarimeter.}",
      journal = {\solphys},
         year = 1980,
        month = feb,
       volume = {65},
       number = {1},
        pages = {111-146},
          doi = {10.1007/BF00151388},
       adsurl = {https://ui.adsabs.harvard.edu/abs/1980SoPh...65..111B}
}

@PHDTHESIS{Becker_1964_phd,
       author = {{Beckers}, Jacques Maurice},
        title = "{A Study of the Fine Structures in the Solar Chromosphere}",
       school = {University of Utrecht, Netherlands},
         year = 1964,
        month = sep,
       adsurl = {https://ui.adsabs.harvard.edu/abs/1964PhDT........83B}
}

@ARTICLE{Beckers_1975_RFs,
       author = {{Beckers}, J.~M. and {Milkey}, R.~W.},
        title = "{The Line Response Function of Stellar Atmospheres and the Effective Depth of Line Formation}",
      journal = {\solphys},
         year = 1975,
        month = aug,
       volume = {43},
       number = {2},
        pages = {289-292},
          doi = {10.1007/BF00152353},
       adsurl = {https://ui.adsabs.harvard.edu/abs/1975SoPh...43..289B}
}

@ARTICLE{Belluzzi_2012_HPRD,
       author = {{Belluzzi}, Luca and {Trujillo Bueno}, Javier and {{\v{S}}t{\v{e}}p{\'a}n}, Ji{\v{r}}{\'\i}},
        title = "{The Scattering Polarization of the Ly{\ensuremath{\alpha}} Lines of H I and He II Taking into Account Partial Frequency Redistribution and J-state Interference Effects}",
      journal = {\apjl},
         year = 2012,
        month = aug,
       volume = {755},
       number = {1},
          eid = {L2},
        pages = {L2},
          doi = {10.1088/2041-8205/755/1/L2},
archivePrefix = {arXiv},
       eprint = {1207.0415},
 primaryClass = {astro-ph.SR},
       adsurl = {https://ui.adsabs.harvard.edu/abs/2012ApJ...755L...2B}
}

@ARTICLE{Benedusi_2021_krylov,
       author = {{Benedusi}, Pietro and {Janett}, Gioele and {Belluzzi}, Luca and {Krause}, Rolf},
        title = "{Numerical solutions to linear transfer problems of polarized radiation. II. Krylov methods and matrix-free implementation}",
      journal = {\aap},
         year = 2021,
        month = nov,
       volume = {655},
          eid = {A88},
        pages = {A88},
          doi = {10.1051/0004-6361/202141238},
archivePrefix = {arXiv},
       eprint = {2110.11873},
 primaryClass = {math.NA},
       adsurl = {https://ui.adsabs.harvard.edu/abs/2021A&A...655A..88B}
}

@ARTICLE{Bianda_2011_fwd,
       author = {{Bianda}, M. and {Ramelli}, R. and {Anusha}, L.~S. and {Stenflo}, J.~O. and {Nagendra}, K.~N. and {Holzreuter}, R. and {Sampoorna}, M. and {Frisch}, H. and {Smitha}, H.~N.},
        title = "{Observations of the forward scattering Hanle effect in the Ca I 4227 {\r{A}} line}",
      journal = {\aap},
         year = 2011,
        month = jun,
       volume = {530},
          eid = {L13},
        pages = {L13},
          doi = {10.1051/0004-6361/201117047},
archivePrefix = {arXiv},
       eprint = {1105.2157},
 primaryClass = {astro-ph.SR},
       adsurl = {https://ui.adsabs.harvard.edu/abs/2011A&A...530L..13B}
}

@book{bjorck1996numerical,
  author    = {Bj{\"o}rck, {\AA}ke},
  title     = {Numerical Methods for Least Squares Problems},
  publisher = {SIAM},
  address   = {Philadelphia, PA},
  year      = {1996},
  isbn      = {0-89871-360-9},
  doi       = {10.1137/1.9781611971484},
  url       = {https://doi.org/10.1137/1.9781611971484}
}

@ARTICLE{Bjorgen_2019_3D,
       author = {{Bj{\o}rgen}, Johan P. and {Leenaarts}, Jorrit and {Rempel}, Matthias and {Cheung}, Mark C.~M. and {Danilovic}, Sanja and {de la Cruz Rodr{\'\i}guez}, Jaime and {Sukhorukov}, Andrii V.},
        title = "{Three-dimensional modeling of chromospheric spectral lines in a simulated active region}",
      journal = {\aap},
         year = 2019,
        month = nov,
       volume = {631},
          eid = {A33},
        pages = {A33},
          doi = {10.1051/0004-6361/201834919},
archivePrefix = {arXiv},
       eprint = {1906.01098},
 primaryClass = {astro-ph.SR},
       adsurl = {https://ui.adsabs.harvard.edu/abs/2019A&A...631A..33B}
}

@book{Blum2012texbook,
  author    = {Karl Blum},
  title     = {Density Matrix Theory and Applications},
  series    = {Springer Series on Atomic, Optical, and Plasma Physics},
  volume    = {64},
  publisher = {Springer Verlag},
  address   = {Berlin, Heidelberg},
  year      = {2012},
  isbn      = {978-3-642-20560-6},
  doi       = {10.1007/978-3-642-20561-3}
}

@BOOK{Bohm_1989_book,
      address = {Cambridge},
    publisher = {Cambridge University Press},
       author = {{B{\"o}hm-Vitense}, E.},
        title = {{Introduction to Stellar Astrophysics. Vol. 2: Stellar Atmospheres}},
         year = 1989,
       volume = {2},
       adsurl = {https://ui.adsabs.harvard.edu/abs/1989isa2.book.....B}
}

@BOOK{Born_1999_book,
      address = {Cambridge},
    publisher = {Cambridge University Press},
      edition = {7th},
       author = {{Born}, Max and {Wolf}, Emil},
        title = "{Principles of Optics}",
         year = 1999,
       adsurl = {https://ui.adsabs.harvard.edu/abs/1999prop.book.....B}
}

@ARTICLE{Borrero_2011_VFISV,
       author = {{Borrero}, J.~M. and {Tomczyk}, S. and {Kubo}, M. and {Socas-Navarro}, H. and {Schou}, J. and {Couvidat}, S. and {Bogart}, R.},
        title = "{VFISV: Very Fast Inversion of the Stokes Vector for the Helioseismic and Magnetic Imager}",
      journal = {\solphys},
         year = 2011,
        month = oct,
       volume = {273},
       number = {1},
        pages = {267-293},
          doi = {10.1007/s11207-010-9515-6},
archivePrefix = {arXiv},
       eprint = {0901.2702},
 primaryClass = {astro-ph.IM},
       adsurl = {https://ui.adsabs.harvard.edu/abs/2011SoPh..273..267B}
}

@ARTICLE{Borrero_2011_photon_noise,
       author = {{Borrero}, J.~M. and {Kobel}, P.},
        title = "{Inferring the magnetic field vector in the quiet Sun. I. Photon noise and selection criteria}",
      journal = {\aap},
         year = 2011,
        month = mar,
       volume = {527},
          eid = {A29},
        pages = {A29},
          doi = {10.1051/0004-6361/201015634},
archivePrefix = {arXiv},
       eprint = {1011.4380},
 primaryClass = {astro-ph.SR},
       adsurl = {https://ui.adsabs.harvard.edu/abs/2011A&A...527A..29B}
}

@ARTICLE{Borrero_2012_QS,
       author = {{Borrero}, J.~M. and {Kobel}, P.},
        title = "{Inferring the magnetic field vector in the quiet Sun. II. Interpreting results from the inversion of Stokes profiles}",
      journal = {\aap},
         year = 2012,
        month = nov,
       volume = {547},
          eid = {A89},
        pages = {A89},
          doi = {10.1051/0004-6361/201118238},
archivePrefix = {arXiv},
       eprint = {1209.4830},
 primaryClass = {astro-ph.SR},
       adsurl = {https://ui.adsabs.harvard.edu/abs/2012A&A...547A..89B}
}

@ARTICLE{Borrero_2014_MEMHD,
       author = {{Borrero}, J.~M. and {Lites}, B.~W. and {Lagg}, A. and {Rezaei}, R. and {Rempel}, M.},
        title = "{Comparison of inversion codes for polarized line formation in MHD simulations. I. Milne-Eddington codes}",
      journal = {\aap},
         year = 2014,
        month = dec,
       volume = {572},
          eid = {A54},
        pages = {A54},
          doi = {10.1051/0004-6361/201424584},
archivePrefix = {arXiv},
       eprint = {1409.3376},
 primaryClass = {astro-ph.SR},
       adsurl = {https://ui.adsabs.harvard.edu/abs/2014A&A...572A..54B}
}

@ARTICLE{Borrero_2019_firtez_I,
       author = {{Borrero}, J.~M. and {Pastor Yabar}, A. and {Rempel}, M. and {Ruiz Cobo}, B.},
        title = "{Combining magnetohydrostatic constraints with Stokes profiles inversions. I. Role of boundary conditions}",
      journal = {\aap},
         year = 2019,
        month = dec,
       volume = {632},
          eid = {A111},
        pages = {A111},
          doi = {10.1051/0004-6361/201936367},
archivePrefix = {arXiv},
       eprint = {1910.14131},
 primaryClass = {astro-ph.SR},
       adsurl = {https://ui.adsabs.harvard.edu/abs/2019A&A...632A.111B}
}

@ARTICLE{Borrero_2021_mhsII,
       author = {{Borrero}, J.~M. and {Pastor Yabar}, A. and {Ruiz Cobo}, B.},
        title = "{Combining magneto-hydrostatic constraints with Stokes profiles inversions. II. Application to Hinode/SP observations}",
      journal = {\aap},
         year = 2021,
        month = mar,
       volume = {647},
          eid = {A190},
        pages = {A190},
          doi = {10.1051/0004-6361/202039927},
archivePrefix = {arXiv},
       eprint = {2101.04394},
 primaryClass = {astro-ph.SR},
       adsurl = {https://ui.adsabs.harvard.edu/abs/2021A&A...647A.190B}
}

@ARTICLE{Borrero_2024_mhsiv,
       author = {{Borrero}, J.~M. and {Pastor Yabar}, A. and {Ruiz Cobo}, B.},
        title = "{Combining magneto-hydrostatic constraints with Stokes profile inversions. IV. Imposing {\ensuremath{\nabla}}{\ensuremath{\cdot}}B = 0 condition}",
      journal = {\aap},
         year = 2024,
        month = jul,
       volume = {687},
          eid = {A155},
        pages = {A155},
          doi = {10.1051/0004-6361/202449572},
archivePrefix = {arXiv},
       eprint = {2404.02045},
 primaryClass = {astro-ph.SR},
       adsurl = {https://ui.adsabs.harvard.edu/abs/2024A&A...687A.155B}
}

@ARTICLE{Borrero_2025_eq,
       author = {{Borrero}, J.~M. and {Pastor Yabar}, A. and {Schmassmann}, M. and {Rempel}, M. and {van Noort}, M. and {Collados}, M.},
        title = "{The role of the Lorentz force in sunspot equilibrium}",
      journal = {\aap},
         year = 2025,
        month = jul,
       volume = {699},
          eid = {A149},
        pages = {A149},
          doi = {10.1051/0004-6361/202554241},
archivePrefix = {arXiv},
       eprint = {2505.23986},
 primaryClass = {astro-ph.SR},
       adsurl = {https://ui.adsabs.harvard.edu/abs/2025A&A...699A.149B}
}

@ARTICLE{Bommier_1997_masterII,
       author = {{Bommier}, V.},
        title = "{Master equation theory applied to the redistribution of polarized radiation, in the weak radiation field limit. II. Arbitrary magnetic field case}",
      journal = {\aap},
         year = 1997,
        month = dec,
       volume = {328},
        pages = {726-751},
       adsurl = {https://ui.adsabs.harvard.edu/abs/1997A&A...328..726B}
}

@Article{Box_1976_SS,
  author = {Box, George E. P.},
  title = {Science and Statistics},
  journal = {Journal of the American Statistical Association},
  year = {1976},
  volume = {71},
  number = {356},
  pages = {791--799},
  doi = {10.1080/01621459.1976.10480949},
  url = {}
}

@ARTICLE{Busa_2001_blanket,
       author = {{Bus{\'a}}, I. and {Andretta}, V. and {Gomez}, M.~T. and {Terranegra}, L.},
        title = "{A method to estimate the effect of line blanketing in NLTE radiative transfer calculations}",
      journal = {\aap},
         year = 2001,
        month = jul,
       volume = {373},
        pages = {993-997},
          doi = {10.1051/0004-6361:20010661},
       adsurl = {https://ui.adsabs.harvard.edu/abs/2001A&A...373..993B}
}

@ARTICLE{Casini_2025_unified,
       author = {{Casini}, R. and {Manso Sainz}, R. and {L{\'o}pez Ariste}, A. and {Kaikati}, N.},
        title = "{A Unifying Polarization Formalism for Electric and Magnetic Multipole Interactions}",
      journal = {\apj},
         year = 2025,
        month = feb,
       volume = {980},
       number = {1},
          eid = {67},
        pages = {67},
          doi = {10.3847/1538-4357/ad7677},
archivePrefix = {arXiv},
       eprint = {2409.01197},
 primaryClass = {physics.atom-ph},
       adsurl = {https://ui.adsabs.harvard.edu/abs/2025ApJ...980...67C}
}

@ARTICLE{Cao_2010_GST,
       author = {{Cao}, W. and {Gorceix}, N. and {Coulter}, R. and {Ahn}, K. and {Rimmele}, T.~R. and {Goode}, P.~R.},
        title = "{Scientific instrumentation for the 1.6 m New Solar Telescope in Big Bear}",
      journal = {Astronomische Nachrichten},
         year = 2010,
        month = jun,
       volume = {331},
       number = {6},
        pages = {636},
          doi = {10.1002/asna.201011390},
       adsurl = {https://ui.adsabs.harvard.edu/abs/2010AN....331..636C}
}

@ARTICLE{Carlsson_1997_RADYN,
       author = {{Carlsson}, Mats and {Stein}, Robert F.},
        title = "{Formation of Solar Calcium H and K Bright Grains}",
      journal = {\apj},
         year = 1997,
        month = may,
       volume = {481},
       number = {1},
        pages = {500-514},
          doi = {10.1086/304043},
       adsurl = {https://ui.adsabs.harvard.edu/abs/1997ApJ...481..500C}
}

@INPROCEEDINGS{Cauzzi_2012_mosaic,
       author = {{Cauzzi}, G. and {Reardon}, K.},
        title = "{The IBIS Mosaic}",
    booktitle = {IAU Special Session},
         year = 2012,
       editor = {{Cauzzi}, G. and {Tritschler}, A. and {Deng}, Y.},
       series = {IAU Special Session},
       volume = {6},
        month = dec,
          eid = {E5.11},
        pages = {E5.11},
       adsurl = {https://ui.adsabs.harvard.edu/abs/2012IAUSS...6E.511C}
}

@ARTICLE{Cavallini_2006_IBIS,
       author = {{Cavallini}, F.},
        title = "{IBIS: A New Post-Focus Instrument for Solar Imaging Spectroscopy}",
      journal = {\solphys},
         year = 2006,
        month = jul,
       volume = {236},
       number = {2},
        pages = {415-439},
          doi = {10.1007/s11207-006-0103-8},
       adsurl = {https://ui.adsabs.harvard.edu/abs/2006SoPh..236..415C}
}

@ARTICLE{Castellanos_2025_bridges,
       author = {{Castellanos Dur{\'a}n}, J.~S. and {Korpi-Lagg}, A. and {Solanki}, S.~K. and {van Noort}, M. and {Milanovic}, N.},
        title = "{Superstrong Magnetic Fields in Sunspot Bipolar Light Bridges}",
      journal = {\apjl},
         year = 2025,
        month = jan,
       volume = {978},
       number = {2},
          eid = {L16},
        pages = {L16},
          doi = {10.3847/2041-8213/ad9f5d},
archivePrefix = {arXiv},
       eprint = {2501.01879},
 primaryClass = {astro-ph.SR},
       adsurl = {https://ui.adsabs.harvard.edu/abs/2025ApJ...978L..16C}
}

@ARTICLE{Carslsson_2019_review,
       author = {{Carlsson}, Mats and {De Pontieu}, Bart and {Hansteen}, Viggo H.},
        title = "{New View of the Solar Chromosphere}",
      journal = {\araa},
         year = 2019,
        month = aug,
       volume = {57},
        pages = {189-226},
          doi = {10.1146/annurev-astro-081817-052044},
       adsurl = {https://ui.adsabs.harvard.edu/abs/2019ARA&A..57..189C}
}

@ARTICLE{Centeno_2018_caIIwfa,
       author = {{Centeno}, Rebecca},
        title = "{On the Weak Field Approximation for Ca 8542 {\r{A}}}",
      journal = {\apj},
         year = 2018,
        month = oct,
       volume = {866},
       number = {2},
          eid = {89},
        pages = {89},
          doi = {10.3847/1538-4357/aae087},
archivePrefix = {arXiv},
       eprint = {1811.03168},
 primaryClass = {astro-ph.SR},
       adsurl = {https://ui.adsabs.harvard.edu/abs/2018ApJ...866...89C}
}

@ARTICLE{CD_2002_review,
       author = {{Christensen-Dalsgaard}, J{\o}rgen},
        title = "{Helioseismology}",
      journal = {Reviews of Modern Physics},
         year = 2002,
        month = nov,
       volume = {74},
       number = {4},
        pages = {1073-1129},
          doi = {10.1103/RevModPhys.74.1073},
archivePrefix = {arXiv},
       eprint = {astro-ph/0207403},
 primaryClass = {astro-ph},
       adsurl = {https://ui.adsabs.harvard.edu/abs/2002RvMP...74.1073C}
}

@BOOK{Chandrasekhar_1950_book,
      address = {Oxford},
    publisher = {Clarendon Press},
       author = {{Chandrasekhar}, Subrahmanyan},
        title = {{Radiative Transfer}},
         year = 1950,
       adsurl = {https://ui.adsabs.harvard.edu/abs/1950ratr.book.....C}
}

@INPROCEEDINGS{CLASP_2016,
       author = {{Narukage}, Noriyuki and {McKenzie}, David E. and {Ishikawa}, Ryoko and {Trujillo-Bueno}, Javier and {De Pontieu}, Bart and {Kubo}, Masahito and {Ishikawa}, Shin-nosuke and {Kano}, Ryouhei and {Suematsu}, Yoshinori and {Yoshida}, Masaki and {Rachmeler}, Laurel A. and {Kobayashi}, Ken and {Cirtain}, Jonathan W. and {Winebarger}, Amy R. and {Asensio Ramos}, Andres and {del Pino Aleman}, Tanausu and {{\v{S}}t{\k{e}}p{\'a}n}, Jiri and {Belluzzi}, Luca and {Larruquert}, Juan Ignacio and {Auch{\`e}re}, Fr{\'e}d{\'e}ric and {Leenaarts}, Jorrit and {Carlsson}, Mattias J.~L.},
        title = "{Chromospheric LAyer SpectroPolarimeter (CLASP2)}",
    booktitle = {Space Telescopes and Instrumentation 2016: Ultraviolet to Gamma Ray},
         year = 2016,
       editor = {{den Herder}, Jan-Willem A. and {Takahashi}, Tadayuki and {Bautz}, Marshall},
       series = {Society of Photo-Optical Instrumentation Engineers (SPIE) Conference Series},
       volume = {9905},
        month = jul,
          eid = {990508},
        pages = {990508},
          doi = {10.1117/12.2232245},
       adsurl = {https://ui.adsabs.harvard.edu/abs/2016SPIE.9905E..08N}
}

@ARTICLE{Danilovic_2016_qs,
       author = {{Danilovic}, S. and {van Noort}, M. and {Rempel}, M.},
        title = "{Internetwork magnetic field as revealed by two-dimensional inversions}",
      journal = {\aap},
         year = 2016,
        month = sep,
       volume = {593},
          eid = {A93},
        pages = {A93},
          doi = {10.1051/0004-6361/201527842},
archivePrefix = {arXiv},
       eprint = {1607.00772},
 primaryClass = {astro-ph.SR},
       adsurl = {https://ui.adsabs.harvard.edu/abs/2016A&A...593A..93D}
}

@ARTICLE{daSilvaSantos2020_alma,
       author = {{da Silva Santos}, J.~M. and {de la Cruz Rodr{\'\i}guez}, J. and {Leenaarts}, J. and {Chintzoglou}, G. and {De Pontieu}, B. and {Wedemeyer}, S. and {Szydlarski}, M.},
        title = "{The multi-thermal chromosphere. Inversions of ALMA and IRIS data}",
      journal = {\aap},
         year = 2020,
        month = feb,
       volume = {634},
          eid = {A56},
        pages = {A56},
          doi = {10.1051/0004-6361/201937117},
archivePrefix = {arXiv},
       eprint = {1912.09886},
 primaryClass = {astro-ph.SR},
       adsurl = {https://ui.adsabs.harvard.edu/abs/2020A&A...634A..56D}
}

@ARTICLE{Dai_2022_flows,
       author = {{Dai}, Biwei and {Seljak}, Uro{\v{s}}},
        title = "{Translation and rotation equivariant normalizing flow (TRENF) for optimal cosmological analysis}",
      journal = {\mnras},
         year = 2022,
        month = oct,
       volume = {516},
       number = {2},
        pages = {2363-2373},
          doi = {10.1093/mnras/stac2010},
archivePrefix = {arXiv},
       eprint = {2202.05282},
 primaryClass = {astro-ph.CO},
       adsurl = {https://ui.adsabs.harvard.edu/abs/2022MNRAS.516.2363D},
}

@ARTICLE{delaCruz_2012_3D1D,
       author = {{de la Cruz Rodr{\'\i}guez}, J. and {Socas-Navarro}, H. and {Carlsson}, M. and {Leenaarts}, J.},
        title = "{Non-local thermodynamic equilibrium inversions from a 3D magnetohydrodynamic chromospheric model}",
      journal = {\aap},
         year = 2012,
        month = jul,
       volume = {543},
          eid = {A34},
        pages = {A34},
          doi = {10.1051/0004-6361/201218825},
archivePrefix = {arXiv},
       eprint = {1205.3171},
 primaryClass = {astro-ph.SR},
       adsurl = {https://ui.adsabs.harvard.edu/abs/2012A&A...543A..34D}
}

@ARTICLE{delaCruz_2013_delo,
       author = {{de la Cruz Rodr{\'\i}guez}, J. and {Piskunov}, N.},
        title = "{DELO-Bezier Formal Solutions of the Polarized Radiative Transfer Equation}",
      journal = {\apj},
         year = 2013,
        month = feb,
       volume = {764},
       number = {1},
          eid = {33},
        pages = {33},
          doi = {10.1088/0004-637X/764/1/33},
archivePrefix = {arXiv},
       eprint = {1212.2737},
 primaryClass = {astro-ph.SR},
       adsurl = {https://ui.adsabs.harvard.edu/abs/2013ApJ...764...33D}
}

@ARTICLE{delaCruz_2017_rv,
       author = {{de la Cruz Rodr{\'\i}guez}, J. and {van Noort}, M.},
        title = "{Radiative Diagnostics in the Solar Photosphere and Chromosphere}",
      journal = {\ssr},
         year = 2017,
        month = sep,
       volume = {210},
       number = {1-4},
        pages = {109-143},
          doi = {10.1007/s11214-016-0294-8},
archivePrefix = {arXiv},
       eprint = {1609.08324},
 primaryClass = {astro-ph.SR},
       adsurl = {https://ui.adsabs.harvard.edu/abs/2017SSRv..210..109D}
}

@ARTICLE{delacruz_2024_spatemp,
       author = {{de la Cruz Rodr{\'\i}guez}, J. and {Leenaarts}, J.},
        title = "{Improved reconstruction of solar magnetic fields from imaging spectropolarimetry through spatio-temporal regularisation}",
      journal = {\aap},
         year = 2024,
        month = may,
       volume = {685},
          eid = {A85},
        pages = {A85},
          doi = {10.1051/0004-6361/202348810},
archivePrefix = {arXiv},
       eprint = {2402.03440},
 primaryClass = {astro-ph.SR},
       adsurl = {https://ui.adsabs.harvard.edu/abs/2024A&A...685A..85D}
}

@ARTICLE{delaCruz_2019_multires,
       author = {{de la Cruz Rodr{\'\i}guez}, J.},
        title = "{A method for global inversion of multi-resolution solar data}",
      journal = {\aap},
         year = 2019,
        month = nov,
       volume = {631},
          eid = {A153},
        pages = {A153},
          doi = {10.1051/0004-6361/201936635},
archivePrefix = {arXiv},
       eprint = {1909.02604},
 primaryClass = {astro-ph.SR},
       adsurl = {https://ui.adsabs.harvard.edu/abs/2019A&A...631A.153D}
}

@ARTICLE{delacruz_2019_stic,
       author = {{de la Cruz Rodr{\'\i}guez}, J. and {Leenaarts}, J. and {Danilovic}, S. and {Uitenbroek}, H.},
        title = "{STiC: A multiatom non-LTE PRD inversion code for full-Stokes solar observations}",
      journal = {\aap},
         year = 2019,
        month = mar,
       volume = {623},
          eid = {A74},
        pages = {A74},
          doi = {10.1051/0004-6361/201834464},
archivePrefix = {arXiv},
       eprint = {1810.08441},
 primaryClass = {astro-ph.SR},
       adsurl = {https://ui.adsabs.harvard.edu/abs/2019A&A...623A..74D}
}

@BOOK{delToroIniesta_2003_book,
      address = {Cambridge},
    publisher = {Cambridge University Press},
       author = {{del Toro Iniesta}, Jos{\'e} Carlos},
        title = "{Introduction to Spectropolarimetry}",
         year = 2003,
       adsurl = {https://ui.adsabs.harvard.edu/abs/2003isp..book.....D}
}

@ARTICLE{delToroIniesta_2016_review,
       author = {{del Toro Iniesta}, Jose Carlos and {Ruiz Cobo}, Basilio},
        title = "{Inversion of the radiative transfer equation for polarized light}",
      journal = {Living Reviews in Solar Physics},
         year = 2016,
        month = dec,
       volume = {13},
       number = {1},
          eid = {4},
        pages = {4},
          doi = {10.1007/s41116-016-0005-2},
archivePrefix = {arXiv},
       eprint = {1610.10039},
 primaryClass = {astro-ph.SR},
       adsurl = {https://ui.adsabs.harvard.edu/abs/2016LRSP...13....4D}
}

@ARTICLE{dePointieu_2014_IRIS,
       author = {{De Pontieu}, B. and {Title}, A.~M. and {Lemen}, J.~R. and {Kushner}, G.~D. and {Akin}, D.~J. and {Allard}, B. and {Berger}, T. and {Boerner}, P. and {Cheung}, M. and {Chou}, C. and {Drake}, J.~F. and {Duncan}, D.~W. and {Freeland}, S. and {Heyman}, G.~F. and {Hoffman}, C. and {Hurlburt}, N.~E. and {Lindgren}, R.~W. and {Mathur}, D. and {Rehse}, R. and {Sabolish}, D. and {Seguin}, R. and {Schrijver}, C.~J. and {Tarbell}, T.~D. and {W{\"u}lser}, J.-P. and {Wolfson}, C.~J. and {Yanari}, C. and {Mudge}, J. and {Nguyen-Phuc}, N. and {Timmons}, R. and {van Bezooijen}, R. and {Weingrod}, I. and {Brookner}, R. and {Butcher}, G. and {Dougherty}, B. and {Eder}, J. and {Knagenhjelm}, V. and {Larsen}, S. and {Mansir}, D. and {Phan}, L. and {Boyle}, P. and {Cheimets}, P.~N. and {DeLuca}, E.~E. and {Golub}, L. and {Gates}, R. and {Hertz}, E. and {McKillop}, S. and {Park}, S. and {Perry}, T. and {Podgorski}, W.~A. and {Reeves}, K. and {Saar}, S. and {Testa}, P. and {Tian}, H. and {Weber}, M. and {Dunn}, C. and {Eccles}, S. and {Jaeggli}, S.~A. and {Kankelborg}, C.~C. and {Mashburn}, K. and {Pust}, N. and {Springer}, L. and {Carvalho}, R. and {Kleint}, L. and {Marmie}, J. and {Mazmanian}, E. and {Pereira}, T.~M.~D. and {Sawyer}, S. and {Strong}, J. and {Worden}, S.~P. and {Carlsson}, M. and {Hansteen}, V.~H. and {Leenaarts}, J. and {Wiesmann}, M. and {Aloise}, J. and {Chu}, K.-C. and {Bush}, R.~I. and {Scherrer}, P.~H. and {Brekke}, P. and {Martinez-Sykora}, J. and {Lites}, B.~W. and {McIntosh}, S.~W. and {Uitenbroek}, H. and {Okamoto}, T.~J. and {Gummin}, M.~A. and {Auker}, G. and {Jerram}, P. and {Pool}, P. and {Waltham}, N.},
        title = "{The Interface Region Imaging Spectrograph (IRIS)}",
      journal = {\solphys},
         year = 2014,
        month = jul,
       volume = {289},
       number = {7},
        pages = {2733-2779},
          doi = {10.1007/s11207-014-0485-y},
archivePrefix = {arXiv},
       eprint = {1401.2491},
 primaryClass = {astro-ph.SR},
       adsurl = {https://ui.adsabs.harvard.edu/abs/2014SoPh..289.2733D}
}

@ARTICLE{delpino_2018_sr4607,
       author = {{del Pino Alem{\'a}n}, T. and {Trujillo Bueno}, J. and {{\v{S}}t{\v{e}}p{\'a}n}, J. and {Shchukina}, N.},
        title = "{A Novel Investigation of the Small-scale Magnetic Activity of the Quiet Sun via the Hanle Effect in the Sr I 4607 {\r{A}} Line}",
      journal = {\apj},
         year = 2018,
        month = aug,
       volume = {863},
       number = {2},
          eid = {164},
        pages = {164},
          doi = {10.3847/1538-4357/aaceab},
archivePrefix = {arXiv},
       eprint = {1806.07293},
 primaryClass = {astro-ph.SR},
       adsurl = {https://ui.adsabs.harvard.edu/abs/2018ApJ...863..164D}
}

@ARTICLE{Derouich_2003_coll,
       author = {{Derouich}, M. and {Sahal-Br{\'e}chot}, S. and {Barklem}, P.~S. and {O'Mara}, B.~J.},
        title = "{Semi-classical theory of collisional depolarization of spectral lines by atomic hydrogen  I. Application to p states of neutral atoms}",
      journal = {\aap},
         year = 2003,
        month = jun,
       volume = {404},
        pages = {763-773},
          doi = {10.1051/0004-6361:20030568},
archivePrefix = {arXiv},
       eprint = {astro-ph/0304115},
 primaryClass = {astro-ph},
       adsurl = {https://ui.adsabs.harvard.edu/abs/2003A&A...404..763D}
}

@ARTICLE{Diaz_Baso_2022_normflows,
       author = {{D{\'\i}az Baso}, C.~J. and {Asensio Ramos}, A. and {de la Cruz Rodr{\'\i}guez}, J.},
        title = "{Bayesian Stokes inversion with normalizing flows}",
      journal = {\aap},
         year = 2022,
        month = mar,
       volume = {659},
          eid = {A165},
        pages = {A165},
          doi = {10.1051/0004-6361/202142018},
archivePrefix = {arXiv},
       eprint = {2108.07089},
 primaryClass = {astro-ph.SR},
       adsurl = {https://ui.adsabs.harvard.edu/abs/2022A&A...659A.165D}
}

@ARTICLE{Diaz_Baso_2025_neuralfields,
       author = {{D{\'\i}az Baso}, C.~J. and {Asensio Ramos}, A. and {de la Cruz Rodr{\'\i}guez}, J. and {da Silva Santos}, J.~M. and {Rouppe van der Voort}, L.},
        title = "{Exploring spectropolarimetric inversions using neural fields: Solar chromospheric magnetic field under the weak-field approximation}",
      journal = {\aap},
         year = 2025,
        month = jan,
       volume = {693},
          eid = {A170},
        pages = {A170},
          doi = {10.1051/0004-6361/202452172},
archivePrefix = {arXiv},
       eprint = {2409.05156},
 primaryClass = {astro-ph.SR},
       adsurl = {https://ui.adsabs.harvard.edu/abs/2025A&A...693A.170D}
}

@ARTICLE{Dunn_1969_DST,
       author = {{Dunn}, Richard B.},
        title = "{Sacramento Peak's New Solar Telescope}",
      journal = {\skytel},
         year = 1969,
        month = dec,
       volume = {38},
        pages = {368},
       adsurl = {https://ui.adsabs.harvard.edu/abs/1969S&T....38..368D}
}

@ARTICLE{Fletcher_2011_SSRV,
       author = {{Fletcher}, L. and {Dennis}, B.~R. and {Hudson}, H.~S. and {Krucker}, S. and {Phillips}, K. and {Veronig}, A. and {Battaglia}, M. and {Bone}, L. and {Caspi}, A. and {Chen}, Q. and {Gallagher}, P. and {Grigis}, P.~T. and {Ji}, H. and {Liu}, W. and {Milligan}, R.~O. and {Temmer}, M.},
        title = "{An Observational Overview of Solar Flares}",
      journal = {\ssr},
         year = 2011,
        month = sep,
       volume = {159},
       number = {1-4},
        pages = {19-106},
          doi = {10.1007/s11214-010-9701-8},
archivePrefix = {arXiv},
       eprint = {1109.5932},
 primaryClass = {astro-ph.SR},
       adsurl = {https://ui.adsabs.harvard.edu/abs/2011SSRv..159...19F}
}

@ARTICLE{frutiger_2000_spinor,
       author = {{Frutiger}, C. and {Solanki}, S.~K. and {Fligge}, M. and {Bruls}, J.~H.~M.~J.},
        title = "{Properties of the solar granulation obtained from the inversion of low spatial resolution spectra}",
      journal = {\aap},
         year = 2000,
        month = jun,
       volume = {358},
        pages = {1109-1121},
       adsurl = {https://ui.adsabs.harvard.edu/abs/2000A&A...358.1109F}
}

@ARTICLE{Grevesse_2007_composition,
       author = {{Grevesse}, N. and {Asplund}, M. and {Sauval}, A.~J.},
        title = "{The Solar Chemical Composition}",
      journal = {\ssr},
         year = 2007,
        month = jun,
       volume = {130},
       number = {1-4},
        pages = {105-114},
          doi = {10.1007/s11214-007-9173-7},
       adsurl = {https://ui.adsabs.harvard.edu/abs/2007SSRv..130..105G}
}

@BOOK{Gray_2005_book,
      address = {Cambridge},
      edition = {3rd},
       author = {{Gray}, David F.},
        title = "{The Observation and Analysis of Stellar Photospheres}",
         year = 2005,
          doi = {10.1017/CBO9781316036570},
          publisher = {Cambridge University Press},
       adsurl = {https://ui.adsabs.harvard.edu/abs/2005oasp.book.....G}
}

@ARTICLE{Heinzel_1995_nlteprom,
       author = {{Heinzel}, P.},
        title = "{Multilevel NLTE radiative transfer in isolated atmospheric structures: implementation of the MALI-technique.}",
      journal = {\aap},
         year = 1995,
        month = jul,
       volume = {299},
        pages = {563},
       adsurl = {https://ui.adsabs.harvard.edu/abs/1995A&A...299..563H}
}

@ARTICLE{Heinzel_2001_2DRT,
       author = {{Heinzel}, P. and {Anzer}, U.},
        title = "{Prominence fine structures in a magnetic equilibrium: Two-dimensional models with multilevel radiative transfer}",
      journal = {\aap},
         year = 2001,
        month = sep,
       volume = {375},
        pages = {1082-1090},
          doi = {10.1051/0004-6361:20010926},
       adsurl = {https://ui.adsabs.harvard.edu/abs/2001A&A...375.1082H}
}

@ARTICLE{Hanle_1924_og,
       author = {{Hanle}, Wilhelm},
        title = "{{\"U}ber magnetische Beeinflussung der Polarisation der Resonanzfluoreszenz}",
      journal = {Zeitschrift fur Physik},
         year = 1924,
        month = dec,
       volume = {30},
       number = {1},
        pages = {93-105},
          doi = {10.1007/BF01331827},
       adsurl = {https://ui.adsabs.harvard.edu/abs/1924ZPhy...30...93H}
}

@ARTICLE{Hogg_2010_fitting,
       author = {{Hogg}, David W. and {Bovy}, Jo and {Lang}, Dustin},
        title = "{Data analysis recipes: Fitting a model to data}",
      journal = {arXiv e-prints},
         year = 2010,
        month = aug,
          eid = {arXiv:1008.4686},
        pages = {arXiv:1008.4686},
          doi = {10.48550/arXiv.1008.4686},
archivePrefix = {arXiv},
       eprint = {1008.4686},
 primaryClass = {astro-ph.IM},
       adsurl = {https://ui.adsabs.harvard.edu/abs/2010arXiv1008.4686H},
          url = {}
}

@ARTICLE{Hofmann_2022_alma,
       author = {{Hofmann}, Ryan A. and {Reardon}, Kevin P. and {Milic}, Ivan and {Molnar}, Momchil E. and {Chai}, Yi and {Uitenbroek}, Han},
        title = "{Evaluating Non-LTE Spectral Inversions with ALMA and IBIS}",
      journal = {\apj},
         year = 2022,
        month = jul,
       volume = {933},
       number = {2},
          eid = {244},
        pages = {244},
          doi = {10.3847/1538-4357/ac6f00},
archivePrefix = {arXiv},
       eprint = {2205.08760},
 primaryClass = {astro-ph.SR},
       adsurl = {https://ui.adsabs.harvard.edu/abs/2022ApJ...933..244H}
}

@ARTICLE{Hoelken_2026_multi,
       author = {{H{\"o}lken}, J. and {van Noort}, M. and {Solanki}, S.~K. and {Feller}, A. and {Doerr}, H.-P. and {Cao}, W. and {Gorceix}, N. and {Kang}, J. and {Ahn}, K. and {Chae}, J.},
        title = "{Toward solar many-line inversions of high-resolution spectropolarimetric data}",
      journal = {\aap},
         year = 2026,
        month = jan,
       volume = {705},
          eid = {A220},
        pages = {A220},
          doi = {10.1051/0004-6361/202556890},
       adsurl = {https://ui.adsabs.harvard.edu/abs/2026A&A...705A.220H}
}

@BOOK{HubenyMihalas_2014_SAIII,
      address = {Princeton, NJ},
    publisher = {Princeton University Press},
       series = {Princeton Series in Astrophysics},
       author = {{Hubeny}, Ivan and {Mihalas}, Dimitri},
        title = "{Theory of Stellar Atmospheres}",
         year = 2014,
       adsurl = {https://ui.adsabs.harvard.edu/abs/2014tsa..book.....H}
}

@ARTICLE{Hummer_1962_PRD,
       author = {{Hummer}, D.~G.},
        title = "{Non-coherent scattering: I. The redistribution function with Doppler broadening}",
      journal = {\mnras},
         year = 1962,
        month = jan,
       volume = {125},
        pages = {21-37},
          doi = {10.1093/mnras/125.1.21},
       adsurl = {https://ui.adsabs.harvard.edu/abs/1962MNRAS.125...21H}
}

@ARTICLE{Iglesias_2019_pol,
       author = {{Iglesias}, Francisco A. and {Feller}, Alex},
        title = "{Instrumentation for solar spectropolarimetry: state of the art and prospects}",
      journal = {Optical Engineering},
         year = 2019,
        month = aug,
       volume = {58},
          eid = {082417},
        pages = {082417},
          doi = {10.1117/1.OE.58.8.082417},
archivePrefix = {arXiv},
       eprint = {1911.01368},
 primaryClass = {astro-ph.IM},
       adsurl = {https://ui.adsabs.harvard.edu/abs/2019OptEn..58h2417I}
}

@book{Jaynes2003Probability,
      address = {Cambridge},
  author = {Jaynes, Edwin T},
  publisher = {Cambridge University Press},
  title = {Probability theory: The logic of science},
  year = 2003
}

@ARTICLE{Jenkins_2024_25D,
       author = {{Jenkins}, Jack M. and {Osborne}, Christopher M.~J. and {Qiu}, Ye and {Keppens}, Rony and {Li}, Chuan},
        title = "{The Bright Rim Prominences according to 2.5D Radiative Transfer}",
      journal = {\apjl},
         year = 2024,
        month = apr,
       volume = {964},
       number = {2},
          eid = {L34},
        pages = {L34},
          doi = {10.3847/2041-8213/ad3423},
archivePrefix = {arXiv},
       eprint = {2403.09931},
 primaryClass = {astro-ph.SR},
       adsurl = {https://ui.adsabs.harvard.edu/abs/2024ApJ...964L..34J}
}

@ARTICLE{Janett_2017_higherorder,
       author = {{Janett}, Gioele and {Steiner}, Oskar and {Belluzzi}, Luca},
        title = "{Formal Solutions for Polarized Radiative Transfer. II. High-order Methods}",
      journal = {\apj},
         year = 2017,
        month = aug,
       volume = {845},
       number = {2},
          eid = {104},
        pages = {104},
          doi = {10.3847/1538-4357/aa7aa3},
archivePrefix = {arXiv},
       eprint = {1709.01280},
 primaryClass = {astro-ph.SR},
       adsurl = {https://ui.adsabs.harvard.edu/abs/2017ApJ...845..104J}
}

@ARTICLE{Janett_2018_formalsolution,
       author = {{Janett}, Gioele and {Steiner}, Oskar and {Belluzzi}, Luca},
        title = "{Formal Solutions for Polarized Radiative Transfer. IV. Numerical Performances in Practical Problems}",
      journal = {\apj},
         year = 2018,
        month = sep,
       volume = {865},
       number = {1},
          eid = {16},
        pages = {16},
          doi = {10.3847/1538-4357/aad4a4},
archivePrefix = {arXiv},
       eprint = {1809.06604},
 primaryClass = {astro-ph.SR},
       adsurl = {https://ui.adsabs.harvard.edu/abs/2018ApJ...865...16J}
}

@ARTICLE{Kianfar_2020_fibrils,
       author = {{Kianfar}, Sepideh and {Leenaarts}, Jorrit and {Danilovic}, Sanja and {de la Cruz Rodr{\'\i}guez}, Jaime and {D{\'\i}az Baso}, Carlos Jos{\'e}},
        title = "{Physical properties of bright Ca II K fibrils in the solar chromosphere}",
      journal = {\aap},
         year = 2020,
        month = may,
       volume = {637},
          eid = {A1},
        pages = {A1},
          doi = {10.1051/0004-6361/202037572},
archivePrefix = {arXiv},
       eprint = {2003.11302},
 primaryClass = {astro-ph.SR},
       adsurl = {https://ui.adsabs.harvard.edu/abs/2020A&A...637A...1K}
}

@ARTICLE{Kaithakkal_2023_ellerman,
       author = {{Kaithakkal}, Anjali J. and {Borrero}, J.~M. and {Pastor Yabar}, A. and {de la Cruz Rodr{\'\i}guez}, J.},
        title = "{A reconnection-driven magnetic flux cancellation and a quiet Sun Ellerman bomb}",
      journal = {\mnras},
         year = 2023,
        month = may,
       volume = {521},
       number = {3},
        pages = {3882-3897},
          doi = {10.1093/mnras/stad759},
archivePrefix = {arXiv},
       eprint = {2303.06331},
 primaryClass = {astro-ph.SR},
       adsurl = {https://ui.adsabs.harvard.edu/abs/2023MNRAS.521.3882K}
}

@ARTICLE{Lagg_2004_HELIXplus,
       author = {{Lagg}, A. and {Woch}, J. and {Krupp}, N. and {Solanki}, S.~K.},
        title = "{Retrieval of the full magnetic vector with the He I multiplet at 1083 nm. Maps of an emerging flux region}",
      journal = {\aap},
         year = 2004,
        month = feb,
       volume = {414},
        pages = {1109-1120},
          doi = {10.1051/0004-6361:20031643},
       adsurl = {https://ui.adsabs.harvard.edu/abs/2004A&A...414.1109L}
}

@article{Levenberg1944,
  author = {Levenberg, Kenneth},
  title = {A method for the solution of certain non-linear problems in least squares},
  journal = {Quarterly of Applied Mathematics},
  volume = {2},
  number = {2},
  pages = {164--168},
  year = {1944}
}

@ARTICLE{Liu_2025_paperI,
       author = {{Liu}, G. and {Mili{\'c}}, I. and {Castellanos Dur{\'a}n}, J.~S. and {Borrero}, J.~M. and {van Noort}, M. and {Kuckein}, C.},
        title = "{Fine-scale opposite-polarity magnetic fields in a solar plage revealed by integral field spectropolarimetry}",
      journal = {\aap},
         year = 2025,
        month = may,
       volume = {697},
          eid = {L7},
        pages = {L7},
          doi = {10.1051/0004-6361/202554498},
archivePrefix = {arXiv},
       eprint = {2505.07561},
 primaryClass = {astro-ph.SR},
       adsurl = {https://ui.adsabs.harvard.edu/abs/2025A&A...697L...7L}
}

@ARTICLE{landideglinnocenti_1977_rfs,
       author = {{Landi Degl'Innocenti}, E. and {Landi Degl'Innocenti}, M.},
        title = "{Response function for magnetic lines.}",
      journal = {\aap},
         year = 1977,
        month = apr,
       volume = {56},
        pages = {111-115},
       adsurl = {https://ui.adsabs.harvard.edu/abs/1977A&A....56..111L}
}

@BOOK{LL_2004_bible,
      address = {Dordrecht},
    publisher = {Kluwer Academic Publishers},
       series = {Astrophysics and Space Science Library},
       author = {{Landi Degl'Innocenti}, E. and {Landolfi}, M.},
        title = "{Polarization in Spectral Lines}",
         year = 2004,
       volume = {307},
          doi = {10.1007/978-1-4020-2415-3},
       adsurl = {https://ui.adsabs.harvard.edu/abs/2004ASSL..307.....L}
}

@book{LandiDeglInnocenti2014book,
  author = {Landi Degl'Innocenti, Egidio},
  title = {Atomic Spectroscopy and Radiative Processes},
  year = {2014},
  publisher = {Springer-Verlag Italia},
  address = {Milano},
  series = {UNITEXT for Physics},
  isbn = {978-8847028074}
}

@article{LandiDeglInnocenti1983_linepolI,
  author    = {Landi Degl'Innocenti, E.},
  title     = {Polarization in Spectral Lines: I. A Unifying Theoretical Approach},
  journal   = {Solar Physics},
  volume    = {85},
  pages     = {3--47},
  year      = {1983},
  doi       = {10.1007/BF00148645}
}

@INPROCEEDINGS{Leenaarts_2009_multi3d,
       author = {{Leenaarts}, J. and {Carlsson}, M.},
        title = "{MULTI3D: A Domain-Decomposed 3D Radiative Transfer Code}",
    booktitle = {The Second Hinode Science Meeting: Beyond Discovery-Toward Understanding},
         year = 2009,
       editor = {{Lites}, B. and {Cheung}, M. and {Magara}, T. and {Mariska}, J. and {Reeves}, K.},
       series = {Astronomical Society of the Pacific Conference Series},
       volume = {415},
        month = dec,
        pages = {87},
       adsurl = {https://ui.adsabs.harvard.edu/abs/2009ASPC..415...87L}
}

@ARTICLE{Leenaarts_2020_rmhd,
       author = {{Leenaarts}, Jorrit},
        title = "{Radiation hydrodynamics in simulations of the solar atmosphere}",
      journal = {Living Reviews in Solar Physics},
         year = 2020,
        month = dec,
       volume = {17},
       number = {1},
          eid = {3},
        pages = {3},
          doi = {10.1007/s41116-020-0024-x},
archivePrefix = {arXiv},
       eprint = {2002.03623},
 primaryClass = {astro-ph.SR},
       adsurl = {https://ui.adsabs.harvard.edu/abs/2020LRSP...17....3L}
}

@ARTICLE{Li_2022_tic,
       author = {{Li}, H. and {del Pino Alem{\'a}n}, T. and {Trujillo Bueno}, J. and {Casini}, R.},
        title = "{TIC: A Stokes Inversion Code for Scattering Polarization with Partial Frequency Redistribution and Arbitrary Magnetic Fields}",
      journal = {\apj},
         year = 2022,
        month = jul,
       volume = {933},
       number = {2},
          eid = {145},
        pages = {145},
          doi = {10.3847/1538-4357/ac745c},
archivePrefix = {arXiv},
       eprint = {2205.15666},
 primaryClass = {astro-ph.SR},
       adsurl = {https://ui.adsabs.harvard.edu/abs/2022ApJ...933..145L}
}

@book{mackay2003information,
  author    = {MacKay, David J. C.},
  title     = {Information Theory, Inference, and Learning Algorithms},
  publisher = {Cambridge University Press},
  address   = {Cambridge, UK},
  year      = {2003},
  isbn      = {978-0-521-64298-1},
  url       = {https://www.inference.org.uk/mackay/itila/}
}

@INPROCEEDINGS{MansoSainz_2003_multilevel2ndkind,
       author = {{Manso Sainz}, R. and {Trujillo Bueno}, J.},
        title = "{A Multilevel Radiative Transfer Program for Modeling Scattering Line Polarization and the Hanle Effect in Stellar Atmospheres}",
    booktitle = {Solar Polarization},
         year = 2003,
       editor = {{Trujillo-Bueno}, Javier and {Sanchez Almeida}, Jorge},
       series = {Astronomical Society of the Pacific Conference Series},
       volume = {307},
        month = jan,
        pages = {251},
       adsurl = {https://ui.adsabs.harvard.edu/abs/2003ASPC..307..251M}
}

@article{Marquardt63,
author = {Marquardt, Donald W.},
title = {An Algorithm for Least-Squares Estimation of Nonlinear Parameters},
journal = {Journal of the Society for Industrial and Applied Mathematics},
volume = {11},
number = {2},
pages = {431-441},
year = {1963},
doi = {10.1137/0111030},

URL = { 
    
        https://doi.org/10.1137/0111030
    
    

},
eprint = { 
    
        https://doi.org/10.1137/0111030
    
    

}

}

@ARTICLE{Mein_1971_rfs,
       author = {{Mein}, Pierre},
        title = "{Inhomogeneities in the Solar Atmosphere from the Ca II Infra-Red Lines}",
      journal = {\solphys},
         year = 1971,
        month = oct,
       volume = {20},
       number = {1},
        pages = {3-18},
          doi = {10.1007/BF00146089},
       adsurl = {https://ui.adsabs.harvard.edu/abs/1971SoPh...20....3M}
}

@ARTICLE{Mein_1988_diffcloud,
       author = {{Mein}, P. and {Mein}, N.},
        title = "{Differential cloud models for solar velocity field measurements}",
      journal = {\aap},
         year = 1988,
        month = sep,
       volume = {203},
       number = {1},
        pages = {162-169},
       adsurl = {https://ui.adsabs.harvard.edu/abs/1988A&A...203..162M}
}

@ARTICLE{Mein_1996_cloudmodel_new,
       author = {{Mein}, N. and {Mein}, P. and {Heinzel}, P. and {Vial}, J.-C. and {Malherbe}, J.~M. and {Staiger}, J.},
        title = "{Cloud model with variable source function for solar H{\ensuremath{\alpha}} structures.}",
      journal = {\aap},
         year = 1996,
        month = may,
       volume = {309},
        pages = {275-283},
       adsurl = {https://ui.adsabs.harvard.edu/abs/1996A&A...309..275M}
}

@BOOK{Mihalas_1978_SAII,
      address = {San Francisco},
    publisher = {W. H. Freeman},
      edition = {2nd},
       author = {{Mihalas}, Dimitri},
        title = {{Stellar Atmospheres}},
         year = 1978,
       adsurl = {https://ui.adsabs.harvard.edu/abs/1978stat.book.....M}
}

@ARTICLE{Jarolim_2023_pinnex,
       author = {{Jarolim}, R. and {Thalmann}, J.~K. and {Veronig}, A.~M. and {Podladchikova}, T.},
        title = "{Probing the solar coronal magnetic field with physics-informed neural networks.}",
      journal = {Nature Astronomy},
         year = 2023,
        month = oct,
       volume = {7},
        pages = {1171-1179},
          doi = {10.1038/s41550-023-02030-9},
       adsurl = {https://ui.adsabs.harvard.edu/abs/2023NatAs...7.1171J}
}

@ARTICLE{Jarolim_2025_pinnme,
       author = {{Jarolim}, R. and {Molnar}, M.~E. and {Tremblay}, B. and {Centeno}, R. and {Rempel}, M.},
        title = "{PINN ME: A Physics-informed Neural Network Framework for Accurate Milne-Eddington Inversions of Solar Magnetic Fields.}",
      journal = {\apjl},
         year = 2025,
        month = jan,
       volume = {985},
          eid = {L7},
        pages = {L7},
          doi = {10.48550/arXiv.2502.13924},
archivePrefix = {arXiv},
       eprint = {2502.13924},
 primaryClass = {astro-ph.SR},
       adsurl = {https://ui.adsabs.harvard.edu/abs/2025ApJ...985L...7J}
}

@ARTICLE{Morosin_2022_heating,
       author = {{Morosin}, R. and {de la Cruz Rodr{\'\i}guez}, J. and {D{\'\i}az Baso}, C.~J. and {Leenaarts}, J.},
        title = "{Spatio-temporal analysis of chromospheric heating in a plage region}",
      journal = {\aap},
         year = 2022,
        month = aug,
       volume = {664},
          eid = {A8},
        pages = {A8},
          doi = {10.1051/0004-6361/202243461},
archivePrefix = {arXiv},
       eprint = {2203.01688},
 primaryClass = {astro-ph.SR},
       adsurl = {https://ui.adsabs.harvard.edu/abs/2022A&A...664A...8M}
}

@ARTICLE{Mueller_2020_SO,
       author = {{M{\"u}ller}, D. and {St. Cyr}, O.~C. and {Zouganelis}, I. and {Gilbert}, H.~R. and {Marsden}, R. and {Nieves-Chinchilla}, T. and {Antonucci}, E. and {Auch{\`e}re}, F. and {Berghmans}, D. and {Horbury}, T.~S. and {Howard}, R.~A. and {Krucker}, S. and {Maksimovic}, M. and {Owen}, C.~J. and {Rochus}, P. and {Rodriguez-Pacheco}, J. and {Romoli}, M. and {Solanki}, S.~K. and {Bruno}, R. and {Carlsson}, M. and {Fludra}, A. and {Harra}, L. and {Hassler}, D.~M. and {Livi}, S. and {Louarn}, P. and {Peter}, H. and {Sch{\"u}hle}, U. and {Teriaca}, L. and {del Toro Iniesta}, J.~C. and {Wimmer-Schweingruber}, R.~F. and {Marsch}, E. and {Velli}, M. and {De Groof}, A. and {Walsh}, A. and {Williams}, D.},
        title = "{The Solar Orbiter mission. Science overview}",
      journal = {\aap},
         year = 2020,
        month = oct,
       volume = {642},
          eid = {A1},
        pages = {A1},
          doi = {10.1051/0004-6361/202038467},
archivePrefix = {arXiv},
       eprint = {2009.00861},
 primaryClass = {astro-ph.SR},
       adsurl = {https://ui.adsabs.harvard.edu/abs/2020A&A...642A...1M}
}

@ARTICLE{Olson_1986_ALO,
       author = {{Olson}, G.~L. and {Auer}, L.~H. and {Buchler}, J.~R.},
        title = "{A rapidly convergent iterative solution of the non-LTE radiation transfer problem.}",
      journal = {\jqsrt},
         year = 1986,
        month = jun,
       volume = {35},
        pages = {431-442},
          doi = {10.1016/0022-4073(86)90030-0},
       adsurl = {https://ui.adsabs.harvard.edu/abs/1986JQSRT..35..431O}
}

@ARTICLE{OK_1987_SC1,
       author = {{Olson}, G.~L. and {Kunasz}, P.~B.},
        title = "{Short characteristic solution of the non-LTE transfer problem by operator perturbation. I. The one-dimensional planar slab.}",
      journal = {\jqsrt},
         year = 1987,
        month = jan,
       volume = {38},
       number = {5},
        pages = {325-336},
          doi = {10.1016/0022-4073(87)90027-6},
       adsurl = {https://ui.adsabs.harvard.edu/abs/1987JQSRT..38..325O}
}

@ARTICLE{Osborne_2025_dexRT,
       author = {{Osborne}, C.~M.~J. and {Sannikov}, A.},
        title = "{Radiance cascades: a novel high-resolution formal solution for multidimensional non-LTE radiative transfer}",
      journal = {RAS Techniques and Instruments},
         year = 2025,
        month = jan,
       volume = {4},
          eid = {rzae062},
        pages = {rzae062},
          doi = {10.1093/rasti/rzae062},
archivePrefix = {arXiv},
       eprint = {2408.14425},
 primaryClass = {astro-ph.SR},
       adsurl = {https://ui.adsabs.harvard.edu/abs/2025RASTI...4...62O}
}

@ARTICLE{Orozco_2007_milos,
       author = {{Orozco Su{\'a}rez}, D. and {Del Toro Iniesta}, J.~C.},
        title = "{The usefulness of analytic response functions}",
      journal = {\aap},
         year = 2007,
        month = feb,
       volume = {462},
       number = {3},
        pages = {1137-1145},
          doi = {10.1051/0004-6361:20066201},
archivePrefix = {arXiv},
       eprint = {1211.1502},
 primaryClass = {astro-ph.SR},
       adsurl = {https://ui.adsabs.harvard.edu/abs/2007A&A...462.1137O}
}

@ARTICLE{Osborne_2019_radynv,
       author = {{Osborne}, Christopher M.~J. and {Armstrong}, John A. and {Fletcher}, Lyndsay},
        title = "{RADYNVERSION: Learning to Invert a Solar Flare Atmosphere with Invertible Neural Networks}",
      journal = {\apj},
         year = 2019,
        month = mar,
       volume = {873},
       number = {2},
          eid = {128},
        pages = {128},
          doi = {10.3847/1538-4357/ab07b4},
archivePrefix = {arXiv},
       eprint = {1901.08626},
 primaryClass = {astro-ph.SR},
       adsurl = {https://ui.adsabs.harvard.edu/abs/2019ApJ...873..128O}
}

@ARTICLE{Okamoto_2018_strong,
       author = {{Okamoto}, Takenori J. and {Sakurai}, Takashi},
        title = "{Super-strong Magnetic Field in Sunspots}",
      journal = {\apjl},
         year = 2018,
        month = jan,
       volume = {852},
       number = {1},
          eid = {L16},
        pages = {L16},
          doi = {10.3847/2041-8213/aaa3d8},
archivePrefix = {arXiv},
       eprint = {1712.08700},
 primaryClass = {astro-ph.SR},
       adsurl = {https://ui.adsabs.harvard.edu/abs/2018ApJ...852L..16O}
}

@ARTICLE{PastorYabar_2019_firtez,
       author = {{Pastor Yabar}, A. and {Borrero}, J.~M. and {Ruiz Cobo}, B.},
        title = "{FIRTEZ-dz. A forward and inverse solver of the polarized radiative transfer equation under Zeeman regime in geometrical scale}",
      journal = {\aap},
         year = 2019,
        month = sep,
       volume = {629},
          eid = {A24},
        pages = {A24},
          doi = {10.1051/0004-6361/201935692},
archivePrefix = {arXiv},
       eprint = {1908.08075},
 primaryClass = {astro-ph.SR},
       adsurl = {https://ui.adsabs.harvard.edu/abs/2019A&A...629A..24P}
}

@inproceedings{paszke2019pytorch,
        pages = {8024--8035},
       editor = {Wallach, H. and Larochelle, H. and Beygelzimer, A. and d'Alch\'{e}-Buc, F. and Fox, E. and Garnett, R.},
author = {Paszke, Adam and Gross, Sam and Massa, Francisco and Lerer, Adam and Bradbury, James and Chanan, Gregory and Killeen, Trevor and Lin, Zeming and Gimelshein, Natalia and Antiga, Luca and Desmaison, Alban and K\"{o}pf, Andreas and Yang, Edward and DeVito, Zach and Raison, Martin and Tejani, Alykhan and Chilamkurthy, Sasank and Steiner, Benoit and Fang, Lu and Bai, Junjie and Chintala, Soumith},
title = {{PyTorch}: an imperative style, high-performance deep learning library},
year = {2019},
publisher = {Curran Associates, Inc.},
address = {Red Hook, NY},
booktitle = {Advances in Neural Information Processing Systems 32 (NeurIPS 2019)},
}

@misc{jax2018github,
         note = {Version 0.3.13},
 howpublished = {\url{https://github.com/jax-ml/jax}},
  author = {James Bradbury and Roy Frostig and Peter Hawkins and Matthew James Johnson and Yash Katariya and Chris Leary and Dougal Maclaurin and George Necula and Adam Paszke and Jake Vander{P}las and Skye Wanderman-{M}ilne and Qiao Zhang},
  title = {{JAX}: composable transformations of {P}ython+{N}um{P}y programs},
  year = {2018},
}

@ARTICLE{Pesnell_2012_SDO,
       author = {{Pesnell}, W. Dean and {Thompson}, B.~J. and {Chamberlin}, P.~C.},
        title = "{The Solar Dynamics Observatory (SDO)}",
      journal = {\solphys},
         year = 2012,
        month = jan,
       volume = {275},
       number = {1-2},
        pages = {3-15},
          doi = {10.1007/s11207-011-9841-3},
       adsurl = {https://ui.adsabs.harvard.edu/abs/2012SoPh..275....3P}
}

@ARTICLE{Peat_2021_inversion,
       author = {{Peat}, A.~W. and {Labrosse}, N. and {Schmieder}, B. and {Barczynski}, K.},
        title = "{Solar prominence diagnostics from non-LTE modelling of Mg II h\&k line profiles}",
      journal = {\aap},
         year = 2021,
        month = sep,
       volume = {653},
          eid = {A5},
        pages = {A5},
          doi = {10.1051/0004-6361/202140907},
archivePrefix = {arXiv},
       eprint = {2106.10351},
 primaryClass = {astro-ph.SR},
       adsurl = {https://ui.adsabs.harvard.edu/abs/2021A&A...653A...5P}
}

@article{Rayleigh1871,
  author = {Rayleigh, Lord},
  title = {On the Light from the Sky, its Polarization and Colour},
  journal = {Philosophical Magazine},
  volume = {41},
  pages = {107--120},
  year = {1871}
}

@INCOLLECTION{Rees_1987nrt,
      address = {Cambridge},
    publisher = {Cambridge University Press},
       author = {{Rees}, David E.},
        title = "{A Gentle Introduction to Polarized Radiative Transfer}",
    booktitle = {Numerical Radiative Transfer},
         year = 1987,
       editor = {{Kalkofen}, Wolfgang},
        pages = {213},
       adsurl = {https://ui.adsabs.harvard.edu/abs/1987nrt..book..213R}
}

@INPROCEEDINGS{Reardon_2023_white,
       author = {{Reardon}, Kevin and {Milic}, Ivan and {da Silva Santos}, Joao and {Diercke}, Andrea and {Fischer}, Catherine and {Hofmann}, Ryan Anthony and {Stauffer}, Johnathan and {Molnar}, Momchil and {Centeno}, Rebecca and {Schad}, Tom and {Uitenbroek}, Han and {Gilly}, Chris R. and {Cauzzi}, Gianna and {Paraschiv}, Alin Razvan and {Yadav}, Rahul and {Crowley}, James and {de Wijn}, Alfred and {West}, Matthew},
        title = "{Spectropolarimetric inversions: Our key to unlocking the secrets of the solar atmosphere}",
    booktitle = {Bulletin of the American Astronomical Society},
         year = 2023,
       volume = {55},
        month = jul,
          eid = {335},
        pages = {335},
          doi = {10.3847/25c2cfeb.ee0486e0},
       adsurl = {https://ui.adsabs.harvard.edu/abs/2023BAAS...55c.335R}
}

@ARTICLE{Rimmele_2011_AO,
       author = {{Rimmele}, Thomas R. and {Marino}, Jose},
        title = "{Solar Adaptive Optics}",
      journal = {Living Reviews in Solar Physics},
         year = 2011,
        month = dec,
       volume = {8},
       number = {1},
          eid = {2},
        pages = {2},
          doi = {10.12942/lrsp-2011-2},
       adsurl = {https://ui.adsabs.harvard.edu/abs/2011LRSP....8....2R}
}

@ARTICLE{Rimmele_2020_DKIST,
       author = {{Rimmele}, Thomas R. and {Warner}, Mark and {Keil}, Stephen L. and {Goode}, Philip R. and {Kn{\"o}lker}, Michael and {Kuhn}, Jeffrey R. and {Rosner}, Robert R. and {McMullin}, Joseph P. and {Casini}, Roberto and {Lin}, Haosheng and {W{\"o}ger}, Friedrich and {von der L{\"u}he}, Oskar and {Tritschler}, Alexandra and {Davey}, Alisdair and {de Wijn}, Alfred and {Elmore}, David F. and {Fehlmann}, Andr{\'e} and {Harrington}, David M. and {Jaeggli}, Sarah A. and {Rast}, Mark P. and {Schad}, Thomas A. and {Schmidt}, Wolfgang and {Mathioudakis}, Mihalis and {Mickey}, Donald L. and {Anan}, Tetsu and {Beck}, Christian and {Marshall}, Heather K. and {Jeffers}, Paul F. and {Oschmann}, Jacobus M. and {Beard}, Andrew and {Berst}, David C. and {Cowan}, Bruce A. and {Craig}, Simon C. and {Cross}, Eric and {Cummings}, Bryan K. and {Donnelly}, Colleen and {de Vanssay}, Jean-Benoit and {Eigenbrot}, Arthur D. and {Ferayorni}, Andrew and {Foster}, Christopher and {Galapon}, Chriselle Ann and {Gedrites}, Christopher and {Gonzales}, Kerry and {Goodrich}, Bret D. and {Gregory}, Brian S. and {Guzman}, Stephanie S. and {Guzzo}, Stephen and {Hegwer}, Steve and {Hubbard}, Robert P. and {Hubbard}, John R. and {Johansson}, Erik M. and {Johnson}, Luke C. and {Liang}, Chen and {Liang}, Mary and {McQuillen}, Isaac and {Mayer}, Christopher and {Newman}, Karl and {Onodera}, Brialyn and {Phelps}, LeEllen and {Puentes}, Myles M. and {Richards}, Christopher and {Rimmele}, Lukas M. and {Sekulic}, Predrag and {Shimko}, Stephan R. and {Simison}, Brett E. and {Smith}, Brett and {Starman}, Erik and {Sueoka}, Stacey R. and {Summers}, Richard T. and {Szabo}, Aimee and {Szabo}, Louis and {Wampler}, Stephen B. and {Williams}, Timothy R. and {White}, Charles},
        title = "{The Daniel K. Inouye Solar Telescope - Observatory Overview}",
      journal = {\solphys},
         year = 2020,
        month = dec,
       volume = {295},
       number = {12},
          eid = {172},
        pages = {172},
          doi = {10.1007/s11207-020-01736-7},
       adsurl = {https://ui.adsabs.harvard.edu/abs/2020SoPh..295..172R}
}

@ARTICLE{RuizCobo_1992_sir,
       author = {{Ruiz Cobo}, B. and {del Toro Iniesta}, J.~C.},
        title = "{Inversion of Stokes Profiles}",
      journal = {\apj},
         year = 1992,
        month = oct,
       volume = {398},
        pages = {375},
          doi = {10.1086/171862},
       adsurl = {https://ui.adsabs.harvard.edu/abs/1992ApJ...398..375R}
}

@ARTICLE{ruizcobo1994_sensitivity,
       author = {{Ruiz Cobo}, B. and {del Toro Iniesta}, J.~C.},
        title = "{On the sensitivity of Stokes profiles to physical quantities.}",
      journal = {\aap},
         year = 1994,
        month = mar,
       volume = {283},
        pages = {129-143},
       adsurl = {https://ui.adsabs.harvard.edu/abs/1994A&A...283..129R}
}

@ARTICLE{ruizcobo_2022_desire,
       author = {{Ruiz Cobo}, B. and {Quintero Noda}, C. and {Gafeira}, R. and {Uitenbroek}, H. and {Orozco Su{\'a}rez}, D. and {P{\'a}ez Ma{\~n}{\'a}}, E.},
        title = "{DeSIRe: Departure coefficient aided Stokes Inversion based on Response functions}",
      journal = {\aap},
         year = 2022,
        month = apr,
       volume = {660},
          eid = {A37},
        pages = {A37},
          doi = {10.1051/0004-6361/202140877},
archivePrefix = {arXiv},
       eprint = {2202.02226},
 primaryClass = {astro-ph.SR},
       adsurl = {https://ui.adsabs.harvard.edu/abs/2022A&A...660A..37R}
}

@ARTICLE{Reardon_2011_ibismagnetic,
       author = {{Reardon}, K.~P. and {Wang}, Y.-M. and {Muglach}, K. and {Warren}, H.~P.},
        title = "{Evidence for Two Separate but Interlaced Components of the Chromospheric Magnetic Field}",
      journal = {\apj},
         year = 2011,
        month = dec,
       volume = {742},
       number = {2},
          eid = {119},
        pages = {119},
          doi = {10.1088/0004-637X/742/2/119},
       adsurl = {https://ui.adsabs.harvard.edu/abs/2011ApJ...742..119R}
}

@BOOK{Rutten_2003_book,
       author = {{Rutten}, Robert J.},
        title = "{Radiative Transfer in Stellar Atmospheres}",
         year = 2003,
         publisher = "Utrecht University lecture notes",
       adsurl = {https://ui.adsabs.harvard.edu/abs/2003rtsa.book.....R}
}

@ARTICLE{Rybicki_1991_ALI,
       author = {{Rybicki}, G.~B. and {Hummer}, D.~G.},
        title = "{An accelerated lambda iteration method for multilevel radiative transfer. I. Non-overlapping lines with background continuum}",
      journal = {\aap},
         year = 1991,
        month = may,
       volume = {245},
        pages = {171-181},
       adsurl = {https://ui.adsabs.harvard.edu/abs/1991A&A...245..171R}
}

@INPROCEEDINGS{Rybicki_1972,
      address = {Boulder, CO},
    publisher = {High Altitude Observatory, NCAR},
       editor = {{Athay}, R.~G. and {House}, L.~L. and {Newkirk}, Jr., G.},
       author = {{Rybicki}, G.~B.},
        title = "{A Novel Approach to the Solution of Multilevel Transfer Problems}",
    booktitle = {Line Formation in the Presence of Magnetic Fields},
         year = 1972,
        month = jan,
        pages = {145},
       adsurl = {https://ui.adsabs.harvard.edu/abs/1972lfpm.conf..145R}
}

@article{RAISSI2019686,
title = {Physics-informed neural networks: A deep learning framework for solving forward and inverse problems involving nonlinear partial differential equations},
journal = {Journal of Computational Physics},
volume = {378},
pages = {686-707},
year = {2019},
issn = {0021-9991},
doi = {https://doi.org/10.1016/j.jcp.2018.10.045},
url = {https://www.sciencedirect.com/science/article/pii/S0021999118307125},
author = {M. Raissi and P. Perdikaris and G.E. Karniadakis}
}

@ARTICLE{Riethmuller_2017_MHD,
       author = {{Riethm{\"u}ller}, T.~L. and {Solanki}, S.~K. and {Barthol}, P. and {Gandorfer}, A. and {Gizon}, L. and {Hirzberger}, J. and {van Noort}, M. and {Blanco Rodr{\'\i}guez}, J. and {Del Toro Iniesta}, J.~C. and {Orozco Su{\'a}rez}, D. and {Schmidt}, W. and {Mart{\'\i}nez Pillet}, V. and {Kn{\"o}lker}, M.},
        title = "{A New MHD-assisted Stokes Inversion Technique}",
      journal = {\apjs},
         year = 2017,
        month = mar,
       volume = {229},
       number = {1},
          eid = {16},
        pages = {16},
          doi = {10.3847/1538-4365/aa5830},
archivePrefix = {arXiv},
       eprint = {1611.05175},
 primaryClass = {astro-ph.SR},
       adsurl = {https://ui.adsabs.harvard.edu/abs/2017ApJS..229...16R}
}

@ARTICLE{Riethmuller_2019_multiline,
       author = {{Riethm{\"u}ller}, T.~L. and {Solanki}, S.~K.},
        title = "{The potential of many-line inversions of photospheric spectropolarimetric data in the visible and near UV}",
      journal = {\aap},
         year = 2019,
        month = feb,
       volume = {622},
          eid = {A36},
        pages = {A36},
          doi = {10.1051/0004-6361/201833379},
archivePrefix = {arXiv},
       eprint = {1812.03757},
 primaryClass = {astro-ph.IM},
       adsurl = {https://ui.adsabs.harvard.edu/abs/2019A&A...622A..36R}
}

@ARTICLE{Scharmer_1985_NLTE,
       author = {{Scharmer}, G.~B. and {Carlsson}, M.},
        title = "{A new approach to multi-level non-LTE radiative transfer problems.}",
      journal = {Journal of Computational Physics},
         year = 1985,
        month = may,
       volume = {59},
        pages = {56-80},
          doi = {10.1016/0021-9991(85)90107-X},
       adsurl = {https://ui.adsabs.harvard.edu/abs/1985JCoPh..59...56S}
}

@INPROCEEDINGS{Scharmer_2003_SST,
       author = {{Scharmer}, Goran B. and {Bjelksjo}, Klas and {Korhonen}, Tapio K. and {Lindberg}, Bo and {Petterson}, Bertil},
        title = "{The 1-meter Swedish solar telescope}",
    booktitle = {Innovative Telescopes and Instrumentation for Solar Astrophysics},
         year = 2003,
       editor = {{Keil}, Stephen L. and {Avakyan}, Sergey V.},
       series = {Society of Photo-Optical Instrumentation Engineers (SPIE) Conference Series},
       volume = {4853},
        month = feb,
        pages = {341-350},
          doi = {10.1117/12.460377},
       adsurl = {https://ui.adsabs.harvard.edu/abs/2003SPIE.4853..341S}
}

@ARTICLE{Scharmer_2008_CRISP,
       author = {{Scharmer}, G.~B. and {Narayan}, G. and {Hillberg}, T. and {de la Cruz Rodriguez}, J. and {L{\"o}fdahl}, M.~G. and {Kiselman}, D. and {S{\"u}tterlin}, P. and {van Noort}, M. and {Lagg}, A.},
        title = "{CRISP Spectropolarimetric Imaging of Penumbral Fine Structure}",
      journal = {\apjl},
         year = 2008,
        month = dec,
       volume = {689},
       number = {1},
        pages = {L69},
          doi = {10.1086/595744},
archivePrefix = {arXiv},
       eprint = {0806.1638},
 primaryClass = {astro-ph},
       adsurl = {https://ui.adsabs.harvard.edu/abs/2008ApJ...689L..69S}
}

@ARTICLE{Schmidt_2012_GREGOR,
       author = {{Schmidt}, W. and {von der L{\"u}he}, O. and {Volkmer}, R. and {Denker}, C. and {Solanki}, S.~K. and {Balthasar}, H. and {Bello Gonz{\'a}lez}, N. and {Berkefeld}, Th. and {Collados}, M. and {Fischer}, A. and {Halbgewachs}, C. and {Heidecke}, F. and {Hofmann}, A. and {Kneer}, F. and {Lagg}, A. and {Nicklas}, H. and {Popow}, E. and {Puschmann}, K.~G. and {Schmidt}, D. and {Sigwarth}, M. and {Sobotka}, M. and {Soltau}, D. and {Staude}, J. and {Strassmeier}, K.~G. and {Waldmann}, T.~A.},
        title = "{The 1.5 meter solar telescope GREGOR}",
      journal = {Astronomische Nachrichten},
         year = 2012,
        month = nov,
       volume = {333},
       number = {9},
        pages = {796-809},
          doi = {10.1002/asna.201211725},
       adsurl = {https://ui.adsabs.harvard.edu/abs/2012AN....333..796S}
}

@ARTICLE{Shchukina_2005_fenlte,
       author = {{Shchukina}, N.~G. and {Trujillo Bueno}, J. and {Asplund}, M.},
        title = "{The Impact of Non-LTE Effects and Granulation Inhomogeneities on the Derived Iron and Oxygen Abundances in Metal-Poor Halo Stars}",
      journal = {\apj},
         year = 2005,
        month = jan,
       volume = {618},
       number = {2},
        pages = {939-952},
          doi = {10.1086/426012},
archivePrefix = {arXiv},
       eprint = {astro-ph/0410475},
 primaryClass = {astro-ph},
       adsurl = {https://ui.adsabs.harvard.edu/abs/2005ApJ...618..939S}
}

@ARTICLE{Skumanich_1987_inv,
       author = {{Skumanich}, A. and {Lites}, B.~W.},
        title = "{Stokes Profile Analysis and Vector Magnetic Fields. I. Inversion of Photospheric Lines}",
      journal = {\apj},
         year = 1987,
        month = nov,
       volume = {322},
        pages = {473},
          doi = {10.1086/165743},
       adsurl = {https://ui.adsabs.harvard.edu/abs/1987ApJ...322..473S}
}

@ARTICLE{SiuTapia_2017_coupled,
       author = {{Siu-Tapia}, A. and {Lagg}, A. and {Solanki}, S.~K. and {van Noort}, M. and {Jur{\v{c}}{\'a}k}, J.},
        title = "{Normal and counter Evershed flows in the photospheric penumbra of a sunspot. SPINOR 2D inversions of Hinode-SOT/SP observations}",
      journal = {\aap},
         year = 2017,
        month = nov,
       volume = {607},
          eid = {A36},
        pages = {A36},
          doi = {10.1051/0004-6361/201730647},
archivePrefix = {arXiv},
       eprint = {1709.07386},
 primaryClass = {astro-ph.SR},
       adsurl = {https://ui.adsabs.harvard.edu/abs/2017A&A...607A..36S}
}

@ARTICLE{Smitha_2020_NLTEFEI,
       author = {{Smitha}, H.~N. and {Holzreuter}, R. and {van Noort}, M. and {Solanki}, S.~K.},
        title = "{The influence of NLTE effects in Fe I lines on an inverted atmosphere. I. 6301 {\r{A}} and 6302 {\r{A}} lines formed in 1D NLTE}",
      journal = {\aap},
         year = 2020,
        month = jan,
       volume = {633},
          eid = {A157},
        pages = {A157},
          doi = {10.1051/0004-6361/201937041},
archivePrefix = {arXiv},
       eprint = {1912.07007},
 primaryClass = {astro-ph.SR},
       adsurl = {https://ui.adsabs.harvard.edu/abs/2020A&A...633A.157S}
}

@ARTICLE{Socas_1998_nlte,
       author = {{Socas-Navarro}, H. and {Ruiz Cobo}, B. and {Trujillo Bueno}, J.},
        title = "{Non-LTE Inversion of Line Profiles}",
      journal = {\apj},
         year = 1998,
        month = nov,
       volume = {507},
       number = {1},
        pages = {470-481},
          doi = {10.1086/306330},
       adsurl = {https://ui.adsabs.harvard.edu/abs/1998ApJ...507..470S}
}

@ARTICLE{Socas_2000_nltezeeman,
       author = {{Socas-Navarro}, H. and {Trujillo Bueno}, J. and {Ruiz Cobo}, B.},
        title = "{Non-LTE Inversion of Stokes Profiles Induced by the Zeeman Effect}",
      journal = {\apj},
         year = 2000,
        month = feb,
       volume = {530},
       number = {2},
        pages = {977-993},
          doi = {10.1086/308414},
       adsurl = {https://ui.adsabs.harvard.edu/abs/2000ApJ...530..977S}
}

@ARTICLE{Stepan_2013_PORTA,
       author = {{{\v{S}}t{\v{e}}p{\'a}n}, Ji{\v{r}}{\'\i} and {Trujillo Bueno}, Javier},
        title = "{PORTA: A three-dimensional multilevel radiative transfer code for modeling the intensity and polarization of spectral lines with massively parallel computers}",
      journal = {\aap},
         year = 2013,
        month = sep,
       volume = {557},
          eid = {A143},
        pages = {A143},
          doi = {10.1051/0004-6361/201321742},
archivePrefix = {arXiv},
       eprint = {1307.4217},
 primaryClass = {astro-ph.SR},
       adsurl = {https://ui.adsabs.harvard.edu/abs/2013A&A...557A.143S}
}

@ARTICLE{Stepan_2016_CaIIscatpol,
       author = {{{\v{S}}t{\v{e}}p{\'a}n}, Ji{\v{r}}{\'\i} and {Trujillo Bueno}, Javier},
        title = "{The Hanle and Zeeman Polarization Signals of the Solar Ca II 8542 {\r{A}} Line}",
      journal = {\apjl},
         year = 2016,
        month = jul,
       volume = {826},
       number = {1},
          eid = {L10},
        pages = {L10},
          doi = {10.3847/2041-8205/826/1/L10},
archivePrefix = {arXiv},
       eprint = {1606.07741},
 primaryClass = {astro-ph.SR},
       adsurl = {https://ui.adsabs.harvard.edu/abs/2016ApJ...826L..10S}
}

@ARTICLE{Stepan_2022_3Dinv,
       author = {{{\v{S}}t{\v{e}}p{\'a}n}, Ji{\v{r}}{\'\i} and {del Pino Alem{\'a}n}, Tanaus{\'u} and {Trujillo Bueno}, Javier},
        title = "{Novel framework for the three-dimensional NLTE inverse problem}",
      journal = {\aap},
         year = 2022,
        month = mar,
       volume = {659},
          eid = {A137},
        pages = {A137},
          doi = {10.1051/0004-6361/202142079},
archivePrefix = {arXiv},
       eprint = {2201.01504},
 primaryClass = {astro-ph.SR},
       adsurl = {https://ui.adsabs.harvard.edu/abs/2022A&A...659A.137S}
}

@ARTICLE{Stepan_2025_3Dnlteerrors,
       author = {{{\v{S}}t{\v{e}}p{\'a}n}, Ji{\v{r}}{\'\i} and {del Pino Alem{\'a}n}, Tanaus{\'u} and {Vicente Ar{\'e}valo}, Andr{\'e}s},
        title = "{The uncertainty of magnetic fields in 3D non-local thermodynamic equilibrium inversions}",
      journal = {\aap},
         year = 2025,
        month = jul,
       volume = {699},
          eid = {A73},
        pages = {A73},
          doi = {10.1051/0004-6361/202554019},
archivePrefix = {arXiv},
       eprint = {2505.09708},
 primaryClass = {astro-ph.SR},
       adsurl = {https://ui.adsabs.harvard.edu/abs/2025A&A...699A..73S}
}

@BOOK{Stenflo_1994_book,
      address = {Dordrecht},
       series = {Astrophysics and Space Science Library},
       author = {{Stenflo}, Jan},
        title = "{Solar Magnetic Fields: Polarized Radiation Diagnostics}",
         year = 1994,
       volume = {189},
       publisher = {Kluwer Academic Publishers},
          doi = {10.1007/978-94-015-8246-9},
       adsurl = {https://ui.adsabs.harvard.edu/abs/1994ASSL..189.....S}
}

@ARTICLE{Solanki_2010_sunrise,
       author = {{Solanki}, S.~K. and {Barthol}, P. and {Danilovic}, S. and {Feller}, A. and {Gandorfer}, A. and {Hirzberger}, J. and {Riethm{\"u}ller}, T.~L. and {Sch{\"u}ssler}, M. and {Bonet}, J.~A. and {Mart{\'\i}nez Pillet}, V. and {del Toro Iniesta}, J.~C. and {Domingo}, V. and {Palacios}, J. and {Kn{\"o}lker}, M. and {Bello Gonz{\'a}lez}, N. and {Berkefeld}, T. and {Franz}, M. and {Schmidt}, W. and {Title}, A.~M.},
        title = "{SUNRISE: Instrument, Mission, Data, and First Results}",
      journal = {\apjl},
         year = 2010,
        month = nov,
       volume = {723},
       number = {2},
        pages = {L127-L133},
          doi = {10.1088/2041-8205/723/2/L127},
archivePrefix = {arXiv},
       eprint = {1008.3460},
 primaryClass = {astro-ph.SR},
       adsurl = {https://ui.adsabs.harvard.edu/abs/2010ApJ...723L.127S}
}

@ARTICLE{Temmer_2021_review,
       author = {{Temmer}, Manuela},
        title = "{Space weather: the solar perspective: An update to Schwenn (2006)}",
      journal = {Living Reviews in Solar Physics},
         year = 2021,
        month = dec,
       volume = {18},
       number = {1},
          eid = {4},
        pages = {4},
          doi = {10.1007/s41116-021-00030-3},
archivePrefix = {arXiv},
       eprint = {2104.04261},
 primaryClass = {astro-ph.SR},
       adsurl = {https://ui.adsabs.harvard.edu/abs/2021LRSP...18....4T}
}

@INPROCEEDINGS{Tziotziou_2007_rev,
       author = {{Tziotziou}, K.},
        title = "{Chromospheric Cloud-Model Inversion Techniques}",
    booktitle = {The Physics of Chromospheric Plasmas},
         year = 2007,
       editor = {{Heinzel}, P. and {Dorotovi{\v{c}}}, I. and {Rutten}, R.~J.},
       series = {Astronomical Society of the Pacific Conference Series},
       volume = {368},
        month = may,
        pages = {217},
          doi = {10.48550/arXiv.0704.1558},
archivePrefix = {arXiv},
       eprint = {0704.1558},
 primaryClass = {astro-ph},
       adsurl = {https://ui.adsabs.harvard.edu/abs/2007ASPC..368..217T}
}

@ARTICLE{TBFB_1995_GS,
       author = {{Trujillo Bueno}, J. and {Fabiani Bendicho}, P.},
        title = "{A Novel Iterative Scheme for the Very Fast and Accurate Solution of Non-LTE Radiative Transfer Problems}",
      journal = {\apj},
         year = 1995,
        month = dec,
       volume = {455},
        pages = {646},
          doi = {10.1086/176612},
       adsurl = {https://ui.adsabs.harvard.edu/abs/1995ApJ...455..646T}
}

@INPROCEEDINGS{TrujilloBueno_2003_review,
       author = {{Trujillo Bueno}, J.},
        title = "{The Generation and Transfer of Polarized Radiation in Stellar Atmospheres}",
    booktitle = {Stellar Atmosphere Modeling},
         year = 2003,
       editor = {{Hubeny}, Ivan and {Mihalas}, Dimitri and {Werner}, Klaus},
       series = {Astronomical Society of the Pacific Conference Series},
       volume = {288},
        month = jan,
        pages = {551},
       adsurl = {https://ui.adsabs.harvard.edu/abs/2003ASPC..288..551T}
}

@INPROCEEDINGS{TrujilloBueno_2006_gentle,
       author = {{Trujillo Bueno}, Javier},
        title = "{A Gentle Introduction to the Physics of Spectral Line Polarization}",
    booktitle = {Solar Physics and Solar Eclipses (SPSE 2006)},
         year = 2006,
       editor = {{Ramelli}, Renzo and {Shalabiea}, Osama and {Saleh}, Ibrahim and {Stenflo}, Jan O.},
        month = jan,
        pages = {77-92},
       adsurl = {https://ui.adsabs.harvard.edu/abs/2006spse.conf...77T}
}

@inproceedings{TrujilloBueno2009_2nd,
  author       = {Trujillo Bueno, J.},
  title        = {The Second Stellar Spectrum and the Non-LTE Problem of the Second Kind},
  booktitle    = {Solar Polarization 5},
  editor       = {Sampoorna, M. and Nagendra, K. N.},
  series       = {AIP Conference Proceedings},
  volume       = {1171},
  pages        = {27--42},
  year         = {2009},
  doi          = {10.1063/1.3266679},
  eprint       = {0911.4669},
  archivePrefix= {arXiv}
}

@ARTICLE{TB_2022ARA&A_review,
       author = {{Trujillo Bueno}, J. and {del Pino Alem{\'a}n}, T.},
        title = "{Magnetic Field Diagnostics in the Solar Upper Atmosphere}",
      journal = {\araa},
         year = 2022,
        month = aug,
       volume = {60},
        pages = {415-453},
          doi = {10.1146/annurev-astro-041122-031043},
       adsurl = {https://ui.adsabs.harvard.edu/abs/2022ARA&A..60..415T}
}

@ARTICLE{Trelles_2021_1.5mic,
       author = {{Trelles Arjona}, J.~C. and {Ruiz Cobo}, B. and {Mart{\'\i}nez Gonz{\'a}lez}, M.~J.},
        title = "{Empirical determination of atomic line parameters of the 1.5 {\ensuremath{\mu}}m spectral region}",
      journal = {\aap},
         year = 2021,
        month = apr,
       volume = {648},
          eid = {A68},
        pages = {A68},
          doi = {10.1051/0004-6361/202038941},
archivePrefix = {arXiv},
       eprint = {2103.04160},
 primaryClass = {astro-ph.SR},
       adsurl = {https://ui.adsabs.harvard.edu/abs/2021A&A...648A..68T}
}

@ARTICLE{Tsuneta_2008_Hinode,
       author = {{Tsuneta}, S. and {Ichimoto}, K. and {Katsukawa}, Y. and {Nagata}, S. and {Otsubo}, M. and {Shimizu}, T. and {Suematsu}, Y. and {Nakagiri}, M. and {Noguchi}, M. and {Tarbell}, T. and {Title}, A. and {Shine}, R. and {Rosenberg}, W. and {Hoffmann}, C. and {Jurcevich}, B. and {Kushner}, G. and {Levay}, M. and {Lites}, B. and {Elmore}, D. and {Matsushita}, T. and {Kawaguchi}, N. and {Saito}, H. and {Mikami}, I. and {Hill}, L.~D. and {Owens}, J.~K.},
        title = "{The Solar Optical Telescope for the Hinode Mission: An Overview}",
      journal = {\solphys},
         year = 2008,
        month = jun,
       volume = {249},
       number = {2},
        pages = {167-196},
          doi = {10.1007/s11207-008-9174-z},
archivePrefix = {arXiv},
       eprint = {0711.1715},
 primaryClass = {astro-ph},
       adsurl = {https://ui.adsabs.harvard.edu/abs/2008SoPh..249..167T}
}

@ARTICLE{Uitenbroek_2001_RH,
       author = {{Uitenbroek}, H.},
        title = "{Multilevel Radiative Transfer with Partial Frequency Redistribution}",
      journal = {\apj},
         year = 2001,
        month = aug,
       volume = {557},
       number = {1},
        pages = {389-398},
          doi = {10.1086/321659},
       adsurl = {https://ui.adsabs.harvard.edu/abs/2001ApJ...557..389U}
}

@ARTICLE{vanNoort_2013_penumbra,
       author = {{van Noort}, M. and {Lagg}, A. and {Tiwari}, S.~K. and {Solanki}, S.~K.},
        title = "{Peripheral downflows in sunspot penumbrae}",
      journal = {\aap},
         year = 2013,
        month = sep,
       volume = {557},
          eid = {A24},
        pages = {A24},
          doi = {10.1051/0004-6361/201321073},
archivePrefix = {arXiv},
       eprint = {1308.0466},
 primaryClass = {astro-ph.SR},
       adsurl = {https://ui.adsabs.harvard.edu/abs/2013A&A...557A..24V}
}

@ARTICLE{vanNoort2022MiHI3,
       author = {{van Noort}, M. and {Doerr}, H. -P.},
        title = "{Data reduction and restoration of spectropolarimetric microlensed hyperspectral imager data}",
      journal = {\aap},
         year = 2022,
        month = dec,
       volume = {668},
          eid = {A151},
        pages = {A151},
          doi = {10.1051/0004-6361/202243466},
       adsurl = {https://ui.adsabs.harvard.edu/abs/2022A&A...668A.151V}
}

@ARTICLE{vanNoort2022A&AMihiInstrument,
       author = {{van Noort}, M. and {Bischoff}, J. and {Kramer}, A. and {Solanki}, S.~K. and {Kiselman}, D.},
        title = "{A prototype of a microlensed hyperspectral imager for solar observations}",
      journal = {\aap},
         year = 2022,
        month = dec,
       volume = {668},
          eid = {A149},
        pages = {A149},
          doi = {10.1051/0004-6361/202243464},
       adsurl = {https://ui.adsabs.harvard.edu/abs/2022A&A...668A.149V}
}

@ARTICLE{vanNoort_2005_momfbd,
       author = {{van Noort}, Michiel and {Rouppe Van Der Voort}, Luc and {L{\"o}fdahl}, Mats G.},
        title = "{Solar Image Restoration By Use Of Multi-frame Blind De-convolution With Multiple Objects And Phase Diversity}",
      journal = {\solphys},
         year = 2005,
        month = may,
       volume = {228},
       number = {1-2},
        pages = {191-215},
          doi = {10.1007/s11207-005-5782-z},
       adsurl = {https://ui.adsabs.harvard.edu/abs/2005SoPh..228..191V}
}

@ARTICLE{vanNoort_2012_spatiallyc,
       author = {{van Noort}, M.},
        title = "{Spatially coupled inversion of spectro-polarimetric image data. I. Method and first results}",
      journal = {\aap},
         year = 2012,
        month = dec,
       volume = {548},
          eid = {A5},
        pages = {A5},
          doi = {10.1051/0004-6361/201220220},
archivePrefix = {arXiv},
       eprint = {1210.4636},
 primaryClass = {astro-ph.IM},
       adsurl = {https://ui.adsabs.harvard.edu/abs/2012A&A...548A...5V}
}

@ARTICLE{vannoort_2017_specrest,
       author = {{van Noort}, M.},
        title = "{Image restoration of solar spectra}",
      journal = {\aap},
         year = 2017,
        month = dec,
       volume = {608},
          eid = {A76},
        pages = {A76},
          doi = {10.1051/0004-6361/201731339},
archivePrefix = {arXiv},
       eprint = {1711.09629},
 primaryClass = {astro-ph.IM},
       adsurl = {https://ui.adsabs.harvard.edu/abs/2017A&A...608A..76V}
}

@ARTICLE{Vukadinovic_2024_globin,
       author = {{Vukadinovi{\'c}}, D. and {Smitha}, H.~N. and {Korpi-Lagg}, A. and {van Noort}, M. and {Castellanos Dur{\'a}n}, J.~S. and {Solanki}, S.~K.},
        title = "{globin: A spectropolarimetric inversion code for the coupled inference of atomic line parameters}",
      journal = {\aap},
         year = 2024,
        month = jun,
       volume = {686},
          eid = {A262},
        pages = {A262},
          doi = {10.1051/0004-6361/202347752},
archivePrefix = {arXiv},
       eprint = {2404.03291},
 primaryClass = {astro-ph.SR},
       adsurl = {https://ui.adsabs.harvard.edu/abs/2024A&A...686A.262V}
}

@ARTICLE{Zeeman_1897_effect,
       author = {{Zeeman}, P.},
        title = "{On the Influence of Magnetism on the Nature of the Light Emitted by a Substance.}",
      journal = {\apj},
         year = 1897,
        month = may,
       volume = {5},
        pages = {332},
          doi = {10.1086/140355},
       adsurl = {https://ui.adsabs.harvard.edu/abs/1897ApJ.....5..332Z}
}

@ARTICLE{Zeuner_2018_fsp,
       author = {{Zeuner}, F. and {Feller}, A. and {Iglesias}, F.~A. and {Solanki}, S.~K.},
        title = "{Detection of spatially structured scattering polarization of Sr I 4607.3 {\r{A}} with the Fast Solar Polarimeter}",
      journal = {\aap},
         year = 2018,
        month = nov,
       volume = {619},
          eid = {A179},
        pages = {A179},
          doi = {10.1051/0004-6361/201833241},
archivePrefix = {arXiv},
       eprint = {1808.06539},
 primaryClass = {astro-ph.SR},
       adsurl = {https://ui.adsabs.harvard.edu/abs/2018A&A...619A.179Z}
}

@ARTICLE{Zeuner_2020ApJ_fspII,
       author = {{Zeuner}, Franziska and {Manso Sainz}, Rafael and {Feller}, Alex and {van Noort}, Michiel and {Solanki}, Sami K. and {Iglesias}, Francisco A. and {Reardon}, Kevin and {Mart{\'\i}nez Pillet}, Valent{\'\i}n},
        title = "{Solar Disk Center Shows Scattering Polarization in the Sr I 4607 {\r{A}} Line}",
      journal = {\apjl},
         year = 2020,
        month = apr,
       volume = {893},
       number = {2},
          eid = {L44},
        pages = {L44},
          doi = {10.3847/2041-8213/ab86b8},
archivePrefix = {arXiv},
       eprint = {2004.03679},
 primaryClass = {astro-ph.SR},
       adsurl = {https://ui.adsabs.harvard.edu/abs/2020ApJ...893L..44Z}
}

@ARTICLE{Zeuner_2025_DKIST,
       author = {{Zeuner}, Franziska and {Belluzzi}, Luca and {Alsina Ballester}, Ernest and {Casini}, Roberto and {Harrington}, David M. and {del Pino Alem{\'a}n}, Tanaus{\'u} and {Trujillo Bueno}, Javier},
        title = "{DKIST resolves sub-arcsec photospheric scattering polarization}",
      journal = {\aap},
         year = 2025,
        month = nov,
       volume = {703},
          eid = {L10},
        pages = {L10},
          doi = {10.1051/0004-6361/202557465},
archivePrefix = {arXiv},
       eprint = {2510.20676},
 primaryClass = {astro-ph.SR},
       adsurl = {https://ui.adsabs.harvard.edu/abs/2025A&A...703L..10Z}
}

@ARTICLE{Yadav_2022_radiative,
       author = {{Yadav}, R. and {de la Cruz Rodr{\'\i}guez}, J. and {Kerr}, G.~S. and {D{\'\i}az Baso}, C.~J. and {Leenaarts}, J.},
        title = "{Radiative losses in the chromosphere during a C-class flare}",
      journal = {\aap},
         year = 2022,
        month = sep,
       volume = {665},
          eid = {A50},
        pages = {A50},
          doi = {10.1051/0004-6361/202243440},
archivePrefix = {arXiv},
       eprint = {2207.02840},
 primaryClass = {astro-ph.SR},
       adsurl = {https://ui.adsabs.harvard.edu/abs/2022A&A...665A..50Y}
}

@ARTICLE{2025A&A...701A..80J,
       author = {{Janett}, Gioele and {Mili{\'c}}, Ivan and {Riva}, Fabio and {Belluzzi}, Luca},
        title = "{An inversion approach to retrieve the vector magnetic field from scattering polarization in strong solar resonance lines}",
      journal = {\aap},
         year = 2025,
        month = sep,
       volume = {701},
          eid = {A80},
        pages = {A80},
          doi = {10.1051/0004-6361/202555911},
       adsurl = {https://ui.adsabs.harvard.edu/abs/2025A&A...701A..80J}
}

@ARTICLE{2024ApJ...976...21D,
       author = {{da Silva Santos}, J.~M. and {Molnar}, M. and {Mili{\'c}}, I. and {Rempel}, M. and {Reardon}, K. and {de la Cruz Rodr{\'\i}guez}, J.},
        title = "{Constraints on Acoustic Wave Energy Fluxes and Radiative Losses in the Solar Chromosphere from Non-LTE Inversions}",
      journal = {\apj},
         year = 2024,
        month = nov,
       volume = {976},
       number = {1},
          eid = {21},
        pages = {21},
          doi = {10.3847/1538-4357/ad81d4},
archivePrefix = {arXiv},
       eprint = {2408.15908},
 primaryClass = {astro-ph.SR},
       adsurl = {https://ui.adsabs.harvard.edu/abs/2024ApJ...976...21D}
}

@ARTICLE{2022A&A...664A.182V,
       author = {{Vukadinovi{\'c}}, D. and {Mili{\'c}}, I. and {Atanackovi{\'c}}, O.},
        title = "{Investigating magnetic field inference from the spectral region around the Mg I b$_{2}$ line using the weak-field approximation}",
      journal = {\aap},
         year = 2022,
        month = aug,
       volume = {664},
          eid = {A182},
        pages = {A182},
          doi = {10.1051/0004-6361/202142015},
archivePrefix = {arXiv},
       eprint = {2205.04236},
 primaryClass = {astro-ph.SR},
       adsurl = {https://ui.adsabs.harvard.edu/abs/2022A&A...664A.182V}
}

@INPROCEEDINGS{2019ASPC..526..179M,
       author = {{Mili{\'c}}, I. and {van Noort}, M.},
        title = "{Response Functions for NLTE Lines}",
    booktitle = {Solar Polarization Workshop 8},
         year = 2019,
       editor = {{Belluzzi}, L. and {Casini}, R. and {Romoli}, M. and {Trujillo Bueno}, J.},
       series = {Astronomical Society of the Pacific Conference Series},
       volume = {526},
        month = may,
        pages = {179},
       adsurl = {https://ui.adsabs.harvard.edu/abs/2019ASPC..526..179M}
}

@ARTICLE{2018A&A...617A..24M,
       author = {{Mili{\'c}}, I. and {van Noort}, M.},
        title = "{Spectropolarimetric NLTE inversion code SNAPI}",
      journal = {\aap},
         year = 2018,
        month = sep,
       volume = {617},
          eid = {A24},
        pages = {A24},
          doi = {10.1051/0004-6361/201833382},
archivePrefix = {arXiv},
       eprint = {1806.08134},
 primaryClass = {astro-ph.SR},
       adsurl = {https://ui.adsabs.harvard.edu/abs/2018A&A...617A..24M}
}

@ARTICLE{2017A&A...601A.100M,
       author = {{Mili{\'c}}, I. and {van Noort}, M.},
        title = "{Line response functions in nonlocal thermodynamic equilibrium. Isotropic case}",
      journal = {\aap},
         year = 2017,
        month = may,
       volume = {601},
          eid = {A100},
        pages = {A100},
          doi = {10.1051/0004-6361/201629980},
       adsurl = {https://ui.adsabs.harvard.edu/abs/2017A&A...601A.100M}
}

@ARTICLE{2012A&A...547A..38M,
       author = {{Mili{\'c}}, I. and {Faurobert}, M.},
        title = "{Hanle diagnostics of weak solar magnetic fields: . Inversion of scattering polarization in C$_{2}$ and MgH molecular lines}",
      journal = {\aap},
         year = 2012,
        month = nov,
       volume = {547},
          eid = {A38},
        pages = {A38},
          doi = {10.1051/0004-6361/201219737},
       adsurl = {https://ui.adsabs.harvard.edu/abs/2012A&A...547A..38M}
}
}



\clearpage

{\ }

\clearpage

{\ }

\newpage

\begin{strip}

{\ }



\naslov{Spektropolarimetrijska dijagnostika donje sunqeve atmosfere: pregled tehnika za inverziju}

\authors{I. Mili\'{c}$^{1,2,3}$}

\vskip3mm

\address{$^1$Institute for Solar Physics (KIS), Georges-K\"{o}hler-Allee 401A, 79110, Freiburg im Breisgau, Germany}

\address{$^2$Department of Astronomy, Faculty of Mathematics,
University of Belgrade\break Studentski trg 16, 11000 Belgrade,
Serbia}

\address{$^3$Astronomical Observatory, Volgina 7, 11060 Belgrade 38, Serbia}

\vskip3mm
\address{E--mail: milic@leibniz-kis.de}


\centerline{{\rrm UDK} \udc}


\vskip1mm


\vskip.7cm

\baselineskip=3.8truemm

\begin{multicols}{2}

{
\rrm

Sunce, naxa matiqna zvezda, ima fundamentalnu ulogu u astrofizici, pre svega zbog svoje blizine. Da bismo razumeli procese koji se dogadjaju u spoljnim slojevima Sunca, ali takodje, indirektno, i u njegovoj unutraxnjosti, vrximo posmatranja visoke prostorne, vremenske i spektralne rezolucije, koja su fokusirana na dijagnostiqki va{\z}ne spektralne linije. Intenzitet i polarizacija elektromagnetnog zraqenja u spektralnim linijama su rezultat zajedniqkog delovanja procesa apsorpcije, emisije i rasejanja, a zatim i Zemanovog i Hanleovog efekta. Spektropolarimetrijske inverzije su metode koje su razvijene za interpretaciju oblika i polarizacije posmatranih spektralnih linija i rekonstrukciju strukture posmatrane Sunqeve atmosfere. 

Cilj ovog rada je da pru{\z}i jedan pedagoxki pregled neophodnih tehnika iz teorije prenosa zraqenja i modernih metoda za dijagnostiku, i da na taj naqin pomogne qitaocu da udje u ovu oblast istra{\z}ivanja i iskoristi izlo{\z}ene koncepte za samostalno istra{\z}ivanje. Drugi cilj ovog rada je da predstavi jedan kritiqan pogled na nove pravce istra{\z}ivanja i budu{\cc}i razvoj spektropolarimetrijskih inverzija. Pre svega, da prodiskutuje tenziju izmedju fiziqke realistiqnosti, numeriqke efikasnosti i uzimanja u obzir instrumentalnih efekata.} 


\end{multicols}


\end{strip}


\end{document}